\documentclass[prd,aps,twocolumn,floatfix,superscriptaddress,nofootinbib]{revtex4-1}

\usepackage[utf8]{inputenc}
\usepackage{amsmath, amsthm, amsfonts, amssymb, latexsym}
\usepackage{graphicx}
\usepackage{color}
\usepackage{subfigure}
\usepackage{url}
\usepackage{psfrag}
\usepackage{mathrsfs}
\usepackage{multirow}
\usepackage[normalem]{ulem}
\usepackage{tikz} 
\usepackage{flushend}
\usepackage{mathtools}
\usepackage{slashed}

\usepackage{enumitem}
\setdescription{font=\normalfont}

\def\p{\partial}

\DeclareMathAlphabet\mathbfcal{OMS}{cmsy}{b}{n}

\usepackage{array}
\newcommand{\PreserveBackslash}[1]{\let\temp=\\#1\let\\=\temp}
\newcolumntype{C}[1]{>{\PreserveBackslash\centering}p{#1}}
\newcolumntype{R}[1]{>{\PreserveBackslash\raggedleft}p{#1}}
\newcolumntype{L}[1]{>{\PreserveBackslash\raggedright}p{#1}}

\def\hlinewd#1{%
\noalign{\ifnum0=`}\fi\hrule \@height #1 %
\futurelet\reserved@a\@xhline}

\usepackage[unicode]{hyperref}
\hypersetup{
    unicode=true,          
    pdftitle={model_CCE_CCM},
    pdfcreator={LaTeX},          
    pdfproducer={LaTeX},         
    pdfkeywords={},
    colorlinks=true,            
    linkcolor  = cyan!70!black,
    citecolor  = cyan!70!black,
    urlcolor   = cyan!70!black,
}

\DeclareSymbolFont{matha}{OML}{txmi}{m}{it}
\DeclareMathSymbol{\varv}{\mathord}{matha}{118}

\graphicspath{ {./images/} }

\begin{document}

\title{Numerical convergence of model Cauchy-characteristic extraction
  and matching}

\author{Thanasis Giannakopoulos}
\affiliation{
  School of Mathematical Sciences, University of Nottingham,
  University Park, Nottingham NG7 2RD, United Kingdom}
\affiliation{
  Nottingham Centre of Gravity, University of Nottingham,
  University Park, Nottingham, NG7 2RD, United Kingdom}

\author{Nigel T. Bishop}
\affiliation{
  Department of Mathematics, Rhodes University,
  Grahamstown 6140, South Africa
}

\author{David Hilditch}
\affiliation{
  Centro de Astrof\'{\i}sica e Gravita\c c\~ao -- CENTRA,
  Departamento de F\'{\i}sica, Instituto Superior T\'ecnico -- IST,
  Universidade de Lisboa -- UL, Av.\ Rovisco Pais 1, 1049-001 Lisboa,
  Portugal}

\author{Denis Pollney}
\affiliation{
  Department of Mathematics, Rhodes University,
  Grahamstown 6140, South Africa
}

\author{Miguel Zilh\~ao}
\affiliation{
  Departamento de Matem\'atica da Universidade de Aveiro and Centre
  for Research and Development in Mathematics and Applications
  (CIDMA), Campus de Santiago, 3810-183 Aveiro, Portugal
}

\begin{abstract}
  Gravitational waves provide a powerful enhancement to our
  understanding of fundamental physics. To make the most of their
  detection we need to accurately model the entire process of their
  emission and propagation toward
  interferometers. Cauchy-characteristic extraction and matching are
  methods to compute gravitational waves at null infinity, a
  mathematical idealization of detector location, from numerical
  relativity simulations. Both methods can in principle contribute to
  modeling by providing highly accurate gravitational waveforms. An
  underappreciated subtlety in realizing this potential is posed by
  the (mere) weak hyperbolicity of the particular PDE systems solved
  in the characteristic formulation of the Einstein field
  equations. This shortcoming results from the popular choice of
  Bondi-like coordinates. So motivated, we construct toy models that
  capture that PDE structure and study Cauchy-characteristic
  extraction and matching with them. Where possible we provide energy
  estimates for their solutions and perform careful numerical norm
  convergence tests to demonstrate the effect of weak hyperbolicity on
  Cauchy-characteristic extraction and matching. Our findings strongly
  indicate that, as currently formulated, Cauchy-characteristic
  matching for the Einstein field equations would provide solutions
  that are, at best, convergent at an order lower than expected for
  the numerical method, and may be unstable. In contrast, under
  certain conditions, the extraction method can provide properly
  convergent solutions. Establishing however that these conditions
  hold for the aforementioned characteristic formulations is still an
  open problem.
\end{abstract}

\maketitle

\section{Introduction} \label{Section:Intro}

\begin{figure}[!t]
  \includegraphics[width=0.4\textwidth]{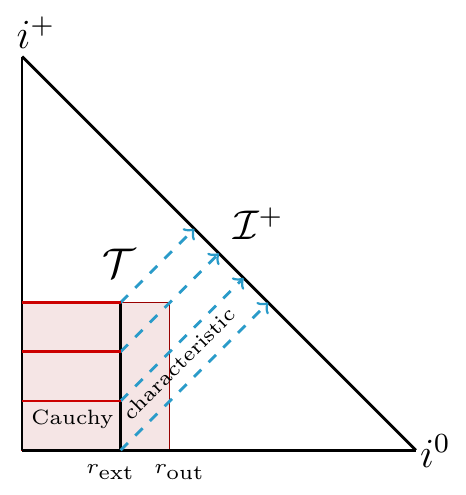}
  \caption{The conformal diagram of part of Minkowski spacetime, where
    spatial~$i^0$, future temporal~$i^+$ and future
    null~$\mathcal{I}^+$ infinities are shown. In gravitational
    waveform modeling, the detector is assumed to be
    at~$\mathcal{I}^+$. As is common in this context, we abuse the
    terminology ``Cauchy'' to denote the IBVP for a domain foliated by
    spacelike hypersurfaces that do not extend to~$i^0$, but are
    truncated to some radius~$r_*$, with~$r_*=r_{\textrm{out}}$ for
    CCE and~$r_*=r_{\textrm{ext}}$ for CCM. By ``characteristic'' we
    refer to the CIBVP for GR. For Cauchy-characteristic extraction
    (CCE) and matching (CCM) information between the Cauchy and
    characteristic domains is communicated via the
    worldtube~$\mathcal{T}$.}
  \label{Fig:CCE_CCM_GR}
\end{figure}

The use of present and improved future gravitational wave (GW)
detectors such as the ground-based advanced LIGO and
Virgo~\cite{AasAlt14ax, AceAlt14}, Kagra \cite{Aku20}, the Cosmic
Explorer~\cite{DwySigBal15}, and the Einstein Telescope~\cite{Mag19},
as well as the space-borne detectors like LISA~\cite{Ama17},
Taiji~\cite{RuaGuoCai20}, and TianQin~\cite{Luo16}, is already
deepening and will continue to deepen our understanding of gravity. To
harness their power, gravitational waveforms of high fidelity are
necessary for the parameter estimation
process~\cite{FouLagLov22}. These waveform models are used to compare
against observational data, and thus to infer the underlying
properties of the sources.

In the modeling process, a GW detector is typically assumed to lie
infinitely far away from the source in an asymptotically flat
spacetime. After emission, GWs propagate toward future null infinity
at the speed of light where they can be detected. There are different
methods to calculate the detected gravitational radiation;
see~\cite{BisRez16} for a review. A widely used technique is the
extrapolation of the Weyl scalar~$\psi_4$ to null infinity, which, in
suitable coordinates~\cite{DuaFenGas22}, can be used to predict the
effects of GWs on a detector. For this method, the initial boundary
value problem (IBVP) that results in the production of GWs is solved,
in a bounded numerical domain with outer
boundary~$r_\textrm{out}$. The value of~$\psi_4$ is then calculated on
worldtubes of different radii, and consequently used to fit a
polynomial expansion, which is finally taken to calculate~$\psi_4$ at
the limit~$r \rightarrow \infty$. This approach however is susceptible
to systematic extrapolation errors, which can be avoided using the
Cauchy-characteristic extraction (CCE) method~\cite{BisGomLeh96a,
  BisGomHol97, BisGomLeh97a, ZloGomHus03, GomBarFri07, BabBisSzi09,
  ReiBisPol09, ReiBisPol09a, Win12, HanSzi14, BarMoxSch19,
  MoxSchTeu20, Ioz21, Mit21, IozBoyDep21, MitMoxSch21, Fou21,
  MoxSchTeu21}. CCE, as illustrated in Fig.~\ref{Fig:CCE_CCM_GR},
allows one to calculate the gravitational radiation at future null
infinity~$\mathcal{I}^+$ directly. In CCE, information from the Cauchy
domain is passed through a worldtube~$\mathcal{T}$ to the
characteristic one, and propagated to~$\mathcal{I}^+$.

Since the evolution of the Cauchy and the characteristic setups are
decoupled, this method can still be affected by errors related to
artificial boundary conditions imposed on the outer boundary of the
Cauchy domain. One way to control these errors is to enlarge the
Cauchy domain, such that the artificial boundary conditions
on~$r_\textrm{out}$ do not affect the solution within a smaller part
of the domain, specifically up to the worldtube~$r_\textrm{ext}$. The
characteristic setup is then used to extract the solution
from~$r_{\textrm{ext}}$ to~$\mathcal{I}^+$. However, the initial data
provided on the initial null hypersurface for the characteristic
initial boundary value problem (CIBVP) very often are not compatible
with the IBVP solution outside~$r_{\textrm{ext}}$, so there is still a
type of error that has not been removed~\cite{TayBoyRei13}.

A way to avoid this type of error as well is to evolve the Cauchy and
the characteristic setups simultaneously, and allow information to
flow via~$\mathcal{T}$ both ways. This approach is often called
Cauchy-characteristic matching (CCM) and can in principle provide
gravitational waveforms of high fidelity~\cite{BisGomHol96a,
  BisGomIsa98, Win12}. Recently, there has been an important effort to
improve and optimize characteristic codes that can perform CCE and
CCM~\cite{MoxSchTeu20, MoxSchTeu21}. However, novel results regarding
the hyperbolicity of general relativity (GR) in the characteristic
setup~\cite{GiaHilZil20, GiaBisHil21} pose pressing questions as to
whether CCE and CCM as currently performed for GR can indeed deliver
on their promise to produce highly accurate waveforms.

Within the framework of numerical relativity, the CIBVP for the
Einstein field equations (EFE) is typically constructed using a
Bondi-like coordinate system~\cite{BonBurMet62a, Win12, CaoHe13}. In
CCE and CCM, the resulting Bondi-like CIBVP is combined with the IBVP
with a strongly or symmetric hyperbolic formulation of GR, such as the
generalized harmonic formulation~\cite{Pre05, LinSchKid05,
  LinMatRin07}. Characteristic codes for GR based on a Bondi-like
formulation have been widely used in numerical simulations. For
instance, the first long-term evolution of a single black hole by the
Binary Black Hole Grand Challenge Alliance~\cite{GomLehMar98a}, was
achieved in such a setup. Many more codes have solved a Bondi-like
CIBVP providing, for instance studies of relativistic
stars~\cite{PapFon99, SieFonMul02}, and gravitational
collapse~\cite{Gar95, CreOliWin19, GunBauHil19, SieFonMue03,
  AlcBarOli21, OliSop16, Oli17b}. This success has led to the belief
that these Bondi-like CIBVPs are well posed.

A partial differential equation (PDE) problem is well posed if it has
a unique solution that depends continuously on the given data in an
appropriate norm. Let us for example consider the first order, linear,
constant coefficient system with real-valued variables and
coefficients
\begin{align}
  \p_t \mathbf{u} = \mathbf{B}^p \p_p \mathbf{u} + \mathbf{B} \mathbf{u}
  \,,
  \label{eq:ref_pde}
\end{align}
where~$\mathbf{u}$ is the state vector,~$\p_p$ denote spatial
derivatives,~$\mathbf{B}^p$ are the principal matrices and the
product~$\mathbf{B} \mathbf{u}$ denotes source terms. 
This system is weakly hyperbolic (WH) if the principal
symbol~$\mathbf{P}^s \equiv \mathbf{B}^p \mathbf{s}_p$ has real
eigenvalues for all unit spatial covectors~$\mathbf{s}_p$, and is
called strongly hyperbolic (SH) if furthermore~$\mathbf{P}^s$ is
uniformly diagonalizable for all~$\mathbf{s}^p$. In addition, if all
the principal matrices~$\mathbf{B}^p$ can be brought simultaneously in
a symmetric form, then the system is called symmetric hyperbolic
(SYMH).

Let us now consider the initial value problem (IVP) for the above
system. Continuous dependence of the solution~$\mathbf{u}$ at all
times~$t$ on the initial data~$f$ can be understood as being able to
provide an estimate of the form
\begin{align}
  ||\mathbf{u}(\cdot ,t)|| \leq K e^{\alpha t} ||f||
  \,,
  \label{ineq:contin_dep}
\end{align}
with real constants~$K \geq 1$ and~$\alpha \in \mathbb{R}$,
and~$||\cdot||$ denoting a suitable norm. The degree of hyperbolicity
of the PDE system determines well-posedness of the problem, with the
IVP for SH systems being (strongly) well posed in the~$L^2$ norm of
the system. On the contrary, the IVP of a system that is only WH but
not SH, is ill posed in the~$L^2$ norm. However, under certain
conditions it may be weakly well posed in a different norm that is not
equivalent to~$L^2$. More details on weak well-posedness can be found
in~\cite{KreLor89, SarTig12}. The essential difference between strong
and weak well-posedness is that the first, in contrast to the second,
is maintained in the presence of lower order perturbations (such as
source terms).

The hyperbolicity analysis of free evolution schemes used in numerical
relativity can be performed by bringing them to the
form~\eqref{eq:ref_pde}, via linearization and the constant
coefficient approach. In~\cite{GiaHilZil20} such an analysis was
presented for the Bondi-Sachs free evolution system. Existence and
uniqueness of solutions to the Bondi-like CIBVP were shown
in~\cite{FriLeh99, GomFri03}. In~\cite{Fri04}, a specific subsystem of
the full system was found symmetric hyperbolic, and consequently
certain norm estimates were provided. However, the hyperbolicity
analyses presented in~\cite{GiaHilZil20, GiaBisHil21} which consider
the full PDE system (in the linear, constant coefficient
approximation), show that it is only weakly hyperbolic. This result is
in contrast to that of~\cite{Fri04} and implies that a Bondi-like
CIBVP is ill posed in~$L^2$ (in the linear, constant coefficient
approximation). It is therefore remarkable that several characteristic
codes can perform long simulations and produce physically sensible
results. This fact, in combination with the existence of symmetric
hyperbolic formulations of GR that involve third order metric
derivatives and use Bondi-like
gauges~\cite{Rac13,CabChrTag14,HilValZha19,Rip21}, suggests that there
may still be some norm that is not equivalent to~$L^2$, which can be
used to demonstrate well-posedness of the standard Bondi-like CI(B)VP
with energy estimates. By standard here we mean that GR is formulated
only via the EFE so that only up to second order metric derivatives
are involved, rather than including the Bianchi identities as
in~\cite{Rac13,CabChrTag14,HilValZha19,Rip21}. However, finding such a
norm and showing even weak well-posedness for these second order
Bondi-like CI(B)VPs is still an open question.

Here, we concern ourselves with the following question: \textit{what
  are the best error estimates we can obtain for CCE and CCM with
  present setups and how can we test them numerically?} To address
this question we focus on simple toy models that have a structure
similar to that of GR in the gauges typically employed. More
specifically, we employ a characteristic model that is WH in a similar
way to free evolution Bondi-like schemes, whereas the Cauchy model is
SYMH like for instance GR in harmonic gauge. For completeness and
comparison, we also consider cases with SYMH CIBVPs and WH IBVPs. The
WH system is such that its IVP is weakly well posed. The paper has the
following structure. In Sec.~\ref{Section:models} we introduce the
models we use for our analysis and provide energy estimates for each
one of them. We also discuss under which conditions we can obtain an
energy estimate for CCE and CCM. In Sec.~\ref{Section:numerics} we
discuss the discretization of the models, and perform numerical
experiments to address convergence of solutions to discrete
approximations of the previously discussed norms. We summarize our
findings in Sec.~\ref{Section:Summary} and present our conclusions in
Sec.~\ref{Section:Conclusions}. Table~\ref{Table:acronyms} contains
the list of all the acronyms used throughout the text.

\begin{table}[tph]
  \begin{tabular}{L{8.5cm}}
    Acronyms
    \\
    \hline
    \vspace{0.01cm}
    CCE: Cauchy-Characteristic Extraction
    \\
    CIBVP: Characteristic Initial Boundary Value Problem
    \\
    CIVP: Characteristic Initial Value Problem
    \\
    CCM: Cauchy-Characteristic Matching
    \\
    EFE: Einstein Field Equations
    \\
    GR: General Relativity
    \\
    GW: Gravitational Wave
    \\
    IBVP: Initial Boundary Value Problem
    \\
    IVP: Initial Value Problem
    \\
    PDE: Partial Differential Equation
    \\
    SH: Strongly Hyperbolic
    \\
    SYMH: Symmetric Hyperbolic
    \\
    WH: Weakly Hyperbolic
    \\
    \hline
    \hline
  \end{tabular}
  \caption{List of acronyms used throughout the text.}
  \label{Table:acronyms}
\end{table}

\section{The models}
\label{Section:models}

In this section we construct toy models and provide energy estimates
for their solutions. For simplicity, we focus on first order, linear,
constant coefficient, homogeneous and real-valued PDEs. We study
energy estimates both for SYMH and WH models for the IBVP and the
CIBVP, as well as for their CCE and CCM. The structure of the WH
models are motivated by popular Bondi-like formulations of GR, and
follows the model used in~\cite{GiaHilZil20}. Working with SYMH and
not just SH systems allows us to mimic the best case scenario in a
current GR CCM implementation, matching between a SYMH and a WH system
for the IBVP and CIBVP respectively. Furthermore, estimates including
boundary terms are simpler to obtain for SYMH systems than for systems
that are only SH. The estimates presented in the section form the
basis of our numerical tests presented in Sec.~\ref{Section:numerics}.

\subsection{The IBVP}
\label{Subsection:cauchy_models}

\begin{figure}[!t]
  \includegraphics[width=0.4\textwidth]{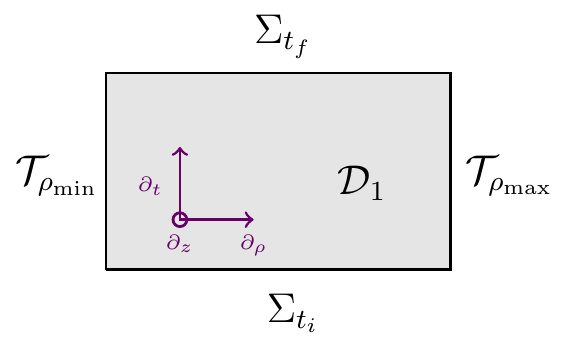}
  \caption{The Cauchy domain~$\mathcal{D}_1$, with boundaries that are
    spacelike~$\Sigma_{t_i}$,~$\Sigma_{t_f}$ and
    timelike~$\mathcal{T}_{\rho_{\textrm{min}}}$,~$\mathcal{T}_{\rho_{\textrm{max}}}$.
    Initial data are provided on~$\Sigma_{t_i}$ and boundary data
    on~$\mathcal{T}_{\rho_{\textrm{min}}}$,~$\mathcal{T}_{\rho_{\textrm{max}}}$
    for the right- and left-moving variables, respectively. The $z$
    direction is compact.}
  \label{Fig:cauchy_domain}
\end{figure}

Our Cauchy PDE model system is
\begin{subequations}
  \begin{align}
    \p_t \phi_1
    & = - \p_\rho \phi_1 + \boxed{\p_z\psi_{\varv1}}
      \label{eq:phi1}
      \,,
    \\
    \p_t \psi_{\varv1}
    & = - \p_\rho \psi_{\varv1}
      + \p_z\phi_1
      \label{eq:psiv1}
      \,,
    \\
    \p_t \psi_1
    & =  \p_\rho \psi_1 + \p_z\psi_1
      \label{eq:psi1}
      \,,
  \end{align}
  \label{eq:cauchy_pde}%
\end{subequations}
with~$\phi_1, \psi_{\varv1}$ the right-moving fields, and~$\psi_1$ the
left-moving. This system is SYMH when the boxed term of
Eq.~\eqref{eq:phi1} is included and only WH otherwise. The Cauchy
domain~$\mathcal{D}_1$ in which we seek a solution is formed
by~$t\in [t_{\textrm{i}},
t_{\textrm{f}}]$,~$\rho \in [\rho_{\textrm{min}},
\rho_{\textrm{max}}]$, and~$z \in [0, 2 \pi)$ with~$z$ taken to be
periodic; see Fig.~\ref{Fig:cauchy_domain} for an illustration. The
given data for the IBVP consist of the initial and boundary data. The
former
are~$\phi_1(t_i,\rho,z)$,~$\psi_{\varv1}(t_i,\rho,z)$,~$\psi_1(t_i,\rho,z)$
on an initial spacelike hypersurface~$\Sigma_{t_i}$, while the latter
are~$\phi_1(t,\rho_{\textrm{min}},z)$,~$\psi_{\varv1}(t,\rho_{\textrm{min}},z)$
on the left boundary~$\mathcal{T}_{\rho_{\textrm{min}}}$ since they
are right moving, and~$\psi_{1}(t,\rho_{\textrm{max}},z)$ on the right
boundary~$\mathcal{T}_{\rho_{\textrm{max}}}$, because it is left
moving. The solution~$\mathbf{u}_1$ to this IBVP at a later
time~$t_f > t_i$ consists then
of~$\phi_1(t_f,\rho,z)$,~$\psi_{\varv1}(t_f,\rho,z)$,~$\psi_1(t_f,\rho,z)$
on the final spacelike hypersurface~$\Sigma_{t_f}$, as well as the
values of the right-moving fields on the right
boundary~$\mathcal{T}_{\rho_{\textrm{max}}}$ and the left-moving on
the left boundary~$\mathcal{T}_{\rho_{\textrm{min}}}$ for
all~$t \in(t_i,t_f]$.

To obtain an energy estimate for the solution to the IBVP of the SYMH
system~\eqref{eq:cauchy_pde} we follow the standard procedure. Let us
start with the~$L^2$ norm
\begin{align}
  || \mathbf{u}_1||^2_{L^2(\Sigma_t)} =
  \int \left( \phi_1^2 + \psi_{\varv1}^2 + \psi_1^2 \right) d\rho\, dz
  \,.
  \label{eq:cauchy_L2_sigma_t}
\end{align}
After acting on it with~$\p_t$, using the right-hand side (rhs) of
Eq.~\eqref{eq:cauchy_pde} to replace~$\p_t \mathbf{u}_1$ and
collecting total derivatives, we obtain
\begin{align*}
  \p_t || \mathbf{u}_1||^2_{L_2(\Sigma_t)}
  &=
    \int \left[ - \p_\rho \phi_1^2 - \p_\rho \psi_{\varv1}^2
    + \p_\rho \psi_1^2 \right.
  \\
  & \left. \qquad \quad
    + 2 \p_z \left( \phi_1 \psi_{\varv1} \right) + \p_z \psi_1^2
    \right] d\rho\, dz
    \,,
\end{align*}
where the total~$\p_z$ terms vanish since the data are assumed to be
periodic in the~$z$ direction throughout. Evaluating the
total~$\p_\rho$ terms and integrating in~$t \in [t_i,t_f]$ we obtain
\begin{align}
  &
    || \mathbf{u}_1||_{L^2(\Sigma_{t_f})}^2
  + \int_{\mathcal{T}_{\rho_{\textrm{max}}}}
  \left( \phi_1^2 + \psi_{\varv 1}^2 \right)
  + \int_{\mathcal{T}_{\rho_{\textrm{min}}}}
  \psi_1^2 
    \nonumber \\
  &
    =
    ||\mathbf{u}_1||_{L^2(\Sigma_{t_i})}^2
  + \int_{\mathcal{T}_{\rho_{\textrm{min}}}}
  \left( \phi_1^2 + \psi_{\varv 1}^2 \right)
  + \int_{\mathcal{T}_{\rho_{\textrm{max}}}}
   \psi_1^2
  \,,
  \label{eq:cauchy_L2_hom_final}
\end{align}
where we have rearranged terms such that the left-hand side (lhs)
includes only the solution and the rhs only given data. This shows
that the solution is completely controlled by the given data in the
norm described by the lhs of Eq.~\eqref{eq:cauchy_L2_hom_final}.

Let us next consider the WH version of~\eqref{eq:cauchy_pde}.  First,
let us note that we fail to obtain an energy estimate in the~$L^2$
norm for the WH case, due to the boxed term in Eq.~\eqref{eq:phi1}.
Instead, motivated by the analysis presented in Sec.~III.B
of~\cite{GiaHilZil20}, we start with the ``lopsided'' or~$q$ norm
\begin{align}
  || \mathbf{u}_1||^2_{q(\Sigma_t)} =
  \int \left[
  \phi_1^2 + \psi_{\varv1}^2 + \psi_1^2
  + \left(\p_z \phi_1 \right)^2
  \right] d\rho\, dz
  \,,
  \label{eq:cauchy_q_sigma_t}
\end{align}
where~$q$ is due to the adoption of the notation of~\cite{KreLor89}
for weak well-posedness and ``lopsided'' refers to the fact that only
the angular derivative of a certain variable is included in the norm,
and not of all variables. This norm leads to
\begin{align*}
  \p_t || \mathbf{u}_1||^2_{q(\Sigma_t)}
  &=
    \int \left[ - \p_\rho \phi_1^2 - \p_\rho \psi_{\varv1}^2
    + \p_\rho \psi_1^2 - \p_\rho \left( \p_z \phi_1\right)^2 \right.
  \\
  & \left. \qquad \quad
    + 2 \psi_{\varv1}  \p_z  \phi_1  + \p_z \psi_1^2
    \right]  d\rho\, dz
    \,,
\end{align*}
where now, due to the WH structure of the system the
term~$ 2 \psi_{\varv1} \p_z \phi_1$ does not form a total derivative.
The lopsided modification of the original~$L^2$ norm was performed
exactly in a way that allows us to control this term. To see this, let
us first integrate in~$t$ and drop the total derivative
term~$\p_z \psi_1^2$ to obtain
\begin{align*}
  &
    || \mathbf{u}_1||_{q(\Sigma_{t_f})}^2
    =
    || \mathbf{u}_1||_{q(\Sigma_{t_i})}^2
  + \int_{\mathcal{T}_{\rho_{\textrm{min}}}}
    \left[
    \phi_1^2 + \psi_{\varv 1}^2 + \left(\p_z \phi_1\right)^2
    - \psi_1^2 
    \right]
     \nonumber \\
  & \quad
    + \int_{\mathcal{T}_{\rho_{\textrm{max}}}}
    \psi_1^2
    - \phi_1^2 - \psi_{\varv 1}^2 - \left(\p_z \phi_1\right)^2
    +
    \int_{\mathcal{D}_1} 2 \psi_{\varv1} \p_z \phi_1
    \,.
\end{align*}
Using
that~$2 \psi_{\varv1} \p_z \phi_1 \leq \phi_1^2 + \psi_{\varv1}^2 +
\psi_1^2 + \left(\p_z \psi_1 \right)^2$ the latter reads as
\begin{align*}
  &
    || \mathbf{u}_1||_{q(\Sigma_{t_f})}^2
    \leq
    || \mathbf{u}_1||_{q(\Sigma_{t_i})}^2
  + \int_{\mathcal{T}_{\rho_{\textrm{min}}}}
    \left[
    \phi_1^2 + \psi_{\varv 1}^2 + \left(\p_z \phi_1\right)^2
    - \psi_1^2 
    \right]
     \nonumber \\
  & \quad
    + \int_{\mathcal{T}_{\rho_{\textrm{max}}}}
    \psi_1^2
    - \phi_1^2 - \psi_{\varv 1}^2 - \left(\p_z \phi_1\right)^2
    +
    \int_{t_i}^{t_f} \int_{\Sigma_{t^\prime}} || \mathbf{u}_1||_{q (\Sigma_{t^\prime})}^2
    \,,
\end{align*}
which after using Gr\"onwall's inequality yields
\begin{align}
  &
    e^{t_i - t_f} || \mathbf{u}_1||_{q(\Sigma_{t_f})}^2
    + \int_{\mathcal{T}_{\rho_{\textrm{min}}}}
    \psi_1^2 
    + \int_{\mathcal{T}_{\rho_{\textrm{max}}}}
     \left[
   \phi_1^2 + \psi_{\varv 1}^2 + \left(\p_z \phi_1\right)^2
    \right]
     \nonumber \\
  & \quad
     \leq
    || \mathbf{u}_1||_{q(\Sigma_{t_i})}^2
  + \int_{\mathcal{T}_{\rho_{\textrm{min}}}}
    \left[
    \phi_1^2 + \psi_{\varv 1}^2 + \left(\p_z \phi_1\right)^2
    \right]
    + \int_{\mathcal{T}_{\rho_{\textrm{max}}}}
    \psi_1^2
    \,,
    \label{eq:cauchy_q_hom_final}
\end{align}
where we have rearranged terms such that the lhs includes only the
solution, and the rhs only the given data, showing that the solution
is completely controlled by the given data in the~$q$
norm~\eqref{eq:cauchy_q_sigma_t}. It is important to remember that
this result is subject to the structure of the source terms, and that
for generic source terms one cannot provide such an energy
estimate. This is the essential difference between strong and weak
well-posedness.

Since the~$q$ norm involves a specific derivative in its integrand, we
find it useful to employ a norm for which one can show energy
estimates for the SYMH IBVP, which also contains derivatives. We
denote by~$H_1$ the following norm:
\begin{align}
  || \mathbf{u}_1||^2_{H_1(\Sigma_t)} &=
    \int \left[
    \phi_1^2 + \psi_{\varv1}^2 + \psi_1^2
    \right.
    \nonumber \\
  & 
    \left.
    +\left(\p_\rho \phi_1 \right)^2
    + \left(\p_\rho \psi_{\varv1} \right)^2
    + \left(\p_\rho \psi_1 \right)^2
    \right.
    \nonumber \\
  & 
    \left. +
    \left(\p_z \phi_1 \right)^2
    + \left(\p_z \psi_{\varv1} \right)^2
    + \left(\p_z \psi_1 \right)^2
    \right] d\rho\, dz
    \,.
    \label{eq:cauchy_H1_sigma_t}
\end{align}
Let us consider the first order system that is formed after we enlarge
the SYMH system~\eqref{eq:cauchy_pde}, by
including~$\p_\rho \mathbf{u}_1$ and~$\p_z \mathbf{u}_1$ as
variables. This system has the same principal structure as the
original SYMH~\eqref{eq:cauchy_pde}, and an energy estimate for its
IBVP can be obtained starting with the~$H_1$
norm~\eqref{eq:cauchy_H1_sigma_t}. Following the same steps that led
to the estimate~\eqref{eq:cauchy_L2_hom_final} in the~$L^2$ norm, one
can show that
\begin{align}
  &
    || \mathbf{u}_1||_{H_1(\Sigma_{t_f})}^2
    + \int_{\mathcal{T}_{\rho_{\textrm{min}}}}
    \left[
    \psi_1^2
    + \left(\p_\rho \psi_1\right)^2 
    + \left(\p_z \psi_1\right)^2 
    \right]
    \nonumber \\
  &
    +\int_{\mathcal{T}_{\rho_{\textrm{max}}}}
    \left[
    \phi_1^2 + \psi_{\varv 1}^2
    +\left(\p_\rho \phi_1 \right)^2
    + \left(\p_\rho \psi_{\varv1} \right)^2
    \right.
    \nonumber
  \\
  &
    \qquad \qquad
    \left.
    + \left(\p_z \phi_1 \right)^2
    + \left(\p_z \psi_{\varv1} \right)^2
    \right]
    \nonumber \\
  &
    =
    || \mathbf{u}_1||_{H_1(\Sigma_{t_i})}^2
    + \int_{\mathcal{T}_{\rho_{\textrm{max}}}}
    \left[
    \psi_1^2 
    + \left(\p_\rho \psi_1\right)^2 
    + \left(\p_z \psi_1\right)^2 
    \right]
    \nonumber \\
  &
    +\int_{\mathcal{T}_{\rho_{\textrm{min}}}}
    \left[
    \phi_1^2 + \psi_{\varv 1}^2
    + \left(\p_\rho \phi_1 \right)^2
    + \left(\p_\rho \psi_{\varv1} \right)^2
    \right.
    \nonumber
  \\
  &
    \qquad \qquad
    \left.
    + \left(\p_z \phi_1 \right)^2
    + \left(\p_z \psi_{\varv1} \right)^2
    \right]
    \,,
  \label{eq:cauchy_H1_hom_final}
\end{align}
where again the lhs involves only the solution, and the rhs the given
data. We can use this result to assess the well-posedness of the IBVP
of the SYMH system~\eqref{eq:cauchy_pde}, in a norm that includes
derivatives. Notice that in comparison to the q
norm~\eqref{eq:cauchy_q_hom_final}, the~$H_1$
norm~\eqref{eq:cauchy_H1_hom_final} contains the same derivatives on
all variables and so is symmetric, rather than lopsided.

\subsection{The CIBVP} \label{Subsection:char_models}

\begin{figure}[!t]
  \includegraphics[width=0.35\textwidth]{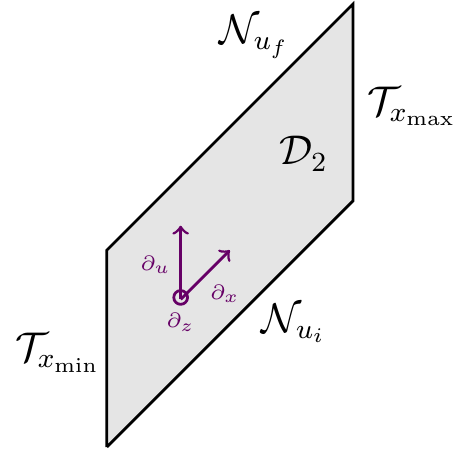}
  \caption{The characteristic domain~$\mathcal{D}_2$, with boundaries
    that are null~$\mathcal{N}_{u_i}$,~$\mathcal{N}_{u_f}$ and
    timelike~$\mathcal{T}_{x_{\textrm{min}}}$,~$\mathcal{T}_{x_{\textrm{max}}}$. Initial
    data for left-moving fields are provided on~$\mathcal{N}_{u_i}$
    and boundary data
    on~$\mathcal{T}_{\rho_{\textrm{min}}}$,~$\mathcal{T}_{\rho_{\textrm{max}}}$
    for the right- and left-moving variables, respectively. The $z$
    direction is compact.}
  \label{Fig:char_domain}
\end{figure}

The characteristic model we consider is
\begin{subequations}
  \begin{align}
    \p_x \phi_2
    & = \boxed{\p_z\psi_{\varv2}}
      \label{eq:phi2}
      \,,
    \\
    \p_x \psi_{\varv2}
    & = \p_z\phi_2
      \label{eq:psiv2}
      \,,
    \\
    \p_u \psi_2
    & =  \frac{1}{2}\p_x \psi_2 + \p_z\psi_2
      \label{eq:psi2}
      \,,
  \end{align}
  \label{eq:char_pde}%
\end{subequations}
where the fields~$\phi_2, \psi_{\varv2}$ are right moving, and
propagate with speed unity and~$\psi_2$ is left moving with
speed~$1/2$. The system has two equations with derivatives only
intrinsic to the characteristic surfaces~$\mathcal{N}_u$ and one that
includes also a derivative transverse to them. The system is SYMH when
the boxed term in Eq.~\eqref{eq:phi2} is included, and only WH
otherwise. If we consider the coordinate transformation
\begin{align*}
  u = t-\rho \,,
  \quad
  x = \rho \,,
\end{align*}
which implies
\begin{align*}
  \p_u = \p_t \,,
  \quad
  \p_x = \p_t + \p_\rho \,,
\end{align*}
the SYMH version of the system~\eqref{eq:char_pde} is the
characteristic version of the Cauchy system~\eqref{eq:cauchy_pde},
provided that
\begin{align*}
  \phi_1
  &= \phi_2 \,,
  \quad
  \psi_{\varv1} = \psi_{\varv2} \,
  \quad
    \psi_1 = \psi_2 \,,
\end{align*}
and similarly for the WH one.

We consider the CIBVP for the characteristic domain~$\mathcal{D}_2$
shown in Fig.~\ref{Fig:char_domain}, that consists
of~$u\in [u_{\textrm{i}},
u_{\textrm{f}}]$,~$x \in [x_{\textrm{min}}, x_{\textrm{max}}]$,
and~$z \in [0, 2 \pi)$ with~$z$ a periodic direction. The given data
are the left-moving field~$\psi_2(u_i,x,z)$ on an initial
characteristic~$\mathcal{N}_{u_i}$, its
value~$\psi_2(u,x_{\textrm{max}},z)$ on the right
boundary~$\mathcal{T}_{x_{\textrm{max}}}$ and the values of the
right-moving
fields~$\phi_2(u,x_{\textrm{min}},z)$,~$\psi_{\varv2}(u,x_{\textrm{min}},z)$
on the left boundary~$\mathcal{T}_{x_{\textrm{min}}}$. The solution
for which we next provide energy estimates consists
of~$\phi_2(u_f,x,z)$ on a surface~$\mathcal{N}_{u_f}$
with~$u_f>u_i$,~$\phi_2(u,x_{\textrm{min}},z)$ on the left
boundary~$\mathcal{T}_{x_{\textrm{min}}}$
and~$\phi_2(u,x_{\textrm{max}},z)$,~$\psi_{\varv2}(u,x_{\textrm{max}},z)$
on the right boundary~$\mathcal{T}_{x_{\textrm{max}}}$.

To obtain energy estimates in the characteristic domain, we can treat
independently the left- and right-moving fields, due to the fact that
the right-moving fields do not have an explicit~$\p_u$ time
derivative. Let us first consider the left-moving field and start from
the energy
\begin{align*}
  || \mathbf{u}_2||^2_{\textrm{left}({\mathcal{N}_u})}
  = \int_{\mathcal{N}_u} \psi_2^2
  \,.
\end{align*}
After acting with~$\p_u$ and replacing with the rhs of
Eq.~\eqref{eq:psi2} we obtain
\begin{align*}
  \p_u || \mathbf{u}_2||^2_{\textrm{left}({\mathcal{N}_u})}
  =
  \int_{\mathcal{N}_u} \p_x \psi_2^2 + \p_z \psi_2^2
  \,.
\end{align*}
By evaluating the total derivatives and integrating
in~$u \in[u_i,u_f]$ we arrive at
\begin{align}
    2 || \mathbf{u}_2||^2_{\textrm{left}({\mathcal{N}_{u_f}})}
    +
    \int_{\mathcal{T}_{x_{\textrm{min}}}} \psi_2^2
    =
    2 || \mathbf{u}_2||^2_{\textrm{left}({\mathcal{N}_{u_i}})}
    +
    \int_{\mathcal{T}_{x_{\textrm{max}}}} \psi_2^2
    \,,
    \label{eq:char_left_estimate}
\end{align}
where we have rearranged terms such that the lhs contains only the
solution and the rhs the given data, as well as multiplied everything
by a factor of 2, to cancel the characteristic speed in the worldtube
integrals. This estimate allows us to completely control the solution
for the left-moving data by given data. If there are source terms in
Eq.~\eqref{eq:psi2} that include the right-moving fields, then the
above calculation does not work. In that case one needs to perform the
calculation for the whole system at once. Potentially, one also needs
to consider a slightly different computational domain such that there
is an integrand involving both left- and right-moving variables and
offer the chance to control them using Gr\"onwall's inequality, as
before. This however is beyond the scope of the current paper.

Next, we consider the right-moving fields, focusing first on the SYMH
case. Let us start by considering the energy
\begin{align*}
  || \mathbf{u}_2||^2_{\textrm{right}({\mathcal{T}_x})} =
  \int_{\mathcal{T}_x} \phi_2^2 + \psi_{\varv2}^2
  \,.
\end{align*}
By acting with~$\p_x$ and replacing with the rhs of
Eq.~\eqref{eq:phi2},~\eqref{eq:psiv2} we obtain
\begin{align*}
  \p_x || \mathbf{u}_2||^2_{\textrm{right}({\mathcal{T}_x})}
  = \int_{\mathcal{T}_x} 2 \p_z \left( \phi_2  \psi_{\varv_2} \right)
  \,,
\end{align*}
where the integrand on the right is a total~$\p_z$ term, and hence
vanishes by periodicity in~$z$. Then, integrating
in~$x \in [x_{\textrm{min}}, x']$ for
some~$x' \in (x_{\textrm{min}}, x_{\textrm{max}}]$, we arrive at the
estimate
\begin{align}
  || \mathbf{u}_2||^2_{\textrm{right}({\mathcal{T}_{x'}})} =
  || \mathbf{u}_2||^2_{\textrm{right}({\mathcal{T}_{x_{\textrm{min}}}})}
  \,,
  \label{eq:char_right_estimate}
\end{align}
where the lhs is the solution and the rhs the given data. As was also
done in a similar calculation in~\cite{GiaHilZil20}, we can take the
supremum of~$|| \mathbf{u}_2||^2_{\textrm{right}({\mathcal{T}_{x'}})}$
within~$ (x_{\textrm{min}}, x_{\textrm{max}}]$ to obtain a better
estimate,
since~$|| \mathbf{u}_2||^2_{\textrm{right}({\mathcal{T}_{x'}})}$ is
not necessarily monotonically increasing with increasing~$x'$. Then,
since Eq.~\eqref{eq:char_right_estimate} is true for
all~$x' \in (x_{\textrm{min}}, x_{\textrm{max}}]$ we obtain
\begin{align}
  \textrm{sup}_{x'}|| \mathbf{u}_2||^2_{\textrm{right}({\mathcal{T}_{x'}})} =
  || \mathbf{u}_2||^2_{\textrm{right}({\mathcal{T}_{x_{\textrm{min}}}})}
  \,.
  \label{ineq:char_right_estimate}
\end{align}

If we consider the WH model~\eqref{eq:char_pde}, then we should
consider the energy
\begin{align*}
  || \mathbf{u}_2||^2_{q-\textrm{right}({\mathcal{T}_x})} =
  \int_{\mathcal{T}_x} \phi_2^2 + \psi_{\varv2}^2 + \left( \p_z\phi_2 \right)^2
  \,,
\end{align*}
in analogy to the $q$ norm~\eqref{eq:cauchy_q_sigma_t}. Following the
same steps as before we can arrive at
\begin{align*}
  &
    || \mathbf{u}_2||^2_{q-\textrm{right}({\mathcal{T}_{x'}})}
  \leq
  || \mathbf{u}_2||^2_{q-\textrm{right}({\mathcal{T}_{x_{\textrm{min}}}})}
  +
    \nonumber \\
  &
    \int_{x_i}^{x_f} || \mathbf{u}_2||^2_{q-\textrm{right}({\mathcal{T}_{x}})}dx
  \,,
\end{align*}
where we have also used
that~$2 \psi_{\varv2} \p_z \phi_2 \leq \phi_2^2 + \psi_{\varv2}^2 +
\left( \p_z\phi_2 \right)^2$. Using again the Gr\"onwall inequality
and noticing
that~$||\mathbf{u}_2||^2_{q-\textrm{right}({\mathcal{T}_{x_{\textrm{min}}}})}$
involves only given (boundary) data, and hence is a constant function
of~$x$ or~$x_f$ (so nondecreasing), we obtain
\begin{align}
   e^{x_i - x_f} \textrm{sup}_{x'}|| \mathbf{u}_2||^2_{q-\textrm{right}({\mathcal{T}_{x'}})}
  \leq
  || \mathbf{u}_2||^2_{q-\textrm{right}({\mathcal{T}_{x_{\textrm{min}}}})}
  \,,
  \label{ineq:char_right_estimate}
\end{align}
where we again took the supremum of~$x'$ to obtain a better estimate
and the lhs contains only the solution and the rhs the given data and
multiplied overall with~$ e^{x_i - x_f}$ for later convenience in the
continuum analysis.

We can then proceed to add the estimates~\eqref{eq:char_left_estimate}
and~\eqref{eq:char_right_estimate} to obtain the estimate for the
homogeneous, symmetric hyperbolic system~\eqref{eq:char_pde}
\begin{align}
  &
    2 || \mathbf{u}_2||^2_{\textrm{left}({\mathcal{N}_{u_f}})}
    +
    \textrm{sup}_{x'} || \mathbf{u}_2||^2_{\textrm{right}({\mathcal{T}_{x'}})}
    +
    \int_{\mathcal{T}_{x_{\textrm{min}}}} \psi_2^2
    =
    \nonumber
  \\
  &
    2 || \mathbf{u}_2||^2_{\textrm{left}({\mathcal{N}_{u_i}})}
    +
    || \mathbf{u}_2||^2_{\textrm{right}({\mathcal{T}_{x_{\textrm{min}}}})}
    +
    \int_{\mathcal{T}_{x_{\textrm{max}}}} \psi_2^2
    \,,
    \label{eq:char_SYMH_L2_estimate}
\end{align}
or add~\eqref{eq:char_left_estimate}
and~\eqref{ineq:char_right_estimate} to obtain an estimate for the WH
characteristic model
\begin{align}
  &
    2 || \mathbf{u}_2||^2_{\textrm{left}({\mathcal{N}_{u_f}})}
    +
    e^{x_i -x_f} \textrm{sup}_{x'} || \mathbf{u}_2||^2_{q-\textrm{right}({\mathcal{T}_{x'}})}
    +
    \int_{\mathcal{T}_{x_{\textrm{min}}}} \psi_2^2
    \nonumber
  \\
  &
    \leq
    2 || \mathbf{u}_2||^2_{\textrm{left}({\mathcal{N}_{u_i}})}
    +
    || \mathbf{u}_2||^2_{q-\textrm{right}({\mathcal{T}_{x_{\textrm{min}}}})}
    +
    \int_{\mathcal{T}_{x_{\textrm{max}}}} \psi_2^2
    \,.
    \label{ineq:char_WH_q_estimate}
\end{align}
Subsequently, we drop the factor of 2 in the lhs of
Eq.~\eqref{eq:char_SYMH_L2_estimate}
and~\eqref{ineq:char_WH_q_estimate} for convenience, but we still
maintain the hypersurface integrals which provide the control of the
solution on that region.

In analogy to the~$H_1$ norm~\eqref{eq:cauchy_H1_sigma_t} for the SYMH
IBVP, we also consider estimates in an~$H_1$ norm adapted to the
CIBVP, again by enlarging the SYMH characteristic
system~\eqref{eq:char_pde} with~$\p_x \mathbf{u}_2$
and~$\p_z \mathbf{u}_2$. In the characteristic~$H_1$ norm one can show
the following estimate for the enlarged system
\begin{align}
  &
    2 || \mathbf{u}_2||^2_{H_1-\textrm{left}({\mathcal{N}_{u_f}})}
    +
    \textrm{sup}_{x'} || \mathbf{u}_2||^2_{H_1-\textrm{right}({\mathcal{T}_{x'}})}
    +
    \nonumber
  \\
  & \qquad
    \int_{\mathcal{T}_{x_{\textrm{min}}}}
    \left[
    \psi_2^2
    + \left(\p_x \psi_2 \right)^2
    + \left(\p_z \psi_2 \right)^2
    \right]
    \nonumber
  \\
  &
    =
    2 || \mathbf{u}_2||^2_{H_1-\textrm{left}({\mathcal{N}_{u_i}})}
    +
    || \mathbf{u}_2||^2_{H_1-\textrm{right}({\mathcal{T}_{x_{\textrm{min}}}})}
    +
    \nonumber
  \\
  &
    \qquad
    \int_{\mathcal{T}_{x_{\textrm{max}}}}
    \left[
    \psi_2^2
    + \left(\p_x \psi_2 \right)^2
    + \left(\p_z \psi_2 \right)^2
    \right]
    \,,
    \label{eq:char_SYMH_H1_estimate}
\end{align}
with
\begin{align*}
  &
    || \mathbf{u}_2||^2_{H_1-\textrm{left}({\mathcal{N}_u})} =
    \int_{\mathcal{N}_u}
    \left[
    \psi_2^2 +
    \left( \p_x \psi_2 \right)^2 +
    \left( \p_z \psi_2 \right)^2
    \right]
    \,,
\end{align*}
and
\begin{align*}
  &
    || \mathbf{u}_2||^2_{H_1-\textrm{right}({\mathcal{T}_x})} =
    \int_{\mathcal{T}_x}
    \left[
    \phi_2^2 + \psi_{\varv2}^2 +
    \right.
    \nonumber
  \\
  &
    \qquad
    \left.
    \left( \p_x \phi_2 \right)^2 +
    \left( \p_x \psi_{\varv2} \right)^2 +
    \left( \p_z \phi_2 \right)^2 +
    \left( \p_z \psi_{\varv2} \right)^2 
    \right]
    \,,
\end{align*}
by simple commutation of partial derivatives.

\subsection{CCE and CCM} \label{Subsection:CCE_CCM}

Using the energy estimates from the previous subsections, we can now
discuss energy estimates for CCE and CCM. For CCE the estimate follows
straightforwardly from the previous results, and if both the IBVP and
CIBVP are individually well posed in some norm, then CCE is too. The
only detail that needs attention for this conclusion to be drawn is to
choose the boundary data on~$\mathcal{T}_0$ such that they are
controlled in the appropriate norm for a WH CIBVP. Since these data
are solutions of the IBVP, this may amount to having enough control
over the derivatives of the solution on~$\mathcal{T}_0$, so as to be
controlled not only in the~$L^2$ but also in the~$H_1$ norm, as for
instance shown in estimate~\eqref{eq:cauchy_H1_hom_final}. However, in
this case we ``lose'' derivatives, in the sense that we control
derivatives of more variables for the boundary data than for the
solution to the WH CIBVP.

\begin{figure}[!t]
  \includegraphics[width=0.4\textwidth]{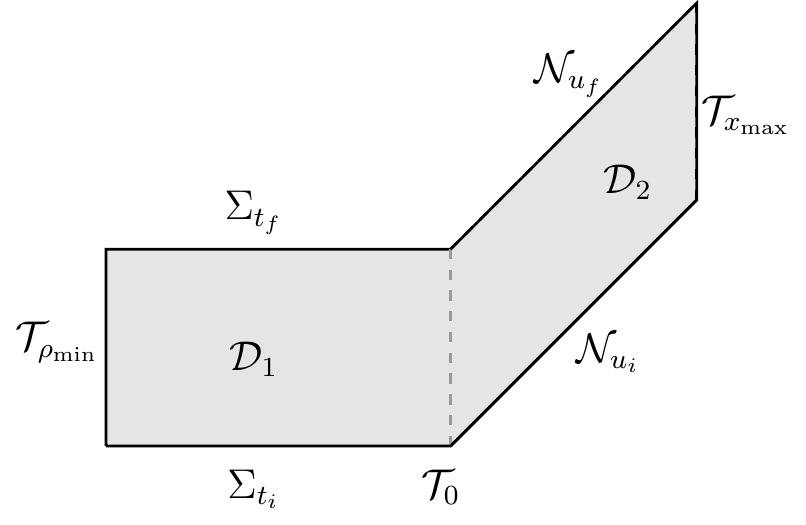}
  \caption{The CCM
    domain~$\mathcal{D}_3 = \mathcal{D}_1 + \mathcal{D}_2$, with
    boundaries the spacelike~$\Sigma_{t_i}$,~$\Sigma_{t_f}$, the
    null~$\mathcal{N}_{u_i}$,~$\mathcal{N}_{u_f}$ and the
    timelike~$\mathcal{T}_{\rho_{\textrm{min}}}$,~$\mathcal{T}_{x_{\textrm{max}}}$. Initial
    data for all fields are provided on~$\Sigma_{t_i}$, as well as
    on~$\mathcal{N}_{u_i}$ only for left-moving fields. Boundary data
    are given on~$\mathcal{T}_{\rho_{\textrm{min}}}$ for right-moving,
    and on~$\mathcal{T}_{x_{\textrm{max}}}$ for left-moving
    fields. Data are communicated between the domains~$\mathcal{D}_1$
    and~$\mathcal{D}_2$ via the
    worldtube~$\mathcal{T}_0 = \mathcal{T}_{\rho_{\textrm{max}}} =
    \mathcal{T}_{x_{\textrm{min}}}$. The $z$ direction is compact.}
  \label{Fig:ccm_domain}
\end{figure}

For the composite CCM problem to be well posed the situation is more
complicated, as the norms in which the IBVP and the CIBVP are well
posed need to be in some sense compatible. As shown
in~\cite{GiaHilZil20} the reason is that otherwise, there is a
remaining integral over the interface~$\mathcal{T}_0$ as shown in
Fig.~\ref{Fig:ccm_domain}. For completeness, we briefly repeat here
this calculation. First we consider matching the two SYMH setups. By
adding the estimates~\eqref{eq:cauchy_L2_hom_final}
and~\eqref{eq:char_SYMH_L2_estimate} and using
that~$\mathbf{u}_1|_{\mathcal{T}_0} = \mathbf{u}_2|_{\mathcal{T}_0}$,
we obtain
\begin{align}
  &
    || \mathbf{u}_1||_{L_2(\Sigma(t_f)}^2
    +
    || \mathbf{u}_2||^2_{\textrm{left}({\mathcal{N}_{u_f}})}
    \label{eq:ccm_SYMH_SYMH_estimate}
  \\
  & \qquad
    + \int_{\mathcal{T}_{\rho_{\textrm{min}}}} \psi_1^2 
    +
    \textrm{sup}_{x'}\int_{\mathcal{T}_{x'}} \phi_2^2 + \psi_{\varv 2}^2
    =
    \nonumber \\
  &
    || \mathbf{u}_1||_{L_2(\Sigma(t_i)}^2
    +
    || \mathbf{u}_2||^2_{\textrm{left}({\mathcal{N}_{u_i}})}
    +
     \int_{\mathcal{T}_{\rho_{\textrm{min}}}}
     \phi_1^2 + \psi_{\varv 1}^2
        +
    \int_{\mathcal{T}_{x_{\textrm{max}}}} \psi_2^2
     \,.
     \nonumber
\end{align}
The latter is the energy estimate for the composite CCM problem when
matching the IBVP and CIBVP of the homogeneous symmetric hyperbolic
Cauchy and characteristic systems~\eqref{eq:cauchy_pde}
and~\eqref{eq:char_pde}, respectively. Notice that, as expected for
the composite CCM problem, there is no worldtube integral
over~$\mathcal{T}_0$. A similar estimate can be obtained using
the~$H_1$ norms and adding the higher derivative
estimates~\eqref{eq:cauchy_H1_hom_final}
and~\eqref{eq:char_SYMH_H1_estimate} for the SYMH IBVP and CIBVP,
respectively.

Let us next consider matching the IBVP of the SYMH Cauchy
system~\eqref{eq:cauchy_pde} to the weakly well-posed CIBVP of the WH
characteristic system~\eqref{eq:char_pde}. In this case we add the
energy estimates~\eqref{eq:cauchy_L2_hom_final}
and~\eqref{ineq:char_WH_q_estimate}, which yields
\begin{align}
  &
    || \mathbf{u}_1||_{L_2(\Sigma(t_f)}^2
    +
    || \mathbf{u}_2||^2_{\textrm{left}({\mathcal{N}_{u_f}})}
    +
    \nonumber \\
  &
    \int_{\mathcal{T}_{\rho_{\textrm{min}}}} \psi_1^2 
    +
    e^{x_i -x_f} \textrm{sup}_{x'}\int_{\mathcal{T}_{x'}}
    \left[ \phi_2^2 + \psi_{\varv2}^2 + \left(\p_z \phi_2\right)^2
    \right]
    \nonumber \\
  &
    \leq
    || \mathbf{u}_1||_{L_2(\Sigma(t_i)}^2
    +
    || \mathbf{u}_2||^2_{\textrm{left}({\mathcal{N}_{u_i}})}
    +
    \nonumber \\
  &
    \int_{\mathcal{T}_{x_{\textrm{max}}}} \psi_2^2
    +
    \int_{\mathcal{T}_{\rho_{\textrm{min}}}}
    \left( \phi_1^2 + \psi_{\varv 1}^2 \right)
    +
    \nonumber \\
  &
    \boxed{
    \int_{\mathcal{T}_0}
    \left(\p_z \phi \right)^2
    }
    \,,
    \label{eq:ccm_SYMH_WH_estimate}
\end{align}
where we have used that
on~$\mathcal{T}_0$,~$\mathbf{u}_1 = \mathbf{u}_2$ in our models. The
boxed term in the rhs of the latter is what prevents us from obtaining
an energy estimate for the composite CCM in this setup and appears
purely due to the incompatibility of the norms in which the IBVP and
the CIBVP are well posed and weakly well posed, respectively. Since
this term is not part of the given data, then the solution is not
completely controlled by the intial and boundary data, and so the
above is not a valid energy estimate for this composite CCM. If
instead of using the~$L^2$ estimate~\eqref{eq:cauchy_L2_hom_final} for
the IBVP, we use the~$H_1$ estimate~\eqref{eq:cauchy_H1_hom_final},
the result would be similar, in that nonvanishing terms including an
integral over the interface~$\mathcal{T}_0$ remain.

Finally, we consider the matching of two WH setups. In this case, we
combine the~$q$ norm estimates~\eqref{eq:cauchy_q_hom_final}
and~\eqref{ineq:char_WH_q_estimate} for the IBVP and CIBVP,
respectively, to obtain
\begin{align}
  &
    e^{t_i -t_f} || \mathbf{u}_1||_{q(\Sigma(t_f)}
    +
    || \mathbf{u}_2||^2_{\textrm{left}({\mathcal{N}_{u_f}})}
    +
    \nonumber \\
  &
    \int_{\mathcal{T}_{\rho_{\textrm{min}}}} \psi_1^2 
    +
    e^{x_i -x_f} \textrm{sup}_{x'}\int_{\mathcal{T}_{x'}}
    \left[ \phi_2^2 + \psi_{\varv2}^2 + \left(\p_z \phi_2\right)^2
    \right]
    \nonumber \\
  &
    \leq
    || \mathbf{u}_1||_{q(\Sigma(t_i)}
    +
    || \mathbf{u}_2||^2_{\textrm{left}({\mathcal{N}_{u_i}})}
    +
    \nonumber \\
  &
    \int_{\mathcal{T}_{x_{\textrm{max}}}} \psi_2^2
    +
    \int_{\mathcal{T}_{\rho_{\textrm{min}}}}
    \left( \phi_1^2 + \psi_{\varv 1}^2 \right)
    \,.
    \label{eq:ccm_WH_WH_estimate}
\end{align}
In this case we see that, due to the compatibility of the norms, the
integral over~$\mathcal{T}_0$ vanishes and the lhs that contains only
the solution, is completely controlled by the rhs, that includes only
given data.

\section{Numerical experiments} \label{Section:numerics}

We present numerical convergence tests for the individual IBVP and
CIBVP, as well as for CCE and CCM, using discrete versions of the
models analyzed in Sec.~\ref{Section:models}. We work with the
homogeneous models, unless explicitly stated otherwise. The
implementation was performed using the~\texttt{Julia} programming
language~\cite{BezEdeKar17} with the following details: 
\begin{enumerate}

\item The numerical domain is defined by
  \begin{align*}
    &
      t \in [0, t_f]\,,
      \quad u \in [0, u_f] \,,
    \\
    &
      \rho \in [-1, 0] \,,
      \quad
      x \in [0,1] \,,
      \quad
      z \in [0, 2\pi)
      \,,
  \end{align*}
  where~$u=t-\rho$, so for~$\rho=0$,~$t_f=u_f$. We denote the grid
  points for~$x$ as~$x_i$ with~$i \in [1, \dots, N]$ and the grid
  spacing as~$h_x \equiv x_2 - x_1$, and so forth.
  
\item All the derivatives that appear on the rhs of the
  systems~\eqref{eq:cauchy_pde} and~\eqref{eq:char_pde} are
  approximated by second order accurate finite difference
  operators. The derivative~$\p_x$ is approximated with a centered
  stencil~$D_{x}f_i=(f_{i+1} - f_{i-1})/(2 h_{x})$ for the internal
  points of the~$x$~grid. The derivative on the first and last points
  is approximated with upwind stencils that match the truncation error
  of the centered one
  \begin{align}
    D_x f_1
    &= \frac{-4 f_1 + 7 f_2 -4 f_3 + f_4}{2 h_x}
      \,,
      \label{eq:boundary_FD2}
    \\
    D_x f_N
    &= \frac{4 f_N - 7 f_{N-1} +4 f_{N-2} - f_{N-3}}{2 h_x}
      \,.
      \nonumber
  \end{align}
  The same operators are used to approximate~$\p_\rho$,
  with~$h_x \rightarrow h_\rho$, whereas for~$\p_z$, the centered
  stencil is used for the first and last points as well, taking
  advantage of the periodicity in~$z$.
  
\item All fields that involve explicit time derivatives, that
  is~$\phi_1\,, \psi_{\varv1}\,, \psi_1\,, \psi_2$, are integrated in
  time using the explicit fourth order Runge-Kutta integration scheme.

\item The fields~$\phi_2\,, \psi_{\varv2}$ that satisfy equations
  intrinsic to the outgoing null hypersurfaces are integrated for every
  timestep from~$x=0$ to~$x=1$ using the trapezoidal rule, which is
  second order accurate.

\item No Kreiss-Oliger artificial dissipation is applied.

\item We choose a fixed timestep~$h_t = 0.25 h_\rho$.
  
\item The initial data
  are~$\phi_1(0,\rho,z)\,, \psi_{\varv1}(0,\rho,z)\,,
  \psi_1(0,\rho,z)$ on the initial spacelike hypersurface~$\Sigma_{t=0}$,
  and $\psi_2(0,x,z)$ on the initial null
  hypersurface~$\mathcal{N}_{u=0}$.
  
\item For the IBVP, the boundary data
  are~$\phi_1(t,-1,z) \,, \psi_{\varv1}(t,-1,z)$ on the
  worldtube~$\mathcal{T}_{\rho=-1}$, and~$\psi_1(t,0,z)$
  on~$\mathcal{T}_{\rho=0}$.

\item For the CIBVP,~$\phi_2(t,0,z)$ and~$\psi_{\varv 2}(t,0,z)$ are
  provided as boundary data on the worldtube~$\mathcal{T}_{x=0}$, as
  well as ~$\psi_2(u,1,z)$ on~$\mathcal{T}_{x=1}$.

\item For CCE, boundary data for the IBVP are given also
  on~$\mathcal{T}_{\rho=0}$ for the left-moving field~$\phi_1$. The
  values of the right-moving fields~$\phi_1\,, \psi_{\varv1}$ are
  propagated from the Cauchy to the characteristic domain
  via~$\mathcal{T}_{\rho=x=0}$, as in CCM.

\item For CCM, on the worldtube (interface)~$\mathcal{T}_{\rho=x=0}$
  the values of the right-moving fields are propagated from the Cauchy
  to the characteristic domain
  via~$\phi_2(u,0,z) =
  \phi_1(t,0,z)$,~$\psi_{\varv2}(u,0,z) = \psi_{\varv1}(t,0,z)$, and
  for the left-moving one from the characteristic to the Cauchy
  via~$\phi_1(t,0,z)=\phi_2(u,0,z)$.

\item For the tests presented in the paper, all the numerical boundary
  data are provided with pure injection, that is if~$g_1(t,z)$ is the
  value of the grid function~$\phi_1$ for~$\rho=-1$,
  then~$\phi_1(t,-1,z) = g_1(t,z)$.  We have also experimented with a
  different method for numerical boundary data, namely by providing
  the time derivative rather than the field itself. The results can be
  found in~\cite{GiaBisHil23_github} and are qualitatively the same.
\end{enumerate}

The main focus of this work is to examine the consequence of (a lack
of) continuous dependence at the continuum level, as
in~\eqref{ineq:contin_dep}, under numerical approximation.  Detecting
lack of convergence for WH systems can be
subtle~\cite{CalPulSar02}. Numerical convergence tests that include
high frequency data are a good choice in order to achieve
this~\cite{CalHinHus05, CaoHil11, GiaHilZil20}. Low frequency data are
not ideal to detect the effect of weak hyperbolicity in numerical
simulations, since the solution may exhibit frequency dependent
exponential or polynomial growth, that may not be evident for small
frequencies in finite simulation time. Thus, the loss of convergence
for an ill-posed problem may not be evident if the given data are not
properly chosen. We run robust stability tests, which use high
frequency given data, and monitor the convergence of the obtained
numerical solutions in the norms presented in
Sec.~\ref{Section:models}.

Since we use uniform grids, we choose the high frequency data to be
random noise of a certain amplitude. The random noise is added on top
of an exact solution to the PDE problems, which in our tests is
zero. We call these \textit{exact convergence tests}, since we know
the exact solution to the PDE problem.
To fix ideas, next we present explicitly only the field~$\phi_1$, but
the discussion is the same for all
fields~$\psi_{\varv1}, \psi_1, \phi_2, \psi_{\varv2}, \psi_2$.
If we denote by~$\phi_1$ the continuum solution and~$\phi_{1h}$ the
numerical at resolution~$h$, and assume that the dominant error for
our discretization is due to the second order finite difference
operators, then we expect the following relation to be true:
\begin{align*}
  \phi_1 = \phi_{1h} + O(h^2)
  \,.
\end{align*}
Given~$\phi_1=0$ in our setup, it follows that
\begin{align*}
  \phi_{1h} = O(h^2)
  \,.
\end{align*}
We denote by~$h_c$ and~$h_m$ the coarse and medium resolutions for
which we solve the same PDE problem, and construct the
\textit{convergence factor}
\begin{align*}
  Q \equiv \frac{h_c^2}{h_m^2} \simeq \frac{||\phi_{1c}||_{h_c}}{|| \phi_{1m}||_{h_m}}
  \,,
\end{align*}
where $\simeq$ here denotes equality up to terms of
order~$O(h_c)$. Notice that the above relation is true only when the
exact solution vanishes. Generically, when this is not true, one needs
to replace~$\phi_{1c} \rightarrow \phi_1 - \phi_{1c}$ in the
expression, and similarly for the medium resolution. Every time we
increase resolution, we halve the grid spacing (for
instance~$h_m = h_c/2$), and therefore the expected convergence factor
is~$Q=4$. We construct and monitor the following quantity during our
simulations
\begin{align}
  C_{\textrm{exact}} \equiv \log_2
  \frac{|| \mathbf{u}_{h_c}||_{h_c}}
  {|| \mathbf{u}_{h_m}||_{h_m}}
  \,,
  \label{eq:exact_conv_rate}
\end{align}
where~$|| \cdot ||_h$ is the discrete approximation to the continuum
norm~$|| \cdot||$. The forms of the state vector~$\mathbf{u}_h$ and
the norm~$|| \cdot ||_h$ depend on the specific discretized PDE
problem solved, and so are clarified in the following
subsections. Given that~$Q=4$, perfect second order convergence
corresponds to~$C_{\textrm{exact}} = 2$ for our exact convergence
tests.

Motivated by the notation of the inequality~\eqref{ineq:contin_dep},
let us denote as~$f_h$ the numerical initial and boundary data, which
we specify as random noise of amplitude~$A_h$ centered around zero. It
is important to choose this amplitude appropriately such
that~$C_{\textrm{exact}}=2$ for all given data.  Since~$f_h \sim A_h$,
for a field that appears without any derivative in the norm under
consideration, we find that when~$A_{h_m} = A_{h_c}/4$,
\begin{align*}
  C_{\textrm{exact}} =
  \log_2 \frac{|| f_{h_c} ||_{h_c}} { || f_{h_m} ||_{h_m}}
  \sim
  \log_2 \frac{ O(A_{h_c})} { O(A_{h_m})}
  = 2
  \,.
\end{align*}
If however the norm includes a derivative of the field, then due to
the finite difference operators used the exact convergence rate
behaves as
\begin{align*}
  \frac{O(A_{h_c})/h_{c}} {O(A_{h_m})/h_{m}}
  \,,
\end{align*}
which for our setup is equal to 2 when we
choose~$A_{h_m} = A_{h_c}/8$. Consequently, our recipe to obtain given
data that have~$C_{\textrm{exact}}=2$ is the following: every time we
double resolution, we drop the noise amplitude of the given data by a
factor of 4 for any field with no derivative in the considered norm,
and 8 for any field with at least one derivative in the considered
norm.

\begin{table*}[tph]
  \centering
  \begin{tabular}{L{6.7cm} C{2.4cm} C{3.3cm} C{2.4cm} C{2.4cm}}
    \hline
    \hline
    \vspace{0.01cm}
    The PDE setup
    &  \vspace{0.01cm}
    $L^2-L^2$ test
    & \vspace{0.01cm}
     $q-q$ test
    &  \vspace{0.01cm}
    $H_1-H_1$ test
    & \vspace{0.01cm}
    $H_1-q$ test \\
    \hline
    \vspace{0.01cm}
    IBVP: SYMH (Fig.~\ref{Fig:IBVP_SYMH_vs_WH_all_tests})
    & \vspace{0.01cm}
    $C_{\textrm{exact}}=2$ (pass)
    & \vspace{0.01cm}
    $C_{\textrm{exact}}=1$ (fail)
    & \vspace{0.01cm}
    $C_{\textrm{exact}}=2$ (pass)
    & \vspace{0.01cm}
    $C_{\textrm{exact}}=2$ (fail) \\  
    IBVP: WH (Fig.~\ref{Fig:IBVP_SYMH_vs_WH_all_tests})
    & $C_{\textrm{exact}}=1$ (fail)
    & $C_{\textrm{exact}}=2$ (pass)
    & $C_{\textrm{exact}}=1$ (fail)
    & $C_{\textrm{exact}}=2$ (fail) \\  
    CIBVP: SYMH (Fig.~\ref{Fig:CIBVP_SYMH_vs_WH_all_tests})
    & $C_{\textrm{exact}}=2$ (pass)
    & $C_{\textrm{exact}}=1$ (fail)
    & $C_{\textrm{exact}}=2$ (pass)
    & $C_{\textrm{exact}}=2$ (fail) \\  
    CIBVP: WH (Fig.~\ref{Fig:CIBVP_SYMH_vs_WH_all_tests})
    & $C_{\textrm{exact}}=1$ (fail)
    & $C_{\textrm{exact}}=2$ (pass)
    & $C_{\textrm{exact}}=1$ (fail)
    & $C_{\textrm{exact}}=2$ (fail) \\  
    CIBVP of CCE: SYMH-WH (Fig.~\ref{Fig:CCE_CIBVP_SYMH_WH_all_tests})
    & $C_{\textrm{exact}}=1$ (fail)
    & $C_{\textrm{exact}}=1$ (fail)
    & $C_{\textrm{exact}}=1$ (fail)
    & $C_{\textrm{exact}}=2$ (pass) \\
    CIBVP of CCE: inhom. SYMH-WH (Fig.~\ref{Fig:inhom_CCM_CCE_comparison_tests234})
    & no test
    & no test
    & no test
    & $C_{\textrm{exact}}=2$ (pass) \\
    CCM: SYMH-SYMH (Fig.~\ref{Fig:CCM_SYMH_SYMH_WH_WH_all_tests})
    & $C_{\textrm{exact}}=2$ (pass)
    & $C_{\textrm{exact}}=1$ (fail)
    & $C_{\textrm{exact}}=2$ (pass)
    & $C_{\textrm{exact}}=2$ (fail) \\  
    CCM: WH-WH (Fig.~\ref{Fig:CCM_SYMH_SYMH_WH_WH_all_tests})
    & $C_{\textrm{exact}}=1$ (fail)
    & $C_{\textrm{exact}}=2$ (pass)
    & $C_{\textrm{exact}}=1$ (fail)
    & $C_{\textrm{exact}}=2$ (fail) \\  
    CCM: SYMH-WH (Fig.~\ref{Fig:CCM_SYMH_WH_all_tests})
    & $C_{\textrm{exact}}=1$ (fail)
    & $C_{\textrm{exact}}=1$ (fail)
    & $C_{\textrm{exact}}=1$ (fail)
    & $C_{\textrm{exact}}=2$ (fail) \\  
    CCM: inhom. SYMH-SYMH (Fig.~\ref{Fig:inhom_CCM_CCE_comparison_tests234})
    & no test
    & no test
    & $C_{\textrm{exact}}=2$ (pass)
    & no test \\
    CCM: inhom. WH-WH (Fig.~\ref{Fig:inhom_CCM_CCE_comparison_tests234})
    & no test
    & $C_{\textrm{exact}}=2$ briefly (fail)
    & no test
    & no test \\
    CCM: inhom. SYMH-WH (Fig.~\ref{Fig:CCM_SYMH_B1_WH_B2_all_tests})
    & no convergence (fail)
    & no convergence \newline(fail)
    & no convergence (fail)
    & no convergence (fail) \\
    \hline
    \hline
  \end{tabular}
  \caption{Summary of the convergence tests presented in
    Sec.~\ref{Section:numerics}. The first column shows the PDE setups
    that we have tested. All the setups are homogeneous, unless stated
    otherwise (``inhom.'' stands for ``inhomogeneous''). The second,
    third and forth columns show convergence tests where the same norm
    is used for the given data and the solution, whereas the last
    column shows a test where the given data are controlled in a norm
    bigger than the one we use to check the convergence of the
    solution. A test is considered as~\textit{pass}
    when~$C_{\textrm{exact}}=2$ for the solution in the same norm as
    the given data, and~\textit{fail} otherwise (see text). Given our
    working definitions of pass and fail, the~$H_1-q$ test should
    always fail by construction. The only case when we
    consider~~$C_{\textrm{exact}}=2$ a pass for this test is for the
    CIBVP part of a SYMH-WH CCE setup. In this case, the~$H_1$
    controlled given data for the SYMH IBVP provide a solution that
    feeds boundary data to the WH CIBVP with enough control over the
    appropriate derivative terms. All our numerical convergence
    results are in line with our expectations from theory.}
  \label{Table:convergence_tests}
\end{table*}

We run tests with given data that exhibit~$C_{\textrm{exact}}=2$ in
the appropriate~$L^2$, $q$, or~$H_1$ norm for the setup under
consideration, and monitor the~$C_{\textrm{exact}}$ of the solution in
its~$L^2$, $q$, or~$H_1$ norm. We consider the test as \textit{passed}
if the solution converges in the same norm as the given data, at the
same order; that is its~$C_{\textrm{exact}}$ tends to~$2$ with
increasing resolution. In any other case we consider the test as
\textit{failed}. In our numerical experiments we see three different
ways in which a test can fail:
\begin{enumerate}
\item $C_\textrm{exact}$ of the solution tends to a value different
  from 2 with increasing resolution, in the same norm as the given
  data.
\item $C_\textrm{exact}$ of the solution tends to 2 with increasing
  resolution, but in a norm different than the given data.
\item $C_\textrm{exact}$ does not tend to any fixed value with
  increasing resolution.
\end{enumerate}
The first and second failing scenarios can actually happen for the
same solution, if we just monitor its convergence in a different
norm.

One argument to support our working definition of a passed and failed
test is the following. Consider inequality~\eqref{ineq:contin_dep} at
the semidiscrete level, for instance at resolution~$h$, namely
\begin{align*}
  || \mathbf{u}_{h}||_{h} \leq K e^{\alpha t} || f_{h} ||_{h}
  \,,
\end{align*}
with~$K, \alpha$ real constants and~$K \geq 1$. For a passed test we
have that
\begin{align*}
  4 \simeq \frac{|| f_{h} ||_{h}}{|| f_{h/2} ||_{h/2}}
  = \frac{|| \mathbf{u}_{h} ||_{h}}{|| \mathbf{u}_{h/2} ||_{h/2}}
  \,,
\end{align*}
where $\simeq$ here denotes equality up to terms of order~$O(h)$,
and therefore
\begin{align*}
  || \mathbf{u}_{h/2}||_{h/2} \leq K e^{\alpha t} || f_{h/2} ||_{h/2}
  \,.
\end{align*}
If this exact convergence rate is maintained for higher resolution
runs, then we can imagine taking the limit of infinite resolution and
recovering inequality~\eqref{ineq:contin_dep} at the continuum. If, in
contrast, during the tests we see the first type of failure, so that
\begin{align*}
  \frac{|| \mathbf{u}_{h} ||_{h}}{|| \mathbf{u}_{h/2} ||_{h/2}} \simeq c
   \neq 4 \simeq \frac{|| f_{h} ||_{h}}{|| f_{h/2} ||_{h/2}} 
  \,,
\end{align*}
with~$c$ some real positive constant, then after doubling the
resolution~$n$ times we obtain
\begin{align*}
  || \mathbf{u}_{h/2^n}||_{h/2^n}
  &
    \leq \left( \frac{4}{c} \right)^n K e^{\alpha t} || f_{h/2^n} ||_{h/2^n}
  \\
  &= \tilde{K} e^{\alpha t} || f_{h/2^n} ||_{h/2^n}
  \,,
\end{align*}
with~$\tilde{K} \equiv \left(4/c\right)^n K$. If~$c<4$ consistently,
then in the limit of infinite
resolution~$n \rightarrow \infty$,~$\tilde{K} \rightarrow \infty$,
whereas if~$c>4$ consistently,~$\tilde{K} \rightarrow 0$
as~$n \rightarrow \infty$. In either case one does not recover the
continuum version of inequality~\eqref{ineq:contin_dep}, which we
interpret as failure of the solution to be controlled by the given
data. In this argument we consider the theoretically expected value
of~$C_{\textrm{exact}}$ for both the solution and the given data, but
in practice, we look for a~$C_{\textrm{exact}}$ that tends to that
value.

The second type of failure is an interesting scenario realized in our
tests, where we obtain the same convergence rate for the solution and
the given data, but only in different norms,
\begin{align*}
  4 \simeq \frac{|| f_{h} ||_{H_1}}{|| f_{h/2} ||_{H_1}}
  = \frac{|| \mathbf{u}_{h} ||_{q}}{|| \mathbf{u}_{h/2} ||_{q}}
  \,,
\end{align*}
where we have dropped the subscript~$ \sim h$ in the different
discrete norms. In this case, one can at most obtain the following
inequality:
\begin{align}
  || \mathbf{u}||_{q} \leq K e^{\alpha t} || f ||_{H_1}
  \,,
  \label{ineq:q_norm_sol_H1_data}
\end{align}
by considering again the limit of infinite resolution, This however is
not the same as inequality~\eqref{ineq:contin_dep}, which is involved
in the textbook definition of well-posedness.  More importantly,
since~$|| \mathbf{u} ||_q$ is smaller than~$|| \mathbf{u} ||_{H_1}$
(controls fewer derivative terms), it is perfectly possible to obtain
a solution~$\mathbf{u}$ with unbounded~$H_1$, but bounded~$q$ norm. In
this case, the blowup would appear in derivatives of the solution that
are included in~$H_1$ but not in the~$q$ norm. We therefore consider
this scenario also a failure.

The lowest resolution we use for our tests is denoted by~$h_0$ and has
$N_\rho = N_x = 17$ and $N_z = 16$. The grid points and grid spacing
for the different resolutions are set according to
\begin{align*}
  &
    N_\rho = N_x = 16^{2 D} + 1
  \,, \quad
  N_z = 16^{2 D}
  \,,
  \\
  &
    h_\rho = h_x= \frac{1}{N_\rho -1}
  \,, \quad
  h_z = \frac{2\pi}{N_z}
  \,,
\end{align*}
and~$h_D$ labels the different resolutions, for instance~$D=0$
corresponds to the lowest resolution~$h_0$. We perform tests
for~$D=0,1,2,3,4$, and compute~$C_{\textrm{exact}}$ only for the
timesteps that are common across all resolutions, in other words at
each timestep with~$h_0$ grid spacing. The code and parameter files
for the convergence tests can be found in~\cite{GiaBisHil23_github}
and the data used to produce the following convergence plots
in~\cite{GiaBisHil23_zenodo}. In the following subsections, all the
norms should be understood as evaluated on the numerical
grid. Table~\ref{Table:convergence_tests} contains the summary of our
numerical experiments.

\subsection{IBVP}
\label{Subsection:ibvp_numerics}

First we examine numerical convergence of the Cauchy setup. The
discrete norms in which we test the convergence of the solution are
\begin{align}
  &
    ||\mathbf{u}_1||^2_{L^2}
    =
    \sum_{\rho,z} \left(\phi_1^2 + \psi_{\varv1}^2 + \psi_1^2 \right) \,h_\rho h_z
    \label{eq:L2_cauchy_numerics}
  \\
  &
    \qquad \qquad  +\sum_{t,\rho=-1, z}  \psi_1^2 \,h_t h_z
    +
    \sum_{t,\rho=0,z}  \left( \phi_1^2 + \psi_{\varv1}^2\right) \,h_t h_z
    \,,
    \nonumber
  \\
  &
    ||\mathbf{u}_1||^2_{q}
    =
    \sum_{\rho,z} \left[\phi_1^2 + \psi_{\varv1}^2 + \psi_1^2
    + \left( D_z \phi_1\right)^2 \right] \,h_\rho h_z
    \label{eq:q_cauchy_numerics}
  \\
  &
    +\sum_{t,\rho=-1, z}  \psi_1^2 \,h_t h_z
    +
    \sum_{t,\rho=0,z}
    \left[ \phi_1^2 + \psi_{\varv1}^2 + \left( D_z \phi_1 \right)^2
    \right] \,h_t h_z
    \,,
    \nonumber
  \\
  &
    ||\mathbf{u}_1||^2_{H_1}
    =
    \sum_{\rho,z} \left[
    \phi_1^2 + \psi_{\varv1}^2 + \psi_1^2 \right.
    \label{eq:H1_cauchy_numerics}
  \\
  & \left. \qquad \qquad \qquad
     +\left(D_\rho \phi_1 \right)^2 + \left( D_\rho \psi_{\varv1} \right)^2
    + \left( D_\rho \psi_1 \right)^2
     \right.
    \nonumber \\
  & \left. \qquad \qquad \qquad
     +\left(D_z \phi_1 \right)^2 + \left( D_z \psi_{\varv1} \right)^2
    + \left( D_z \psi_1 \right)^2
    \right] \,h_\rho h_z
    \nonumber
  \\
  &
    \qquad
    +\sum_{t,\rho=-1, z}
    \left[ \psi_1^2 + \left( D_\rho \psi_1 \right)^2
    + \left( D_z \psi_1 \right)^2  \right]
    \,h_t h_z
    \nonumber
  \\
  &
    \quad
    +\sum_{t,\rho=0,z}
    \left[ \phi_1^2 + \psi_{\varv1}^2 +
    \left( D_\rho \phi_1 \right)^2 + \left( D_\rho \psi_{\varv1} \right)^2
    \right.
    \nonumber
  \\
  &
    \qquad \qquad
    \left.
    + \left( D_z \phi_1 \right)^2 + \left( D_z \psi_{\varv1} \right)^2
    \right] \,h_t h_z
    \,.
    \nonumber
\end{align}

In our tests, we choose given data that converge in either
the~$L^2$,~$q$, or~$H_1$ norms. Based on the energy estimates given
Sec.~\ref{Subsection:cauchy_models}, these can be obtained by
considering
Eqs.~\eqref{eq:L2_cauchy_numerics}-\eqref{eq:H1_cauchy_numerics}, but
where the spacelike hypersurface sums are evaluated only for the
initial data, and the worldtube sums on the left boundary for
right-moving fields and on the right boundary for left-moving fields,
i.e., the opposite of what is shown. We perform four different tests:
\begin{enumerate}
\item $L^2-L^2$ test: convergence of the solution in the~$L^2$
  norm~\eqref{eq:L2_cauchy_numerics}, for~$L^2$ given data.
\item $q-q$ test: convergence of the solution in the~$q$
  norm~\eqref{eq:q_cauchy_numerics}, for~$q$ given data.
\item $H_1-H_1$ test: convergence of the solution in the~$H_1$
  norm~\eqref{eq:H1_cauchy_numerics}, for~$H_1$ given data.
\item $H_1-q$ test: convergence of the solution in the~$q$
  norm~\eqref{eq:q_cauchy_numerics}, for~$H_1$ given data.
\end{enumerate}

The~$L^2-L^2$,~$q-q$, and~$H_1-H_1$ tests examine, at the discrete
level, continuous dependence of the solution on the given data as
described by inequality~\eqref{ineq:contin_dep}. In contrast,
the~$H_1-q$ test examines the discrete version of
inequality~\eqref{ineq:q_norm_sol_H1_data}. If the latter is
satisfied, it shows that a solution is controlled by given data that
converge in a bigger norm. In our particular setup, the~$q$
norm~\eqref{eq:q_cauchy_numerics} is smaller than
the~$H_1$~\eqref{eq:H1_cauchy_numerics} in the sense that it allows
control over fewer derivative terms. It then becomes clear that a
blowup in a derivative of the solution can still happen, but fail to
be detected by the~$q$ norm. We consider this test because it can
explain why CCE involving a SYMH IBVP and a WH CIBVP can result in
well-posed PDE problems that exhibit the expected numerical
convergence. But the~$H_1-q$ test does not address continuous
dependence as understood in the standard sense for well-posedness as
in inequality~\eqref{ineq:contin_dep}.
\begin{figure*}[tph]
\includegraphics[width=1\textwidth]{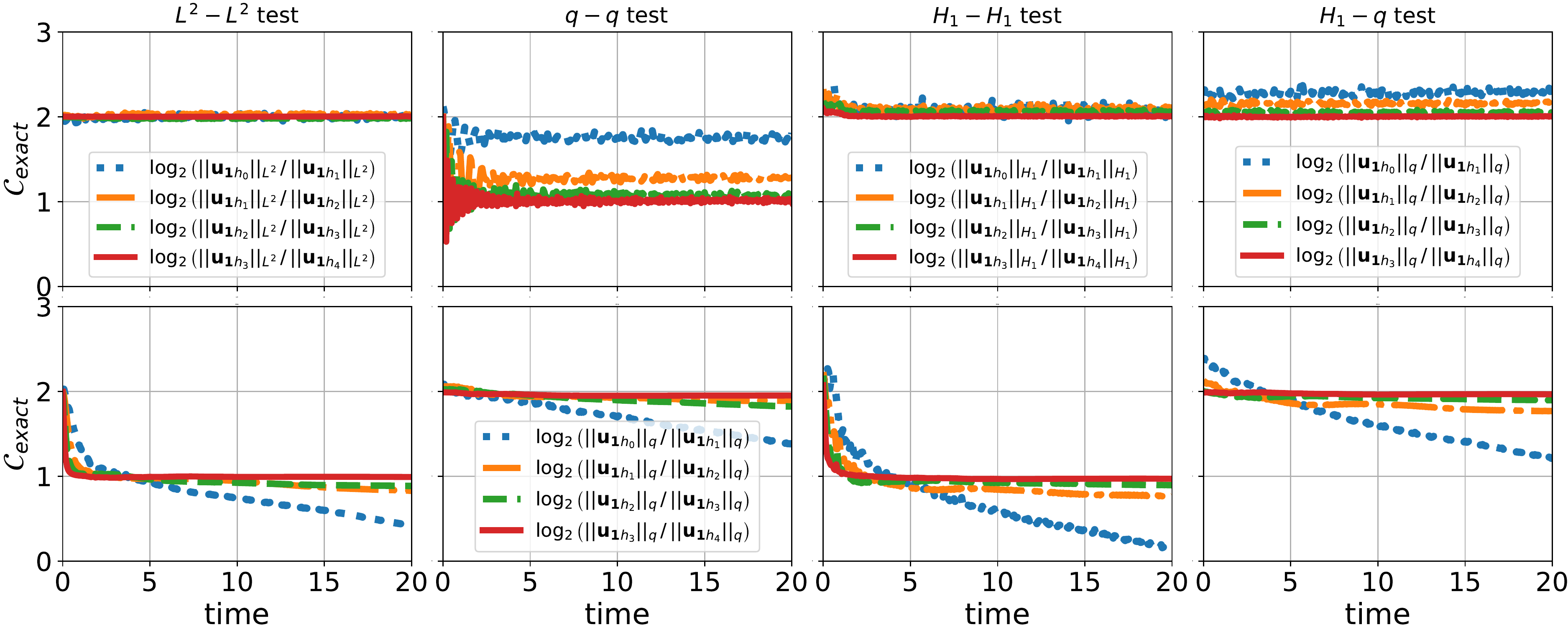}
\caption{The exact convergence rate~$\mathcal{C}_{\textrm{exact}}$ is
  shown for the IBVP, for all the tests. The top row refers to the
  SYMH setup, and the bottom to the WH one. The left column presents
  the results for the~$L^2-L^2$ test, the second to left for the~$q-q$
  test, and so on.  The results of the~$L^2-L^2$,~$q-q$, and~$H_1-H_1$
  tests are as expected by theory, that is convergence in the~$L^2$
  and~$H_1$ for the SYMH case and in the $q$ norm for the WH one. The
  convergence in the~$q$ norm seen in the~$H_1-q$ test is merely a
  result of the fact that $H_1$~\eqref{eq:H1_cauchy_numerics} is a
  bigger norm than $q$~\eqref{eq:q_cauchy_numerics}, in the sense that
  it controls more derivative terms. We only monitor the $\p_z \phi_1$
  derivative term, which still converges appropriately for the SYMH
  case and due to its size in comparison to the nonderivate terms,
  dominates the $q$ norm. We also see good convergence of the solution
  to the WH IBVP in the~$H_1-q$ test, because we ignore the derivative
  terms that do not exhibit convergence. The plots for the~$H_1-H_1$
  and~$H_1-q$ tests use data from exactly the same runs.}
  \label{Fig:IBVP_SYMH_vs_WH_all_tests}
\end{figure*}

The findings of our numerical tests are shown in
Fig.~\ref{Fig:IBVP_SYMH_vs_WH_all_tests}. The top row of the figure
shows~$\mathcal{C}_{\textrm{exact}}$ for a SYMH IBVP setup, whereas
the bottom is for a WH one. Each column refers to one test, starting
from the left and moving to the right with
the~$L^2-L^2$,~$q-q$,~$H_1-H_1$, and finally the~$H_1-q$ test. The
first three provide the expected convergence results, that is, for the
SYMH case, good empirical convergence of the solution in~$L^2$
and~$H_1$ norms and loss of convergence in the~$q$ norm. In contrast
in the WH case, we see~$\mathcal{C}_{\textrm{exact}} =2$ only for
the~$q-q$ test. The~$H_1-q$ test returns interesting and potentially
confusing results: both solutions to the SYMH and WH setups exhibit
good convergence in the~$q$ norm for given data controlled in
their~$H_1$ norm. This can be understood in the following terms:
\begin{enumerate}
\item SYMH: by monitoring the $q$ norm, we take into account only
  the~$\p_z \phi_1$ derivative of the solution. This part still
  converges appropriately, and due to its relative size in comparison
  to the nonderivative terms, dominates the convergence rate. This
  does not mean that the~$H_1-q$ test provides numerical evidence for
  well-posedness of the SYMH IBVP in the $q$ norm, which would be the
  case if the~$q-q$ test was passed.
\item WH: when we monitor the $q$ norm, we ignore all the derivatives
  that appear in~$H_1$ and not in~$q$, which are responsible for the
  loss of convergence. We only monitor the~$\p_z \phi_1$ term, which
  converges properly and due to its relative size, dominates the
  nonderivative terms of the~$q$ norm. Since we are considering the
  homogeneous case, the loss of convergence of the nonmonitored
  derivative terms does not propagate to~$\p_z \phi_1$ due to the lack
  of a coupling among them. Finally, proper convergence of the given
  data in the $H_1$ norm implies the same in the~$q$ norm, since~$H_1$
  is bigger than~$q$.
\end{enumerate}

When we see loss of convergence in these tests we
observe~$C_{\textrm{exact}}=1$, and since this is not the expected
rate of convergence for our numerical scheme, we consider these tests
as failed. However, it is often the case in physically interesting
simulations that the convergence order of the solution is lower than
that of the numerical scheme used (for instance shockwave capturing or
schemes that include interpolation). Here, we insist in considering
these tests as failed, because we do not have additional elements in
the scheme that can drop the order of convergence (the data are smooth
plus the artificially exaggerated numerical error, no interpolations)
and more importantly we try to make contact with the continuum energy
estimates. As a final remark on this point, we should mention that it
is not clear why~$C_{\textrm{exact}}=1$ in these cases. A possibility
is that it relates to the bulk integrals in the continuum analysis
that prevent us from obtaining energy estimates for the scenarios of
the failed tests. Obtaining however a concrete answer requires more
testing and is beyond the scope of this work.

\subsubsection{Synopsis} The SYMH IBVP exhibits second order
convergence in the~$L^2$ and~$H_1$ norms for given data controlled in
the same norms, and second order convergence in the~$q$ norm only when
the given data are controlled in the~$H_1$ norm. The WH IBVP shows
second order convergence in the~$q$ norm for given data controlled in
either their~$q$ and~$H_1$ norms. These results are inline with the
expectations from theory.

\subsection{CIBVP}
\label{Subsection:cibvp_numerics}

We repeat the tests for the CIBVP, monitoring the convergence of the
solution in the following discrete norms
\begin{align}
  &
    ||\mathbf{u}_2||^2_{L^2}
    =
    \sum_{x,z} \psi_2^2 \,h_x h_z
    +
    \sum_{u,x=0, z} \psi_2^2 \,h_u h_z
    \label{eq:L2_char_numerics}
  \\
  &
    \qquad \qquad  
    +\textrm{max}_x \sum_{u,z}  \left( \phi_2^2 + \psi_{\varv2}^2\right) \,h_u h_z
    \,,
    \nonumber  
\end{align}
\begin{align}
  &
    ||\mathbf{u}_2||^2_{q}
    =
    \sum_{x,z} \psi_2^2 \,h_x h_z
    +
    \sum_{u,x=0, z}  \psi_2^2 \,h_u h_z
    \label{eq:q_char_numerics}
  \\
  &
    \qquad \qquad
    +\textrm{max}_x \sum_{u,z}
    \left[ \phi_2^2 + \psi_{\varv2}^2 + \left( D_z \phi_2 \right)^2
    \right] \,h_u h_z
    \,,
    \nonumber
\end{align}
and
\begin{align}
  &
    ||\mathbf{u}_2||^2_{H_1}
    =
    \sum_{x,z} \left[
    \phi_2^2 + \left(D_x \psi_2 \right)^2 + \left( D_z \psi_2 \right)^2
    \right] \,h_x h_z
    \label{eq:H1_char_numerics}
  \\
  &
    \qquad+
    \sum_{u,x=-0, z}
    \left[ \psi_2^2 + \left( D_x \psi_2 \right)^2
    + \left( D_z \psi_2 \right)^2  \right]
    \,h_u h_z
    \nonumber
  \\
  &
    \quad+
    \textrm{max}_x \sum_{u,z}
    \left[ \phi_2^2 + \psi_{\varv2}^2 +
    \left( D_x \phi_2 \right)^2 + \left( D_x \psi_{\varv2} \right)^2
    \right.
    \nonumber
  \\
  &
    \qquad \qquad
    \left.+
    \left( D_z \phi_2 \right)^2 + \left( D_z \psi_{\varv2} \right)^2
    \right] \,h_u h_z
    \,.
    \nonumber
\end{align}

\begin{figure*}[tph]
\includegraphics[width=1\textwidth]{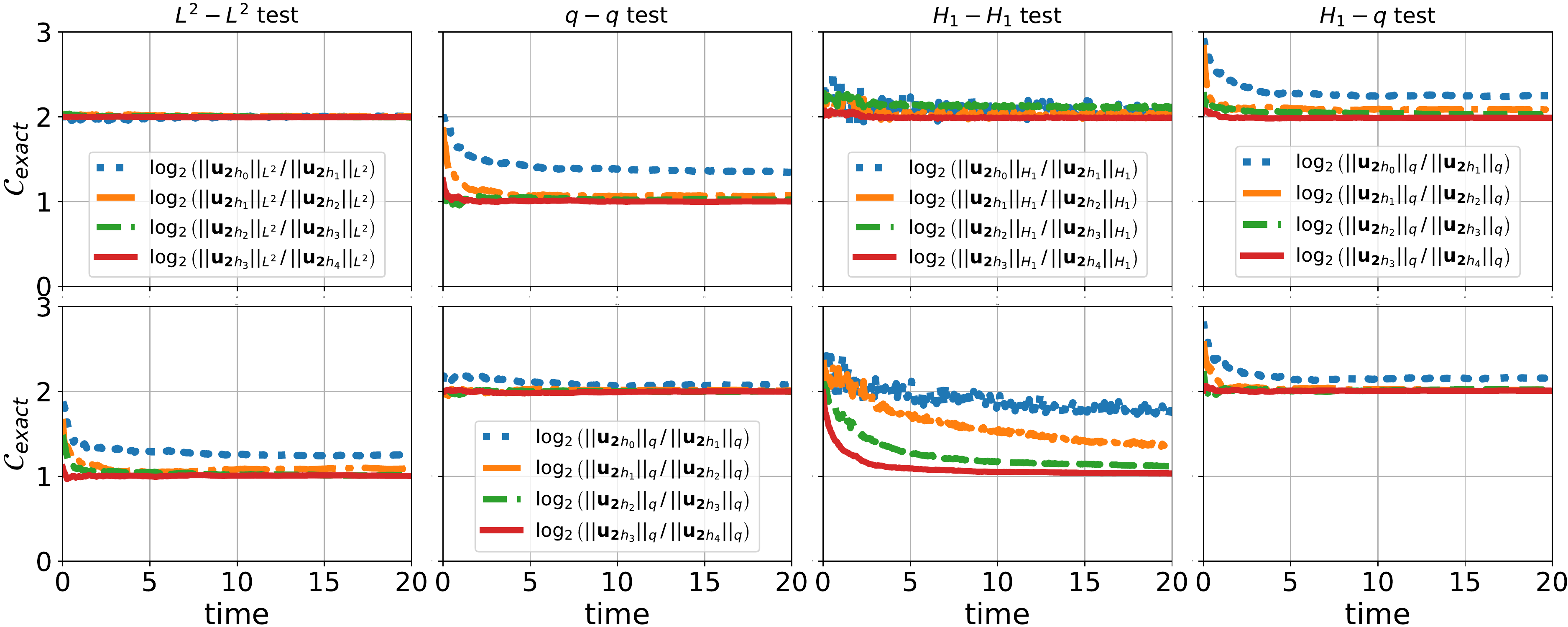}
\caption{Exact convergence rate~$\mathcal{C}_{\textrm{exact}}$ for the
  CIBVP solution, for our tests in the different norms. The SYMH setup
  is shown in top row and the WH one in the bottom. Each column shows
  the results of each test. The~$L^2-L^2$,~$q-q$, and~$H_1-H_1$ tests
  exhibit the expected convergence in the~$L^2$ and~$H_1$ for SYMH and
  in the $q$ norm for WH. Convergence in the~$q$ norm seen in
  the~$H_1-q$ test is again a result of the fact that
  $H_1$~\eqref{eq:H1_char_numerics} is bigger than the
  $q$~\eqref{eq:q_char_numerics} norm, as in the IBVP. The plots for
  the~$H_1-H_1$ and~$H_1-q$ tests use data from exactly the same
  runs.}
  \label{Fig:CIBVP_SYMH_vs_WH_all_tests}
\end{figure*}

\subsubsection{Synopsis} The results are qualitatively the same as for
the IBVP, and are summarized in
Fig.~\ref{Fig:CIBVP_SYMH_vs_WH_all_tests}. We again find the expected
convergence rate in $L^2$ and $H_1$ for the SYMH and in the $q$ norm
for the WH setup. The~$H_1-q$ test provides apparent convergence in
the~$q$ norm for both setups, which is understood exactly in the same
way as in the IBVP.

\subsection{CCE}
\label{Subsection:cce_numerics}

The conclusions drawn from the previous tests for the IBVP and CIBVP shape our
expectations for CCE. In this setup, convergence of the numerical
solution is still examined for the individual IBVP and CIBVP parts of
the CCE scheme. The important difference now is that part of the IBVP
solution serves as boundary data (for the right-moving fields) for the
CIBVP. When performing these tests for CCE we expect the following
results:
\begin{enumerate}
\item CCE between a SYMH IBVP and a SYMH CIBVP: both IBVP and CIBVP
  solutions converge in~$L^2$ and~$H_1$, but not in the~$q$
  norm. The~$H_1-q$ test has the same qualitative results as in
  Secs.~\ref{Subsection:ibvp_numerics}
  and~\ref{Subsection:cibvp_numerics}.
\item WH-WH CCE: Both IBVP and CIBVP converge only in the $q$ norm,
  but not in~$L^2$ or~$H_1$. The~$H_1-q$ test shows again that both
  solutions converge in their $q$ norms for the $H_1$ given data.
\item SYMH-WH CCE: The solution to the IBVP exhibits the same behavior
  as in Sec.~\ref{Subsection:ibvp_numerics}. The solution to the CIBVP
  can converge properly only in the~$q$ norm. The boundary data to the
  CIBVP are part of the solution to the IBVP and so we need a solution
  to the IBVP that converges properly in the~$H_1$ norm, such that the
  boundary data to the CIBVP also converge properly in the~$q$ norm
  (neglecting the derivatives that are in~$H_1$ but not in~$q$). The
  aforementioned setup is understood in the~$H_1-q$ test. In addition,
  choosing characteristic initial data that converge in~$L^2$ or~$H_1$
  does not affect the convergence of the CIBVP solution in the~$q$
  norm. The reason is that the derivative term~$\p_z \phi_2$ dominates
  the characteristic $q$ norm~\eqref{eq:q_char_numerics}. For our
  tests we drop the amplitude of~$\phi_2(u=0,x,z)$ by a factor of 8
  every time the grid spacing is halved, that is compatible with
  $H_1$.
\end{enumerate}

\begin{figure}[th]
\includegraphics[width=0.5\textwidth]{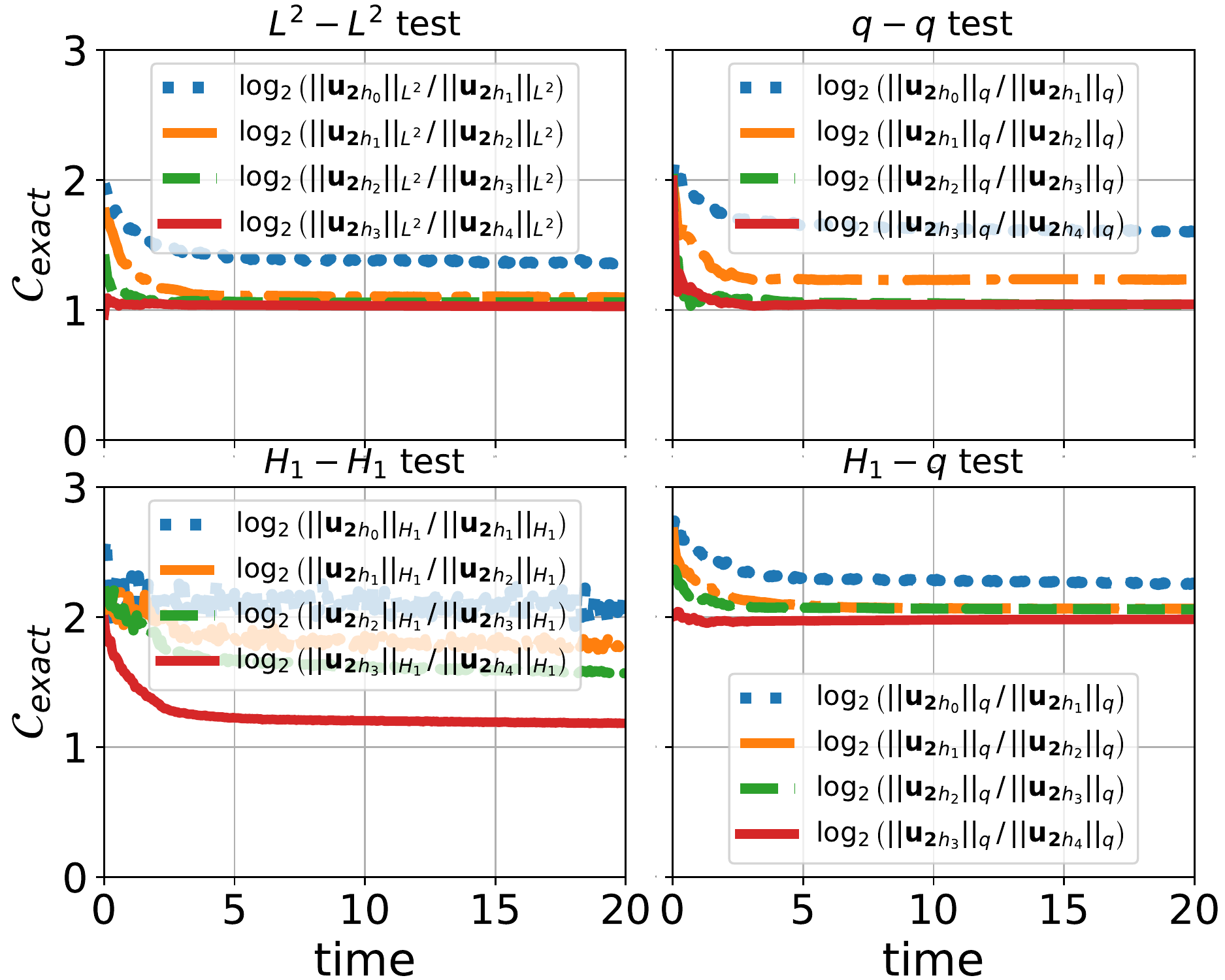}
\caption{$\mathcal{C}_{\textrm{exact}}$ for all our tests, for the
  CIBVP part of a CCE setup between a SYMH IBVP and a WH CIBVP. As
  expected, only the~$H_1-q$ test provides good convergence. In this
  case, the~$H_1$ given data provide an $H_1$ controlled solution to
  the IBVP, part of which serves as boundary data for the WH
  CIBVP. That control is more than enough to have $q$ controlled given
  data for the CIBVP, and hence, the solution to the CIBVP is
  controlled in the $q$ norm.}
  \label{Fig:CCE_CIBVP_SYMH_WH_all_tests}
\end{figure}

Repeating these tests for CCE, our expectations were verified. The
SYMH-WH scenario is the relevant one for current implementations of
CCE in GR simulations, and therefore we present only these results in
Fig.~\ref{Fig:CCE_CIBVP_SYMH_WH_all_tests}. We see that indeed the
solution to the WH CIBVP setup exhibits the appropriate convergence
rate only when~$H_1$ data are given for the IBVP. The solution to that
IBVP provides boundary data for the CIBVP that have~$H_1$ control,
which is more than the necessary control in the~$q$ norm. This result
is compatible with that of the~$H_1-q$ test in
Sec.~\ref{Subsection:cibvp_numerics}.

\subsubsection{Synopsis} CCE between a SYMH IBVP and a weakly
well-posed WH CIBVP, can only provide a characteristic solution with
second order convergence in its~$q$ norm, if the solution of the SYMH
IBVP is controlled in its~$H_1$ norm. This is inline with our
expectations from theory.

\subsection{CCM}
\label{Subsection:ccm_numerics}

\begin{figure*}[tph]
  \includegraphics[width=1\textwidth]{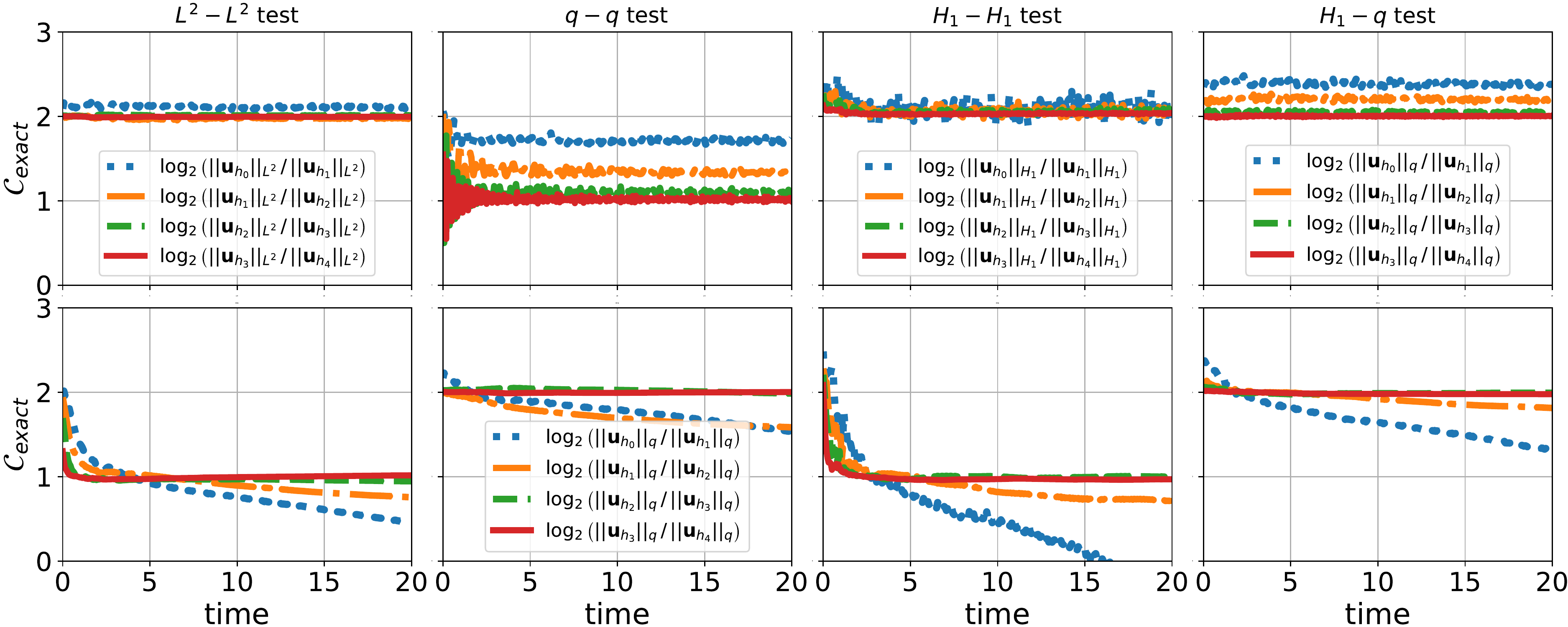}
  \caption{The exact convergence rate of the CCM solution, for
    SYMH-SYMH (top) and WH-WH (bottom). The first column refers to
    the~$L^2-L^2$ test, the second to the~$q-q$ test, and so
    on. The~$L^2-L^2$,~$q-q$, and~$H_1-H_1$ tests show the expected
    convergence in the~$L^2$ and~$H_1$ for SYMH-SYMH and in the $q$
    norm for WH-WH. The convergence in the~$q$ norm seen in
    the~$H_1-q$ test is again a result of the fact that
    $H_1$~\eqref{eq:ccm_H1_norm_numerics} is bigger than
    $q$~\eqref{eq:ccm_q_norm_numerics}, as in the IBVP. The plots for
    the~$H_1-H_1$ and~$H_1-q$ tests use data from exactly the same
    runs.}
  \label{Fig:CCM_SYMH_SYMH_WH_WH_all_tests}
\end{figure*}

The discrete norms for which we run our tests for CCM are
\begin{align}
  &
    ||\mathbf{u}||^2_{L^2}
    =
    \sum_{\rho,z} \left(\phi_1^2 + \psi_{\varv1}^2 + \psi_1^2 \right) \,h_\rho h_z
    +
    \sum_{x,z} \psi_2^2 \,h_x h_z
    \nonumber
  \\
  &
    +\sum_{t,\rho=-1, z}  \psi_1^2 \,h_t h_z
    +
    \textrm{max}_x \sum_{u,x,z}  \left( \phi_2^2 + \psi_{\varv2}^2\right) \,h_u h_z
    \,.
    \label{eq:ccm_L2_norm_numerics}
\end{align}
\begin{align}
  &
    ||\mathbf{u}||^2_{q}
    =
    \sum_{\rho,z} \left(\phi_1^2 + \psi_{\varv1}^2 + \psi_1^2 \right) \,h_\rho h_z
    +
    \sum_{x,z} \psi_2^2 \,h_x h_z
    \label{eq:ccm_q_norm_numerics}
  \\
  &
    +\sum_{t,\rho=-1, z}  \psi_1^2 \,h_t h_z
    +
    \textrm{max}_x \sum_{u,x,z}
    \left[ \phi_2^2 + \psi_{\varv2}^2 + \left(D_z \phi_2 \right)^2 \right] \,h_u h_z
    \,,
    \nonumber
\end{align}
\begin{align}
  &
    ||\mathbf{u}||^2_{H_1}
    =
    \sum_{\rho,z}
    \left[
    \phi_1^2 + \psi_{\varv1}^2 + \psi_1^2
    + \left(D_\rho \phi_1\right)^2
    + \left(D_\rho \psi_{\varv1} \right)^2
    \right.
    \nonumber
  \\
  &
    \left.
    +\left(D_\rho \psi_1 \right)^2
    + \left(D_z \phi_1\right)^2
    + \left(D_z \psi_{\varv1} \right)^2
    + \left(D_z \psi_1 \right)^2
    \right]
    \,h_\rho h_z
    \nonumber
  \\
  & 
    + \sum_{x,z}
    \left[
    \psi_2^2 + \left( D_x \psi_2 \right)^2
    + \left( D_z \psi_2 \right)^2
    \right]
    \,h_x h_z
    \nonumber
  \\
  & 
    +\sum_{t,\rho=-1, z}
    \left[ \psi_1^2 + \left( D_\rho \psi_1\right)^2
    + \left( D_z \psi_1\right)^2
    \right] \,h_t h_z
    \nonumber
  \\
  &
    +\textrm{max}_x \sum_{u,x,z}
    \left[
    \phi_2^2 + \psi_{\varv2}^2 +
    \left(D_x \phi_2\right)^2 + \left( D_x \psi_{\varv2} \right)^2
    \right.
    \nonumber
  \\
  &
    \qquad \qquad \qquad
    \left.
    +\left(D_z \phi_2\right)^2 + \left( D_z \psi_{\varv2} \right)^2
    \right] \,h_u h_z
    \,.
  \label{eq:ccm_H1_norm_numerics}
\end{align}
First, let use consider matching PDEs with the same degree of
hyperbolicity, that is SYMH to SYMH or WH to WH. The convergence
results of these setups are shown in
Fig.~\ref{Fig:CCM_SYMH_SYMH_WH_WH_all_tests}, with SYMH-SYMH at the
top and WH-WH the bottom row. As expected by the energy estimates of
Sec.~\ref{Subsection:CCE_CCM}, we see good convergence for SYMH-SYMH
in the~$L^2-L^2$ and~$H_1-H_1$ tests that is in
norms~$L^2$~\eqref{eq:ccm_L2_norm_numerics}
and~$H_1$~\eqref{eq:ccm_H1_norm_numerics} respectively, and in
the~$q-q$ test for the WH-WH case, i.e. norm
$q$~\eqref{eq:ccm_q_norm_numerics}. In the~$H_1-q$ test we see
convergence of the solution in the~$q$
norm~\eqref{eq:ccm_q_norm_numerics} for~$H_1$ given data, in line with
the results of the same test for the IBVP and CIBVP. The explanation
of this phenomenon is the same as earlier, that is for the SYMH-SYMH
setup we see the convergence of the~$\p_z \phi_2$ term that dominates
the norm, whereas in the WH-WH setup we ignore the rest of the
derivative terms that do not converge.

\begin{figure}[th]
  \includegraphics[width=0.5\textwidth]{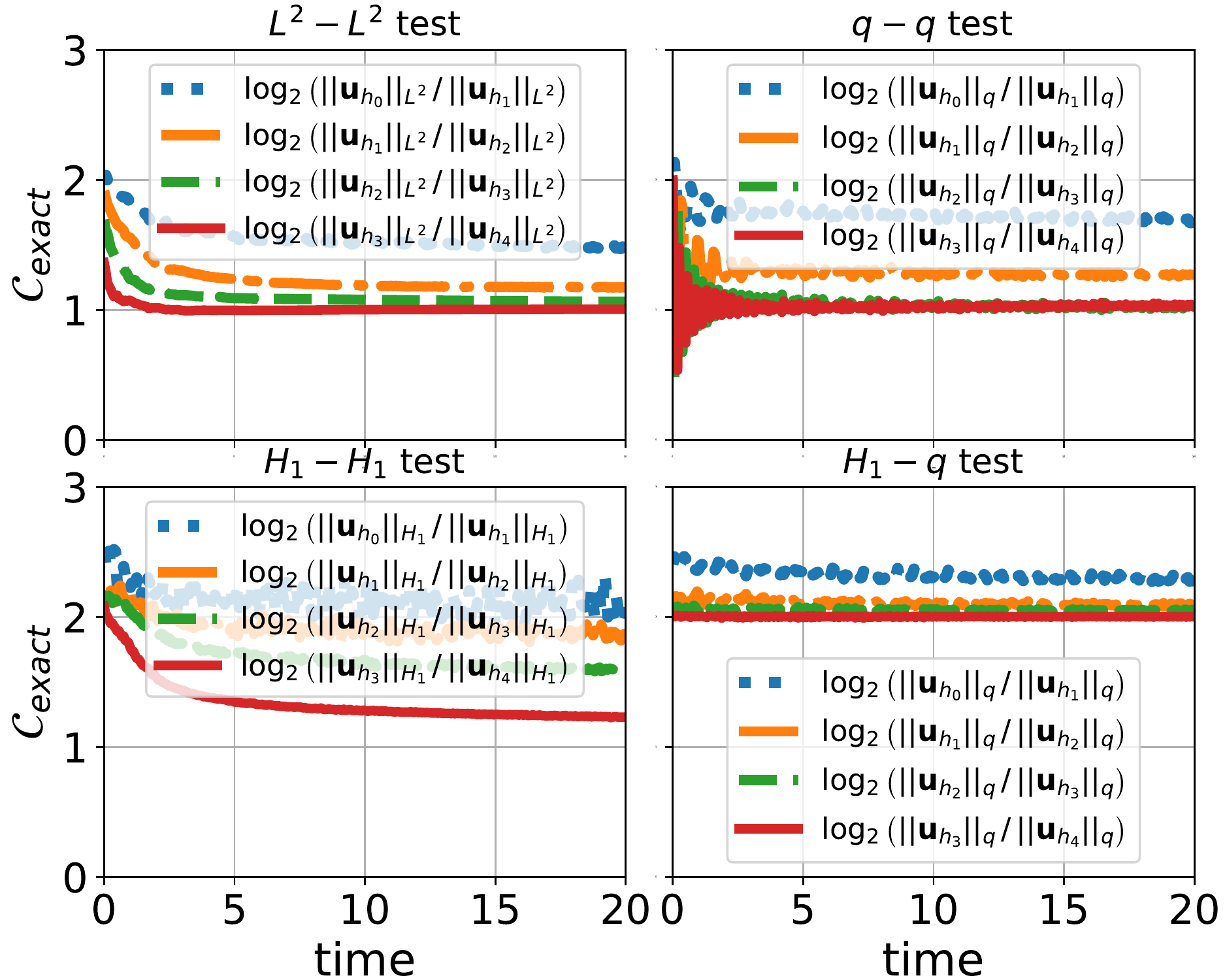}
  \caption{$\mathcal{C}_{\textrm{exact}}$ for a SYMH-WH CCM
    setup. The~$L^2-L^2$ test is shown in top left, the~$q-q$ in top
    right, the~$H_1-H_1$ in bottom left, and the~$H_1-q$ test in
    bottom right. The first three tests fail to exhibit the
    appropriate convergence rate, as expected for this
    setup. The~$H_1-q$ test manifests convergence in the $q$
    norm~\eqref{eq:ccm_q_norm_numerics}, again due to the fact
    that~$H_1$~\eqref{eq:ccm_H1_norm_numerics} is a bigger norm. The
    fact that the behavior seen is similar to that of the CIBVP part
    of CCE, in Fig.~\ref{Fig:CCE_CIBVP_SYMH_WH_all_tests}, is a result
    of treating the homogeneous case. In
    Fig.~\ref{Fig:CCM_SYMH_B1_WH_B2_all_tests} we see that the
    behavior is different for an inhomogeneous SYMH-WH CCM setup.}
  \label{Fig:CCM_SYMH_WH_all_tests}
\end{figure}

Next we consider the matching of a SYMH IBVP to a WH CIBVP, the actual
case for CCM in GR as currently performed. The convergence plots are
presented in Fig.~\ref{Fig:CCM_SYMH_WH_all_tests}. We see that the
solution does not exhibit good convergence in any norm for
the~$L^2-L^2$,~$q-q$, and~$H_1-H_1$ tests, which are the ones that
reflect numerically continuous dependence on given data. Only the~
$H_1-q$ test shows good convergence of the solution in the~$q$ norm
for~$H_1$ given data. This test only provides numerical support for
inequality~\eqref{ineq:q_norm_sol_H1_data} rather
than~\eqref{ineq:contin_dep}.  Comparing
Fig.~\ref{Fig:CCE_CIBVP_SYMH_WH_all_tests} with
Fig.~\ref{Fig:CCM_SYMH_WH_all_tests}, we see the same qualitative
behavior, namely good convergence only for the~$H_1-q$ test. The
latter refers to CCM and the former the CIBVP part of CCE, both for a
SYMH IBVP and a WH CIBVP. The difference between CCM and CCE is only
the additional propagation of information between the characteristic
and Cauchy domain for the left-moving field for CCM. Since we are
considering only the homogeneous case here, we see that there is no
structure in the composite system that couples the right-moving
variables of the characteristic domain that are responsible for weak
hyperbolicity, to the left-moving one. Consequently, the qualitative
behavior in the convergence tests for CCM and CCE is very similar. The
important difference between CCE and CCM is that the~$H_1-q$ test can
actually provide good boundary data for the WH CIBVP in the CCE case,
because then the IBVP and CIBVP are still treated individually
regarding well-posedness. In contrast, for CCM the whole problem has
to be treated simultaneously and so the~$H_1-q$ test is not an
indicator for well-posedness in this case. To clarify the difference
between CCE and CCM further, we consider next the inhomogeneous
setup~\eqref{eq:IBVP_B1} and \eqref{eq:CIBVP_B2}.

\subsubsection{An inhomogeneous example}
\label{Subsubsection:ccm_numerics:inhomogeneous}

We now consider an inhomogeneous IBVP and CIBVP, in which sources are
added in such a way that right- and left-moving variables are coupled
when CCM is performed. The inhomogeneous IBVP (labeled B1) is:
\begin{subequations}
  \begin{align}
  \p_t \phi_1
  & = - \p_\rho \phi_1 + \boxed{\p_z\psi_{\varv1}}
    + \psi_1
    \label{eq:IBVP_B1_phi1}
    \,,
  \\
  \p_t \psi_{\varv1}
  & = - \p_\rho \psi_{\varv1}
    + \p_z\phi_1
    \label{eq:IBVP_B1_psiv1}
    \,,
  \\
  \p_t \psi_1
  & =  \p_\rho \psi_1 + \p_z\psi_1
    \label{eq:IBVP_B1_psi1}
    \,,
  \end{align}
  \label{eq:IBVP_B1}%
\end{subequations}
where the boxed term controls the hyperbolicity of the system as
before. The inhomogeneous CIBVP (labeled B2) is
\begin{subequations}
  \begin{align}
  \p_x \phi_2
  & =  \boxed{\p_z\psi_{\varv2}}
    \label{eq:CIBVP_B2_phi2}
    \,,
  \\
  \p_x \psi_{\varv2}
  & = \p_z\phi_2
    \label{eq:CIBVP_B2_psiv2}
    \,,
  \\
  \p_u \psi_2
  & = \frac{1}{2} \p_\rho \psi_2 + \p_z\psi_2
    + \psi_{\varv2}
    \label{eq:CIBVP_B2_psi2}
    \,.
  \end{align}
  \label{eq:CIBVP_B2}%
\end{subequations}
When the boxed term is included the systems are SYMH, and when omitted
they are only WH. In comparison to the earlier homogeneous case, we
made an addition of source terms that is minimal in the following
sense: in the characteristic setup, the left-moving part is affected
by the right-moving one through the sources, whereas in the Cauchy
part, the opposite is true, that is the right-moving part is affected
by the left-moving one through the sources. Notice that the
characteristic setup has still a nested structure as is desirable in a
model for GR. The important difference in comparison to the
homogeneous SYMH-WH CCM case is that now the right-moving
characteristic variables that form a nontrivial Jordan block in the
angular principal part, can affect the left-moving characteristic
variable, which then influences the solution to the SYMH IBVP. The
latter eventually provides inappropriate boundary data for the
right-moving variables of the CIBVP, enhancing the effect of weak
hyperbolicity. This coupling provides the aforementioned interaction
loop between the left- and right-moving variables when CCM is
performed, but not for CCE, since then boundary data for the IBVP are
not given by the solution to the CIBVP.

\begin{figure*}[tph]
  \includegraphics[width=1\textwidth]{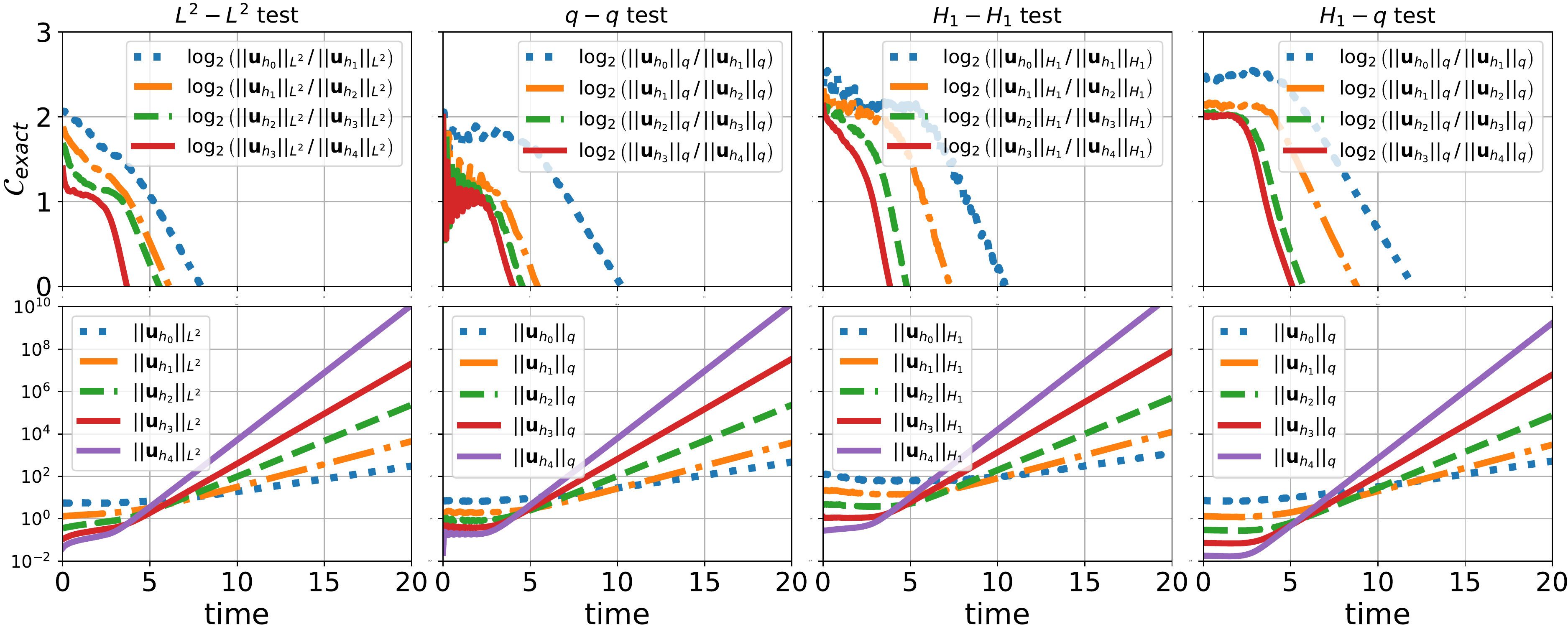}
  \caption{Our convergence tests for the CCM setup SYMH B1-WH B2. The
    top row shows the exact convergence rate, whereas the bottom the
    norms that are used to calculate it. The y axis for the bottom
    plots is $\log_{10}$. We clearly see that for all
    tests,~$\mathcal{C}_{\textrm{exact}}$ moves away from the
    theoretical value faster with increasing resolution. This is a
    clear sign of loss of convergence. Furthermore, we see in the
    bottom row that the norms used in the calculation
    of~$\mathcal{C}_{\textrm{exact}}$ grow exponentially and with
    increasing rate for bigger resolution.}
  \label{Fig:CCM_SYMH_B1_WH_B2_all_tests}
\end{figure*}

In Fig.~\ref{Fig:CCM_SYMH_B1_WH_B2_all_tests} we present all our tests
for CCM performed with the SYMH B1 IBVP and the WH B2 CIBVP. In the
top row we see the exact convergence
rate~$\mathcal{C}_{\textrm{exact}}$, and at the bottom the respective
norms that are included in the calculation
of~$\mathcal{C}_{\textrm{exact}}$. In this case we do not see
convergence in any of the tests, which can be explained by the fact
that the respective norms increase in time, with a growth rate that
increases with resolution. Therefore, the ratio between two norms
calculated at different resolutions cannot be the desired constant in
time. Notice that the way these tests fail is rather different from
the way some earlier tests failed. Here we see complete loss of
convergence, which is often called an instability---even though the
simulation does not stop abruptly---whereas earlier we observed
convergence at an order lower than the theoretically expected, or in a
smaller norm than that of the given data.
  
\begin{figure*}[tph]
  \includegraphics[width=1\textwidth]{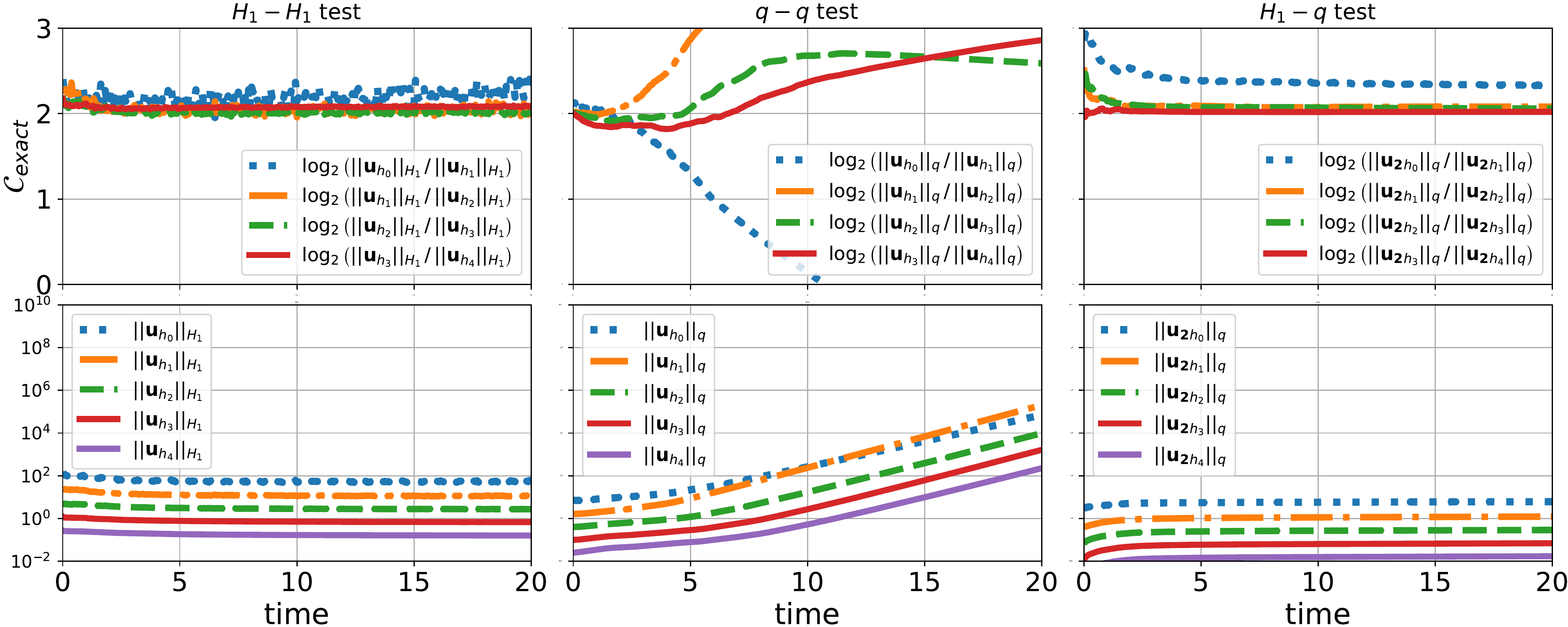}
  \caption{For comparison to the convergence of the CCM setup SYMH
    B1-WH B2 shown in Fig.~\ref{Fig:CCM_SYMH_B1_WH_B2_all_tests}, we
    present here the following (from left to right): the~$H_1-H_1$
    test for SYMH B1-SYMH B2, the~$q-q$ test for WH B1-WH B2, and
    the~$H_1-q$ test for the CCE setup SYMH B1-WH B2 (plotting the
    CIBVP norms).}
  \label{Fig:inhom_CCM_CCE_comparison_tests234}
\end{figure*}

For comparison, we perform the~$H_1-H_1$ test for CCM between SYMH B1
and SYMH B2, the~$q-q$ test for CCM with WH B1-WH B2, as well as
the~$H_1-q$ test for CCE for SYMH B1-WH B2. We chose the specific
tests and combinations of PDEs, because these provided good
convergence results earlier when monitoring derivative terms in the
norms. The results are shown in
Fig.~\ref{Fig:inhom_CCM_CCE_comparison_tests234}. More specifically,
CCM between the inhomogeneous SYMH setups exhibits the expected
convergence in the~$H_1$ norm, whereas for WH setups we see good
convergence in the~$q$ norm only for a brief period. The latter should
not be surprising, since standard numerical methods as the ones
utilized here are developed having in mind PDE problems that are
strongly well posed, such as strongly or symmetric hyperbolic. These
results indicate that the main source of this rapid loss of
convergence is the different degree of hyperbolicity between the IBVP
and CIBVP in CCM. In addition, we see good convergence in~$q$ norm for
the CIBVP part of CCE with SYMH B1 IBVP and WH B2 CIBVP, in line with
earlier CCE results. Based on these findings, we deduce that having
the influence of the bad part of the WH CIBVP solution affecting the
SYMH IBVP is exactly what causes the severe loss of convergence and
the rapid growth of the respective norms. When this influence is
prevented by doing only CCE, this exponential growth of the norms is
not present. For completeness, we have performed convergence tests for
CCM with SYMH B1 IBVP and WH B2 CIBVP, with smooth given data. In
complete contrast to the random noise data, the smooth data tests
exhibit convergence in the~$L^2$ norm. Note that these are not exact,
but self convergence tests, since we do not know the exact solution of
the tested smooth setup.

\subsubsection{Synopsis} CCM provides the expected second order
convergence for the solution of the composite problem, when it is
performed between an IBVP and CIBVP with the same degree of
hyperbolicity, that is SYMH or WH.  In the case of SYMH IBVP matched
with WH CIBVP, we considered two cases. In the homogeneous case, most
convergence tests failed with the convergence rate reduced from second
to first order; whereas in the inhomogeneous case all tests failed
with the solution growing at an exponential rate that increases with
increased resolution.

\section{Summary and Discussion}
\label{Section:Summary}

Modeling gravitational waves at future null infinity accurately is a
challenge that has to be met to obtain waveforms of high fidelity.
One possibility is to successfully develop a CCM scheme for GR. Such a
scheme would combine the ability to capture in detail the physics of
the strong gravity regime in the Cauchy setup, with the accurate
propagation of gravitational waves due to the foliation with null
hypersurfaces of the characteristic setup. CCM does not suffer from
artificial boundary conditions on the outer boundary of the Cauchy
domain, as for instance a CCE scheme. The characteristic setup is not
the only way to obtain accurate gravitational waveform models at
future null infinity. Alternative strategies include hyperboloidal
compactification~\cite{SteFri82, MonRin08, Zen10, BarSarBuc11,
  VanHusHil14, HilHarBug16, AnsMac16, VanHus17} and the use of the
conformal EFEs~\cite{Hub99, DouFra16, DouFraSte19}.

In this paper we focused on mathematical properties of CCE and CCM
schemes and their impact on numerical simulations. Our motivation
comes from recent progress in understanding the mathematical structure
of Einstein's field equations in the characteristic setup, in which
the gauge freedom of GR is typically fixed by choosing Bondi-like
coordinates which are built upon outgoing (in the case of CCE and CCM)
null geodesics. In~\cite{GiaHilZil20}, certain characteristic PDE
systems of GR were shown to be only weakly hyperbolic in some
Bondi-like coordinates, and subsequently in~\cite{GiaBisHil21} this
structure was attributed to the choice of a Bondi-like gauge. Based on
these results, it becomes clear that current CCE and CCM schemes of GR
involve a strongly or symmetric hyperbolic system for the IBVP and the
WH Bondi-like one in the CIBVP. Since PDE problems based on WH systems
are generically not well posed, it is natural to ask whether these
composite CCE and CCM setups are well posed, and to what extent should
a solution based on them be trusted in making accurate predictions for
gravitational waveforms.

To address this question we constructed toy models and utilized them
as models for GR CCE and CCM. To address the effect of weak
hyperbolicity we studied weakly and symmetric hyperbolic models for
both the IBVP and the CIBVP, comparing different combinations of IBVP
and CIBVP in CCE and CCM. The weakly hyperbolic structure of the CIBVP
mimics that of the EFE in a Bondi-like gauge. Importantly, the WH
system is constructed in such a way that it has a weakly well-posed
IBVP and CIBVP. This assumption is the best possible scenario for the
Bondi-like CIBVP of the EFE, involving up to second order metric
derivatives. Given that there are symmetric hyperbolic characteristic
formulations of GR in a Bondi-like gauge that include third order
metric derivatives, one might hope to show weak well-posedness for the
characteristic setups commonly used in numerical relativity. To the
best of our knowledge, such a result is yet to be obtained.

For our toy models, we first studied well-posedness at the continuum
level, where we considered the homogeneous case and provided energy
estimates for the IBVP, CIBVP, CCE and CCM. We discussed energy
estimates in norms with ($q$ and~$H_1$) and without ($L^2$) derivative
terms, and adapted to the IBVP, CIBVP, CCE and CCM setups. A similar
analysis was performed in~\cite{GiaHilZil20} only for the~$L^2$
norm. As in that earlier work, we found that CCM cannot provide a
well-posed PDE problem when matching a WH with a SYMH PDE problem. On
the contrary, CCE can be weakly well posed if the WH problem is weakly
well posed. The essential difference is in the communication of the
solution between the Cauchy and the characteristic domains. In terms
of the energy estimates, this is translated into a worldtube integral
over the interface between the Cauchy and the characteristic domains,
that is not controlled by the given data.

Based on the continuum analysis, we performed several numerical
convergence tests, with a second order accurate implementation. Our
main goal was to address continuous dependence of the solution on the
given data, in discrete approximations to the~$L^2$, $q$, and~$H_1$
norms previously constructed. Our first step was to verify the
expected norm convergence for the individual IBVP and CIBVP. We saw
the expected convergence for the SYMH setups in the~$L^2$ and~$H_1$
norms and for the WH in the~$q$ norm, for both the IBVP and CIBVP. An
interesting scenario was realized when the given data were controlled
in the~$H_1$ norm and the convergence of the solution was tested in
the~$q$ norm, which is appropriate only for the WH setup. In this
scenario, we observed second order convergence for both the SYMH and
WH cases. Our explanation for this phenomenon is that the~$H_1$ norm
provides more control over derivatives than~$q$, so the given data are
appropriate for both SYMH and WH setups. Furthermore, when monitoring
only the~$q$ norm of the solution for the WH case we ignore the
derivative terms that are responsible for the loss of convergence we
see in~$H_1$. For the SYMH case, all derivative terms converge
well. In the~$q$ norm, we only monitor the convergence of one of them,
which dominates the norm. This scenario is crucial in understanding
the reason that CCE between a SYMH IBVP and a WH CIBVP is numerically
well behaved. The key point is that for~$H_1$ controlled given data,
the solution to the SYMH IBVP is also controlled in the same
norm. This solution then provides boundary data for the WH CIBVP with
enough control over derivative terms and so the solution to that PDE
problem converges well in the~$q$ norm.

The expected convergence was also seen for CCM between IBVP and CIBVP
with the same degree of hyperbolicity, that is when matching SYMH to
SYMH or WH to WH. In the first case, we could obtain good convergence
for the solution in the~$L^2$ and~$H_1$ norms, for given data
controlled in the same norms, respectively, whereas in the second, the
same was true in the~$q$ norm. When we analyzed the convergence of the
solution in the~$q$ norm for~$H_1$ given data, we observed the same
scenario as earlier, namely second order convergence in the~$q$ norm
for both SYMH-SYMH and WH-WH setups. Our understanding of this
behavior is exactly the same as for the individual IBVP and CIBVP,
with the only difference that in this case the various norms
considered were adapted to the composite CCM PDE problem.

When matching a SYMH IBVP to a WH CIBVP we did not recover good second
order convergence in any of the~$L^2$, $q$, or~$H_1$ norms for the
solution, when the given data were controlled in the same norm. This
result is in line with our expectation from the preceding continuum
analysis. It is worth mentioning that as earlier, we were able to see
good second order convergence in the~$q$ norm when the given data were
controlled in the~$H_1$. Even though this behavior is qualitatively
the same as for SYMH-WH CCE, the interpretation in terms of continuous
dependence of the solution on the given data is different. In CCM, the
composite IBVP-CIBVP PDE problem is treated as one and therefore the
solution is controlled in a smaller norm than the given data (some
derivative terms are not controlled). In CCE, this was not an issue,
because we could verify convergence of the IBVP solution in the~$H_1$
norm, for~$H_1$ given data, and for the CIBVP solution in the~$q$ norm
for~$H_1$ given data, where the latter include also given data
controlled in the~$q$ norm.

Our homogeneous models exhibit the same qualitative behavior for
SYMH-WH CCE and CCM due to the lack of source terms. To demonstrate
this, we considered an inhomogeneous example in which for the CIBVP we
added a source term such that the right-moving variables which are
responsible for weak hyperbolicity affect the left-moving variable,
and in IBVP the left-moving variable couples to the first right-moving
one. In this way, there is an interaction loop between left- and
right-moving variables that is closed only when we consider CCM but
not CCE. Our understanding is that when this loop is closed, it allows
the part of the CIBVP solution that is influenced by the WH structure
to affect in turn the IBVP solution, which consequently provides bad
boundary data to the CIBVP and drives the numerical solution to
explode. In our numerical experiments we see severe loss of
convergence for the inhomogeneous SYMH-WH CCM setup in all norms and
for all given data, even for the combination of~$H_1$ given data
and~$q$ norm of the solution. In fact, we observe that the norms
involved in the calculation of the convergence rate grow exponentially
with time and faster for increasing resolution, which is a clear sign
for loss of convergence. For comparison, we tested the inhomogeneous
case for SYMH-SYMH and WH-WH CCM. In the first case we found good
second order convergence in the CCM~$L^2$ and~$H_1$ norms for the
solution, whereas in the second we saw second order convergence in
the~$q$ norm only for a brief initial period, the cause of which is
presently unclear to us. Finally, to assess that indeed the two-way
communication between the Cauchy and the characteristic domain is the
crucial point in this setup, we tested the inhomogeneous SYMH-WH CCE
setup for $H_1$ given data. In this case, we find second order
convergence for the inhomogeneous WH CIBVP, exactly as in the
inhomogeneous case. We interpret these results as support of our
understanding of the mechanism that drives this loss of convergence,
as given above.

We should highlight that we used high frequency given data in all our
tests. More specifically, we used random noise with amplitude that
dropped appropriately with increasing resolution. This type of data is
useful in detecting WH in numerical experiments. In fact, we were able
to see second order convergence in the~$L^2$ norm for our
inhomogeneous SYMH-WH CCM model, for smooth given data
(see~\cite{GiaBisHil23_github}). We understand that random noise high
frequency data may not be appropriate for nonuniform grids, as for
instance in spectral methods. In that case, one has to construct a
different type of high frequency given data. A possible suggestion
could be a sinusoidal of some high but resolved frequency, that
increases with increasing resolution, and an amplitude which is
dropped with increasing resolution, mimicking the expected behavior of
numerical error.

\section{Conclusions}
\label{Section:Conclusions}

The take home message from this analysis for current CCE setups in GR
is that if the solution to the Cauchy setup is sufficiently controlled
(in terms of derivatives) and provided that the WH CIBVP is weakly
well posed (a strong assumption that should be firmly established)
then, depending on the discretization, the numerical solution of this
CCE should converge to the true solution in the limit of infinite
resolution. This is another way to say that error estimates on such a
solution can be trusted to an arbitrary small level (given the
necessary computational resources) to provide highly accurate
gravitational waveforms. Ultimately any CCE scheme is still subject to
artificial boundary conditions, for which a detailed consideration is
beyond our present scope.

Another lesson we take from this study is that CCM for the EFE as
currently performed between a SYMH IBVP and WH CIBVP cannot be
guaranteed to provide numerical approximations that converge to the
true solution of the problem in the limit of infinite resolution. We
do not expect the results to be changed if symmetric is replaced with
strong hyperbolicity for the IBVP. Developing a strongly, if not
symmetric hyperbolic characteristic formulation of the EFE seems to be
necessary to move toward an implementation of GR CCM that is
(strongly) well posed and properly convergent.

The supporting data for this paper, including the rest of the
convergence plots and the pointwise convergence tests, are openly
available from~\cite{GiaBisHil23_github, GiaBisHil23_zenodo}.

\section{Acknowledgments}

T.G.\ acknowledges support from the STFC Consolidated Grant No.\
ST/V005596/1. D.H. was supported in part by FCT (Portugal) Project
No. UIDB/00099/2020. M.Z.\ acknowledges financial support by the
Center for Research and Development in Mathematics and Applications
(CIDMA) through the Portuguese Foundation for Science and Technology
(FCT -- Fundação para a Ciência e a Tecnologia)---references
UIDB/04106/2020 and UIDP/04106/2020---as well as FCT Projects No.
2022.00721.CEECIND, No. CERN/FIS-PAR/0027/2019,
No. PTDC/FIS-AST/3041/2020, No. CERN/FIS-PAR/0024/2021, and
No. 2022.04560.PTDC. This work has further been supported by the
European Horizon Europe staff exchange (SE) program
HORIZON-MSCA-2021-SE-01 Grant No.\ NewFunFiCO-101086251.

\bibliography{refs}

\begin{thebibliography}{75}%
\makeatletter
\providecommand \@ifxundefined [1]{%
 \@ifx{#1\undefined}
}%
\providecommand \@ifnum [1]{%
 \ifnum #1\expandafter \@firstoftwo
 \else \expandafter \@secondoftwo
 \fi
}%
\providecommand \@ifx [1]{%
 \ifx #1\expandafter \@firstoftwo
 \else \expandafter \@secondoftwo
 \fi
}%
\providecommand \natexlab [1]{#1}%
\providecommand \enquote  [1]{``#1''}%
\providecommand \bibnamefont  [1]{#1}%
\providecommand \bibfnamefont [1]{#1}%
\providecommand \citenamefont [1]{#1}%
\providecommand \href@noop [0]{\@secondoftwo}%
\providecommand \href [0]{\begingroup \@sanitize@url \@href}%
\providecommand \@href[1]{\@@startlink{#1}\@@href}%
\providecommand \@@href[1]{\endgroup#1\@@endlink}%
\providecommand \@sanitize@url [0]{\catcode `\\12\catcode `\$12\catcode
  `\&12\catcode `\#12\catcode `\^12\catcode `\_12\catcode `\%12\relax}%
\providecommand \@@startlink[1]{}%
\providecommand \@@endlink[0]{}%
\providecommand \url  [0]{\begingroup\@sanitize@url \@url }%
\providecommand \@url [1]{\endgroup\@href {#1}{\urlprefix }}%
\providecommand \urlprefix  [0]{URL }%
\providecommand \Eprint [0]{\href }%
\providecommand \doibase [0]{http://dx.doi.org/}%
\providecommand \selectlanguage [0]{\@gobble}%
\providecommand \bibinfo  [0]{\@secondoftwo}%
\providecommand \bibfield  [0]{\@secondoftwo}%
\providecommand \translation [1]{[#1]}%
\providecommand \BibitemOpen [0]{}%
\providecommand \bibitemStop [0]{}%
\providecommand \bibitemNoStop [0]{.\EOS\space}%
\providecommand \EOS [0]{\spacefactor3000\relax}%
\providecommand \BibitemShut  [1]{\csname bibitem#1\endcsname}%
\let\auto@bib@innerbib\@empty
\bibitem [{\citenamefont {Aasi}\ \emph {et~al.}(2015)\citenamefont {Aasi} \emph
  {et~al.}}]{AasAlt14ax}%
  \BibitemOpen
  \bibfield  {author} {\bibinfo {author} {\bibfnamefont {J.}~\bibnamefont
  {Aasi}} \emph {et~al.} (\bibinfo {collaboration} {LIGO Scientific}),\
  }\href@noop {} {\bibfield  {journal} {\bibinfo  {journal} {Class. Quant.
  Grav.}\ }\textbf {\bibinfo {volume} {32}},\ \bibinfo {pages} {074001}
  (\bibinfo {year} {2015})},\ \Eprint {http://arxiv.org/abs/1411.4547}
  {arXiv:1411.4547 [gr-qc]} \BibitemShut {NoStop}%
\bibitem [{\citenamefont {Acernese}\ \emph {et~al.}(2015)\citenamefont
  {Acernese} \emph {et~al.}}]{AceAlt14}%
  \BibitemOpen
  \bibfield  {author} {\bibinfo {author} {\bibfnamefont {F.}~\bibnamefont
  {Acernese}} \emph {et~al.} (\bibinfo {collaboration} {VIRGO}),\ }\href
  {\doibase 10.1088/0264-9381/32/2/024001} {\bibfield  {journal} {\bibinfo
  {journal} {Class. Quant. Grav.}\ }\textbf {\bibinfo {volume} {32}},\ \bibinfo
  {pages} {024001} (\bibinfo {year} {2015})},\ \Eprint
  {http://arxiv.org/abs/1408.3978} {arXiv:1408.3978 [gr-qc]} \BibitemShut
  {NoStop}%
\bibitem [{\citenamefont {Akutsu}\ \emph {et~al.}(2020)\citenamefont {Akutsu}
  \emph {et~al.}}]{Aku20}%
  \BibitemOpen
  \bibfield  {author} {\bibinfo {author} {\bibfnamefont {T.}~\bibnamefont
  {Akutsu}} \emph {et~al.} (\bibinfo {collaboration} {KAGRA}),\ }\href
  {\doibase 10.1093/ptep/ptaa120} {\  (\bibinfo {year} {2020}),\
  10.1093/ptep/ptaa120},\ \Eprint {http://arxiv.org/abs/2008.02921}
  {arXiv:2008.02921 [gr-qc]} \BibitemShut {NoStop}%
\bibitem [{\citenamefont {Dwyer}\ \emph {et~al.}(2015)\citenamefont {Dwyer},
  \citenamefont {Sigg}, \citenamefont {Ballmer}, \citenamefont {Barsotti},
  \citenamefont {Mavalvala},\ and\ \citenamefont {Evans}}]{DwySigBal15}%
  \BibitemOpen
  \bibfield  {author} {\bibinfo {author} {\bibfnamefont {S.}~\bibnamefont
  {Dwyer}}, \bibinfo {author} {\bibfnamefont {D.}~\bibnamefont {Sigg}},
  \bibinfo {author} {\bibfnamefont {S.~W.}\ \bibnamefont {Ballmer}}, \bibinfo
  {author} {\bibfnamefont {L.}~\bibnamefont {Barsotti}}, \bibinfo {author}
  {\bibfnamefont {N.}~\bibnamefont {Mavalvala}}, \ and\ \bibinfo {author}
  {\bibfnamefont {M.}~\bibnamefont {Evans}},\ }\href {\doibase
  10.1103/PhysRevD.91.082001} {\bibfield  {journal} {\bibinfo  {journal} {Phys.
  Rev. D}\ }\textbf {\bibinfo {volume} {91}},\ \bibinfo {pages} {082001}
  (\bibinfo {year} {2015})}\BibitemShut {NoStop}%
\bibitem [{\citenamefont {Maggiore}\ \emph {et~al.}(2020)\citenamefont
  {Maggiore} \emph {et~al.}}]{Mag19}%
  \BibitemOpen
  \bibfield  {author} {\bibinfo {author} {\bibfnamefont {M.}~\bibnamefont
  {Maggiore}} \emph {et~al.},\ }\href {\doibase 10.1088/1475-7516/2020/03/050}
  {\bibfield  {journal} {\bibinfo  {journal} {JCAP}\ }\textbf {\bibinfo
  {volume} {03}},\ \bibinfo {pages} {050} (\bibinfo {year} {2020})},\ \Eprint
  {http://arxiv.org/abs/1912.02622} {arXiv:1912.02622 [astro-ph.CO]}
  \BibitemShut {NoStop}%
\bibitem [{\citenamefont {Amaro-Seoane}\ \emph {et~al.}(2017)\citenamefont
  {Amaro-Seoane}, \citenamefont {Audley}, \citenamefont {Babak}, \citenamefont
  {Baker}, \citenamefont {Barausse}, \citenamefont {Bender}, \citenamefont
  {Berti}, \citenamefont {Binetruy}, \citenamefont {Born}, \citenamefont
  {Bortoluzzi}, \citenamefont {Camp}, \citenamefont {Caprini}, \citenamefont
  {Cardoso}, \citenamefont {Colpi}, \citenamefont {Conklin}, \citenamefont
  {Cornish}, \citenamefont {Cutler}, \citenamefont {Danzmann}, \citenamefont
  {Dolesi}, \citenamefont {Ferraioli}, \citenamefont {Ferroni}, \citenamefont
  {Fitzsimons}, \citenamefont {Gair}, \citenamefont {Bote}, \citenamefont
  {Giardini}, \citenamefont {Gibert}, \citenamefont {Grimani}, \citenamefont
  {Halloin}, \citenamefont {Heinzel}, \citenamefont {Hertog}, \citenamefont
  {Hewitson}, \citenamefont {Holley-Bockelmann}, \citenamefont {Hollington},
  \citenamefont {Hueller}, \citenamefont {Inchauspe}, \citenamefont {Jetzer},
  \citenamefont {Karnesis}, \citenamefont {Killow}, \citenamefont {Klein},
  \citenamefont {Klipstein}, \citenamefont {Korsakova}, \citenamefont {Larson},
  \citenamefont {Livas}, \citenamefont {Lloro}, \citenamefont {Man},
  \citenamefont {Mance}, \citenamefont {Martino}, \citenamefont {Mateos},
  \citenamefont {McKenzie}, \citenamefont {McWilliams}, \citenamefont {Miller},
  \citenamefont {Mueller}, \citenamefont {Nardini}, \citenamefont {Nelemans},
  \citenamefont {Nofrarias}, \citenamefont {Petiteau}, \citenamefont {Pivato},
  \citenamefont {Plagnol}, \citenamefont {Porter}, \citenamefont {Reiche},
  \citenamefont {Robertson}, \citenamefont {Robertson}, \citenamefont {Rossi},
  \citenamefont {Russano}, \citenamefont {Schutz}, \citenamefont {Sesana},
  \citenamefont {Shoemaker}, \citenamefont {Slutsky}, \citenamefont {Sopuerta},
  \citenamefont {Sumner}, \citenamefont {Tamanini}, \citenamefont {Thorpe},
  \citenamefont {Troebs}, \citenamefont {Vallisneri}, \citenamefont {Vecchio},
  \citenamefont {Vetrugno}, \citenamefont {Vitale}, \citenamefont {Volonteri},
  \citenamefont {Wanner}, \citenamefont {Ward}, \citenamefont {Wass},
  \citenamefont {Weber}, \citenamefont {Ziemer},\ and\ \citenamefont
  {Zweifel}}]{Ama17}%
  \BibitemOpen
  \bibfield  {author} {\bibinfo {author} {\bibfnamefont {P.}~\bibnamefont
  {Amaro-Seoane}}, \bibinfo {author} {\bibfnamefont {H.}~\bibnamefont
  {Audley}}, \bibinfo {author} {\bibfnamefont {S.}~\bibnamefont {Babak}},
  \bibinfo {author} {\bibfnamefont {J.}~\bibnamefont {Baker}}, \bibinfo
  {author} {\bibfnamefont {E.}~\bibnamefont {Barausse}}, \bibinfo {author}
  {\bibfnamefont {P.}~\bibnamefont {Bender}}, \bibinfo {author} {\bibfnamefont
  {E.}~\bibnamefont {Berti}}, \bibinfo {author} {\bibfnamefont
  {P.}~\bibnamefont {Binetruy}}, \bibinfo {author} {\bibfnamefont
  {M.}~\bibnamefont {Born}}, \bibinfo {author} {\bibfnamefont {D.}~\bibnamefont
  {Bortoluzzi}}, \bibinfo {author} {\bibfnamefont {J.}~\bibnamefont {Camp}},
  \bibinfo {author} {\bibfnamefont {C.}~\bibnamefont {Caprini}}, \bibinfo
  {author} {\bibfnamefont {V.}~\bibnamefont {Cardoso}}, \bibinfo {author}
  {\bibfnamefont {M.}~\bibnamefont {Colpi}}, \bibinfo {author} {\bibfnamefont
  {J.}~\bibnamefont {Conklin}}, \bibinfo {author} {\bibfnamefont
  {N.}~\bibnamefont {Cornish}}, \bibinfo {author} {\bibfnamefont
  {C.}~\bibnamefont {Cutler}}, \bibinfo {author} {\bibfnamefont
  {K.}~\bibnamefont {Danzmann}}, \bibinfo {author} {\bibfnamefont
  {R.}~\bibnamefont {Dolesi}}, \bibinfo {author} {\bibfnamefont
  {L.}~\bibnamefont {Ferraioli}}, \bibinfo {author} {\bibfnamefont
  {V.}~\bibnamefont {Ferroni}}, \bibinfo {author} {\bibfnamefont
  {E.}~\bibnamefont {Fitzsimons}}, \bibinfo {author} {\bibfnamefont
  {J.}~\bibnamefont {Gair}}, \bibinfo {author} {\bibfnamefont {L.~G.}\
  \bibnamefont {Bote}}, \bibinfo {author} {\bibfnamefont {D.}~\bibnamefont
  {Giardini}}, \bibinfo {author} {\bibfnamefont {F.}~\bibnamefont {Gibert}},
  \bibinfo {author} {\bibfnamefont {C.}~\bibnamefont {Grimani}}, \bibinfo
  {author} {\bibfnamefont {H.}~\bibnamefont {Halloin}}, \bibinfo {author}
  {\bibfnamefont {G.}~\bibnamefont {Heinzel}}, \bibinfo {author} {\bibfnamefont
  {T.}~\bibnamefont {Hertog}}, \bibinfo {author} {\bibfnamefont
  {M.}~\bibnamefont {Hewitson}}, \bibinfo {author} {\bibfnamefont
  {K.}~\bibnamefont {Holley-Bockelmann}}, \bibinfo {author} {\bibfnamefont
  {D.}~\bibnamefont {Hollington}}, \bibinfo {author} {\bibfnamefont
  {M.}~\bibnamefont {Hueller}}, \bibinfo {author} {\bibfnamefont
  {H.}~\bibnamefont {Inchauspe}}, \bibinfo {author} {\bibfnamefont
  {P.}~\bibnamefont {Jetzer}}, \bibinfo {author} {\bibfnamefont
  {N.}~\bibnamefont {Karnesis}}, \bibinfo {author} {\bibfnamefont
  {C.}~\bibnamefont {Killow}}, \bibinfo {author} {\bibfnamefont
  {A.}~\bibnamefont {Klein}}, \bibinfo {author} {\bibfnamefont
  {B.}~\bibnamefont {Klipstein}}, \bibinfo {author} {\bibfnamefont
  {N.}~\bibnamefont {Korsakova}}, \bibinfo {author} {\bibfnamefont {S.~L.}\
  \bibnamefont {Larson}}, \bibinfo {author} {\bibfnamefont {J.}~\bibnamefont
  {Livas}}, \bibinfo {author} {\bibfnamefont {I.}~\bibnamefont {Lloro}},
  \bibinfo {author} {\bibfnamefont {N.}~\bibnamefont {Man}}, \bibinfo {author}
  {\bibfnamefont {D.}~\bibnamefont {Mance}}, \bibinfo {author} {\bibfnamefont
  {J.}~\bibnamefont {Martino}}, \bibinfo {author} {\bibfnamefont
  {I.}~\bibnamefont {Mateos}}, \bibinfo {author} {\bibfnamefont
  {K.}~\bibnamefont {McKenzie}}, \bibinfo {author} {\bibfnamefont {S.~T.}\
  \bibnamefont {McWilliams}}, \bibinfo {author} {\bibfnamefont
  {C.}~\bibnamefont {Miller}}, \bibinfo {author} {\bibfnamefont
  {G.}~\bibnamefont {Mueller}}, \bibinfo {author} {\bibfnamefont
  {G.}~\bibnamefont {Nardini}}, \bibinfo {author} {\bibfnamefont
  {G.}~\bibnamefont {Nelemans}}, \bibinfo {author} {\bibfnamefont
  {M.}~\bibnamefont {Nofrarias}}, \bibinfo {author} {\bibfnamefont
  {A.}~\bibnamefont {Petiteau}}, \bibinfo {author} {\bibfnamefont
  {P.}~\bibnamefont {Pivato}}, \bibinfo {author} {\bibfnamefont
  {E.}~\bibnamefont {Plagnol}}, \bibinfo {author} {\bibfnamefont
  {E.}~\bibnamefont {Porter}}, \bibinfo {author} {\bibfnamefont
  {J.}~\bibnamefont {Reiche}}, \bibinfo {author} {\bibfnamefont
  {D.}~\bibnamefont {Robertson}}, \bibinfo {author} {\bibfnamefont
  {N.}~\bibnamefont {Robertson}}, \bibinfo {author} {\bibfnamefont
  {E.}~\bibnamefont {Rossi}}, \bibinfo {author} {\bibfnamefont
  {G.}~\bibnamefont {Russano}}, \bibinfo {author} {\bibfnamefont
  {B.}~\bibnamefont {Schutz}}, \bibinfo {author} {\bibfnamefont
  {A.}~\bibnamefont {Sesana}}, \bibinfo {author} {\bibfnamefont
  {D.}~\bibnamefont {Shoemaker}}, \bibinfo {author} {\bibfnamefont
  {J.}~\bibnamefont {Slutsky}}, \bibinfo {author} {\bibfnamefont {C.~F.}\
  \bibnamefont {Sopuerta}}, \bibinfo {author} {\bibfnamefont {T.}~\bibnamefont
  {Sumner}}, \bibinfo {author} {\bibfnamefont {N.}~\bibnamefont {Tamanini}},
  \bibinfo {author} {\bibfnamefont {I.}~\bibnamefont {Thorpe}}, \bibinfo
  {author} {\bibfnamefont {M.}~\bibnamefont {Troebs}}, \bibinfo {author}
  {\bibfnamefont {M.}~\bibnamefont {Vallisneri}}, \bibinfo {author}
  {\bibfnamefont {A.}~\bibnamefont {Vecchio}}, \bibinfo {author} {\bibfnamefont
  {D.}~\bibnamefont {Vetrugno}}, \bibinfo {author} {\bibfnamefont
  {S.}~\bibnamefont {Vitale}}, \bibinfo {author} {\bibfnamefont
  {M.}~\bibnamefont {Volonteri}}, \bibinfo {author} {\bibfnamefont
  {G.}~\bibnamefont {Wanner}}, \bibinfo {author} {\bibfnamefont
  {H.}~\bibnamefont {Ward}}, \bibinfo {author} {\bibfnamefont {P.}~\bibnamefont
  {Wass}}, \bibinfo {author} {\bibfnamefont {W.}~\bibnamefont {Weber}},
  \bibinfo {author} {\bibfnamefont {J.}~\bibnamefont {Ziemer}}, \ and\ \bibinfo
  {author} {\bibfnamefont {P.}~\bibnamefont {Zweifel}},\ }\href {\doibase
  10.48550/ARXIV.1702.00786} {\enquote {\bibinfo {title} {Laser interferometer
  space antenna},}\ } (\bibinfo {year} {2017}),\ \Eprint
  {http://arxiv.org/abs/1702.00786} {arXiv:1702.00786 [astro-ph.IM]}
  \BibitemShut {NoStop}%
\bibitem [{\citenamefont {Ruan}\ \emph {et~al.}(2020)\citenamefont {Ruan},
  \citenamefont {Guo}, \citenamefont {Cai},\ and\ \citenamefont
  {Zhang}}]{RuaGuoCai20}%
  \BibitemOpen
  \bibfield  {author} {\bibinfo {author} {\bibfnamefont {W.-H.}\ \bibnamefont
  {Ruan}}, \bibinfo {author} {\bibfnamefont {Z.-K.}\ \bibnamefont {Guo}},
  \bibinfo {author} {\bibfnamefont {R.-G.}\ \bibnamefont {Cai}}, \ and\
  \bibinfo {author} {\bibfnamefont {Y.-Z.}\ \bibnamefont {Zhang}},\ }\href
  {\doibase 10.1142/S0217751X2050075X} {\bibfield  {journal} {\bibinfo
  {journal} {Int. J. Mod. Phys. A}\ }\textbf {\bibinfo {volume} {35}},\
  \bibinfo {pages} {2050075} (\bibinfo {year} {2020})},\ \Eprint
  {http://arxiv.org/abs/1807.09495} {arXiv:1807.09495 [gr-qc]} \BibitemShut
  {NoStop}%
\bibitem [{\citenamefont {Luo}\ \emph {et~al.}(2016)\citenamefont {Luo} \emph
  {et~al.}}]{Luo16}%
  \BibitemOpen
  \bibfield  {author} {\bibinfo {author} {\bibfnamefont {J.}~\bibnamefont
  {Luo}} \emph {et~al.} (\bibinfo {collaboration} {TianQin}),\ }\href {\doibase
  10.1088/0264-9381/33/3/035010} {\bibfield  {journal} {\bibinfo  {journal}
  {Class. Quant. Grav.}\ }\textbf {\bibinfo {volume} {33}},\ \bibinfo {pages}
  {035010} (\bibinfo {year} {2016})},\ \Eprint
  {http://arxiv.org/abs/1512.02076} {arXiv:1512.02076 [astro-ph.IM]}
  \BibitemShut {NoStop}%
\bibitem [{\citenamefont {Foucart}\ \emph {et~al.}(2022)\citenamefont
  {Foucart}, \citenamefont {Laguna}, \citenamefont {Lovelace}, \citenamefont
  {Radice},\ and\ \citenamefont {Witek}}]{FouLagLov22}%
  \BibitemOpen
  \bibfield  {author} {\bibinfo {author} {\bibfnamefont {F.}~\bibnamefont
  {Foucart}}, \bibinfo {author} {\bibfnamefont {P.}~\bibnamefont {Laguna}},
  \bibinfo {author} {\bibfnamefont {G.}~\bibnamefont {Lovelace}}, \bibinfo
  {author} {\bibfnamefont {D.}~\bibnamefont {Radice}}, \ and\ \bibinfo {author}
  {\bibfnamefont {H.}~\bibnamefont {Witek}},\ }\href@noop {} {\  (\bibinfo
  {year} {2022})},\ \Eprint {http://arxiv.org/abs/2203.08139} {arXiv:2203.08139
  [gr-qc]} \BibitemShut {NoStop}%
\bibitem [{\citenamefont {Bishop}\ and\ \citenamefont
  {Rezzolla}(2016)}]{BisRez16}%
  \BibitemOpen
  \bibfield  {author} {\bibinfo {author} {\bibfnamefont {N.~T.}\ \bibnamefont
  {Bishop}}\ and\ \bibinfo {author} {\bibfnamefont {L.}~\bibnamefont
  {Rezzolla}},\ }\href {\doibase 10.1007/s41114-016-0001-9} {\bibfield
  {journal} {\bibinfo  {journal} {Living Reviews in Relativity}\ }\textbf
  {\bibinfo {volume} {19}},\ \bibinfo {pages} {2} (\bibinfo {year}
  {2016})}\BibitemShut {NoStop}%
\bibitem [{\citenamefont {Duarte}\ \emph {et~al.}(2022)\citenamefont {Duarte},
  \citenamefont {Feng}, \citenamefont {Gasperin},\ and\ \citenamefont
  {Hilditch}}]{DuaFenGas22}%
  \BibitemOpen
  \bibfield  {author} {\bibinfo {author} {\bibfnamefont {M.}~\bibnamefont
  {Duarte}}, \bibinfo {author} {\bibfnamefont {J.~C.}\ \bibnamefont {Feng}},
  \bibinfo {author} {\bibfnamefont {E.}~\bibnamefont {Gasperin}}, \ and\
  \bibinfo {author} {\bibfnamefont {D.}~\bibnamefont {Hilditch}},\ }\href
  {\doibase 10.1088/1361-6382/ac89c5} {\bibfield  {journal} {\bibinfo
  {journal} {Class. Quant. Grav.}\ }\textbf {\bibinfo {volume} {39}},\ \bibinfo
  {pages} {215003} (\bibinfo {year} {2022})},\ \Eprint
  {http://arxiv.org/abs/2205.09405} {arXiv:2205.09405 [gr-qc]} \BibitemShut
  {NoStop}%
\bibitem [{\citenamefont {Bishop}\ \emph
  {et~al.}(1996{\natexlab{a}})\citenamefont {Bishop}, \citenamefont {G\'omez},
  \citenamefont {Lehner},\ and\ \citenamefont {Winicour}}]{BisGomLeh96a}%
  \BibitemOpen
  \bibfield  {author} {\bibinfo {author} {\bibfnamefont {N.~T.}\ \bibnamefont
  {Bishop}}, \bibinfo {author} {\bibfnamefont {R.}~\bibnamefont {G\'omez}},
  \bibinfo {author} {\bibfnamefont {L.}~\bibnamefont {Lehner}}, \ and\ \bibinfo
  {author} {\bibfnamefont {J.}~\bibnamefont {Winicour}},\ }\href {\doibase
  10.1103/PhysRevD.54.6153} {\bibfield  {journal} {\bibinfo  {journal} {Phys.
  Rev. D}\ }\textbf {\bibinfo {volume} {54}},\ \bibinfo {pages} {6153}
  (\bibinfo {year} {1996}{\natexlab{a}})}\BibitemShut {NoStop}%
\bibitem [{\citenamefont {Bishop}\ \emph
  {et~al.}(1997{\natexlab{a}})\citenamefont {Bishop}, \citenamefont
  {G{\'o}mez}, \citenamefont {Holvorcem}, \citenamefont {Matzner},
  \citenamefont {Papadopoulos},\ and\ \citenamefont {Winicour}}]{BisGomHol97}%
  \BibitemOpen
  \bibfield  {author} {\bibinfo {author} {\bibfnamefont {N.~T.}\ \bibnamefont
  {Bishop}}, \bibinfo {author} {\bibfnamefont {R.}~\bibnamefont {G{\'o}mez}},
  \bibinfo {author} {\bibfnamefont {P.~R.}\ \bibnamefont {Holvorcem}}, \bibinfo
  {author} {\bibfnamefont {R.~A.}\ \bibnamefont {Matzner}}, \bibinfo {author}
  {\bibfnamefont {P.}~\bibnamefont {Papadopoulos}}, \ and\ \bibinfo {author}
  {\bibfnamefont {J.}~\bibnamefont {Winicour}},\ }\href@noop {} {\bibfield
  {journal} {\bibinfo  {journal} {J. Comput. Phys.}\ }\textbf {\bibinfo
  {volume} {136}},\ \bibinfo {pages} {236} (\bibinfo {year}
  {1997}{\natexlab{a}})}\BibitemShut {NoStop}%
\bibitem [{\citenamefont {Bishop}\ \emph
  {et~al.}(1997{\natexlab{b}})\citenamefont {Bishop}, \citenamefont
  {G{\'o}mez}, \citenamefont {Lehner}, \citenamefont {Maharaj},\ and\
  \citenamefont {Winicour}}]{BisGomLeh97a}%
  \BibitemOpen
  \bibfield  {author} {\bibinfo {author} {\bibfnamefont {N.~T.}\ \bibnamefont
  {Bishop}}, \bibinfo {author} {\bibfnamefont {R.}~\bibnamefont {G{\'o}mez}},
  \bibinfo {author} {\bibfnamefont {L.}~\bibnamefont {Lehner}}, \bibinfo
  {author} {\bibfnamefont {M.}~\bibnamefont {Maharaj}}, \ and\ \bibinfo
  {author} {\bibfnamefont {J.}~\bibnamefont {Winicour}},\ }\href@noop {}
  {\bibfield  {journal} {\bibinfo  {journal} {Phys. Rev. D}\ }\textbf {\bibinfo
  {volume} {56}},\ \bibinfo {pages} {6298} (\bibinfo {year}
  {1997}{\natexlab{b}})},\ \Eprint {http://arxiv.org/abs/gr-qc/9708065}
  {gr-qc/9708065} \BibitemShut {NoStop}%
\bibitem [{\citenamefont {Zlochower}\ \emph {et~al.}(2003)\citenamefont
  {Zlochower}, \citenamefont {G{\'o}mez}, \citenamefont {Husa}, \citenamefont
  {Lehner},\ and\ \citenamefont {Winicour}}]{ZloGomHus03}%
  \BibitemOpen
  \bibfield  {author} {\bibinfo {author} {\bibfnamefont {Y.}~\bibnamefont
  {Zlochower}}, \bibinfo {author} {\bibfnamefont {R.}~\bibnamefont
  {G{\'o}mez}}, \bibinfo {author} {\bibfnamefont {S.}~\bibnamefont {Husa}},
  \bibinfo {author} {\bibfnamefont {L.}~\bibnamefont {Lehner}}, \ and\ \bibinfo
  {author} {\bibfnamefont {J.}~\bibnamefont {Winicour}},\ }\href@noop {}
  {\bibfield  {journal} {\bibinfo  {journal} {Phys. Rev. D}\ }\textbf {\bibinfo
  {volume} {68}},\ \bibinfo {pages} {084014} (\bibinfo {year}
  {2003})}\BibitemShut {NoStop}%
\bibitem [{\citenamefont {Gomez}\ \emph {et~al.}(2007)\citenamefont {Gomez},
  \citenamefont {Barreto},\ and\ \citenamefont {Frittelli}}]{GomBarFri07}%
  \BibitemOpen
  \bibfield  {author} {\bibinfo {author} {\bibfnamefont {R.}~\bibnamefont
  {Gomez}}, \bibinfo {author} {\bibfnamefont {W.}~\bibnamefont {Barreto}}, \
  and\ \bibinfo {author} {\bibfnamefont {S.}~\bibnamefont {Frittelli}},\ }\href
  {\doibase 10.1103/PhysRevD.76.124029} {\bibfield  {journal} {\bibinfo
  {journal} {Phys. Rev. D}\ }\textbf {\bibinfo {volume} {76}},\ \bibinfo
  {pages} {124029} (\bibinfo {year} {2007})},\ \Eprint
  {http://arxiv.org/abs/0711.0564} {arXiv:0711.0564 [gr-qc]} \BibitemShut
  {NoStop}%
\bibitem [{\citenamefont {Babiuc}\ \emph {et~al.}(2009)\citenamefont {Babiuc},
  \citenamefont {Bishop}, \citenamefont {Szil\'agyi},\ and\ \citenamefont
  {Winicour}}]{BabBisSzi09}%
  \BibitemOpen
  \bibfield  {author} {\bibinfo {author} {\bibfnamefont {M.~C.}\ \bibnamefont
  {Babiuc}}, \bibinfo {author} {\bibfnamefont {N.~T.}\ \bibnamefont {Bishop}},
  \bibinfo {author} {\bibfnamefont {B.}~\bibnamefont {Szil\'agyi}}, \ and\
  \bibinfo {author} {\bibfnamefont {J.}~\bibnamefont {Winicour}},\ }\href
  {\doibase 10.1103/PhysRevD.79.084011} {\bibfield  {journal} {\bibinfo
  {journal} {Phys. Rev. D}\ }\textbf {\bibinfo {volume} {79}},\ \bibinfo
  {pages} {084011} (\bibinfo {year} {2009})}\BibitemShut {NoStop}%
\bibitem [{\citenamefont {Reisswig}\ \emph {et~al.}(2009)\citenamefont
  {Reisswig}, \citenamefont {Bishop}, \citenamefont {Pollney},\ and\
  \citenamefont {Szilagyi}}]{ReiBisPol09}%
  \BibitemOpen
  \bibfield  {author} {\bibinfo {author} {\bibfnamefont {C.}~\bibnamefont
  {Reisswig}}, \bibinfo {author} {\bibfnamefont {N.~T.}\ \bibnamefont
  {Bishop}}, \bibinfo {author} {\bibfnamefont {D.}~\bibnamefont {Pollney}}, \
  and\ \bibinfo {author} {\bibfnamefont {B.}~\bibnamefont {Szilagyi}},\ }\href
  {\doibase 10.1103/PhysRevLett.103.221101} {\bibfield  {journal} {\bibinfo
  {journal} {Phys. Rev. Lett.}\ }\textbf {\bibinfo {volume} {103}},\ \bibinfo
  {pages} {221101} (\bibinfo {year} {2009})},\ \Eprint
  {http://arxiv.org/abs/0907.2637} {arXiv:0907.2637 [gr-qc]} \BibitemShut
  {NoStop}%
\bibitem [{\citenamefont {Reisswig}\ \emph {et~al.}(2010)\citenamefont
  {Reisswig}, \citenamefont {Bishop}, \citenamefont {Pollney},\ and\
  \citenamefont {Szilagyi}}]{ReiBisPol09a}%
  \BibitemOpen
  \bibfield  {author} {\bibinfo {author} {\bibfnamefont {C.}~\bibnamefont
  {Reisswig}}, \bibinfo {author} {\bibfnamefont {N.}~\bibnamefont {Bishop}},
  \bibinfo {author} {\bibfnamefont {D.}~\bibnamefont {Pollney}}, \ and\
  \bibinfo {author} {\bibfnamefont {B.}~\bibnamefont {Szilagyi}},\ }\href
  {\doibase 10.1088/0264-9381/27/7/075014} {\bibfield  {journal} {\bibinfo
  {journal} {Class.Quant.Grav.}\ }\textbf {\bibinfo {volume} {27}},\ \bibinfo
  {pages} {075014} (\bibinfo {year} {2010})},\ \Eprint
  {http://arxiv.org/abs/0912.1285} {arXiv:0912.1285 [gr-qc]} \BibitemShut
  {NoStop}%
\bibitem [{\citenamefont {Winicour}(2012)}]{Win12}%
  \BibitemOpen
  \bibfield  {author} {\bibinfo {author} {\bibfnamefont {J.}~\bibnamefont
  {Winicour}},\ }\href {http://www.livingreviews.org/lrr-2012-2} {\bibfield
  {journal} {\bibinfo  {journal} {Living Rev. Relativity}\ }\textbf {\bibinfo
  {volume} {15}},\ \bibinfo {pages} {2} (\bibinfo {year} {2012})},\ \bibinfo
  {note} {[Online article]}\BibitemShut {NoStop}%
\bibitem [{\citenamefont {Handmer}\ and\ \citenamefont
  {Szilagyi}(2015)}]{HanSzi14}%
  \BibitemOpen
  \bibfield  {author} {\bibinfo {author} {\bibfnamefont {C.~J.}\ \bibnamefont
  {Handmer}}\ and\ \bibinfo {author} {\bibfnamefont {B.}~\bibnamefont
  {Szilagyi}},\ }\href {\doibase 10.1088/0264-9381/32/2/025008} {\bibfield
  {journal} {\bibinfo  {journal} {Class. Quant. Grav.}\ }\textbf {\bibinfo
  {volume} {32}},\ \bibinfo {pages} {025008} (\bibinfo {year} {2015})},\
  \Eprint {http://arxiv.org/abs/1406.7029} {arXiv:1406.7029 [gr-qc]}
  \BibitemShut {NoStop}%
\bibitem [{\citenamefont {Barkett}\ \emph {et~al.}(2020)\citenamefont
  {Barkett}, \citenamefont {Moxon}, \citenamefont {Scheel},\ and\ \citenamefont
  {Szil\'agyi}}]{BarMoxSch19}%
  \BibitemOpen
  \bibfield  {author} {\bibinfo {author} {\bibfnamefont {K.}~\bibnamefont
  {Barkett}}, \bibinfo {author} {\bibfnamefont {J.}~\bibnamefont {Moxon}},
  \bibinfo {author} {\bibfnamefont {M.~A.}\ \bibnamefont {Scheel}}, \ and\
  \bibinfo {author} {\bibfnamefont {B.}~\bibnamefont {Szil\'agyi}},\ }\href
  {\doibase 10.1103/PhysRevD.102.024004} {\bibfield  {journal} {\bibinfo
  {journal} {Phys. Rev. D}\ }\textbf {\bibinfo {volume} {102}},\ \bibinfo
  {pages} {024004} (\bibinfo {year} {2020})},\ \Eprint
  {http://arxiv.org/abs/1910.09677} {arXiv:1910.09677 [gr-qc]} \BibitemShut
  {NoStop}%
\bibitem [{\citenamefont {Moxon}\ \emph {et~al.}(2020)\citenamefont {Moxon},
  \citenamefont {Scheel},\ and\ \citenamefont {Teukolsky}}]{MoxSchTeu20}%
  \BibitemOpen
  \bibfield  {author} {\bibinfo {author} {\bibfnamefont {J.}~\bibnamefont
  {Moxon}}, \bibinfo {author} {\bibfnamefont {M.~A.}\ \bibnamefont {Scheel}}, \
  and\ \bibinfo {author} {\bibfnamefont {S.~A.}\ \bibnamefont {Teukolsky}},\
  }\href {\doibase 10.1103/PhysRevD.102.044052} {\bibfield  {journal} {\bibinfo
   {journal} {Phys. Rev. D}\ }\textbf {\bibinfo {volume} {102}},\ \bibinfo
  {pages} {044052} (\bibinfo {year} {2020})},\ \Eprint
  {http://arxiv.org/abs/2007.01339} {arXiv:2007.01339 [gr-qc]} \BibitemShut
  {NoStop}%
\bibitem [{\citenamefont {Iozzo}\ \emph
  {et~al.}(2021{\natexlab{a}})\citenamefont {Iozzo} \emph {et~al.}}]{Ioz21}%
  \BibitemOpen
  \bibfield  {author} {\bibinfo {author} {\bibfnamefont {D.~A.~B.}\
  \bibnamefont {Iozzo}} \emph {et~al.},\ }\href {\doibase
  10.1103/PhysRevD.103.124029} {\bibfield  {journal} {\bibinfo  {journal}
  {Phys. Rev. D}\ }\textbf {\bibinfo {volume} {103}},\ \bibinfo {pages}
  {124029} (\bibinfo {year} {2021}{\natexlab{a}})},\ \Eprint
  {http://arxiv.org/abs/2104.07052} {arXiv:2104.07052 [gr-qc]} \BibitemShut
  {NoStop}%
\bibitem [{\citenamefont {Mitman}\ \emph {et~al.}(2021)\citenamefont {Mitman}
  \emph {et~al.}}]{Mit21}%
  \BibitemOpen
  \bibfield  {author} {\bibinfo {author} {\bibfnamefont {K.}~\bibnamefont
  {Mitman}} \emph {et~al.},\ }\href {\doibase 10.1103/PhysRevD.103.024031}
  {\bibfield  {journal} {\bibinfo  {journal} {Phys. Rev. D}\ }\textbf {\bibinfo
  {volume} {103}},\ \bibinfo {pages} {024031} (\bibinfo {year} {2021})},\
  \Eprint {http://arxiv.org/abs/2011.01309} {arXiv:2011.01309 [gr-qc]}
  \BibitemShut {NoStop}%
\bibitem [{\citenamefont {Iozzo}\ \emph
  {et~al.}(2021{\natexlab{b}})\citenamefont {Iozzo}, \citenamefont {Boyle},
  \citenamefont {Deppe}, \citenamefont {Moxon}, \citenamefont {Scheel},
  \citenamefont {Kidder}, \citenamefont {Pfeiffer},\ and\ \citenamefont
  {Teukolsky}}]{IozBoyDep21}%
  \BibitemOpen
  \bibfield  {author} {\bibinfo {author} {\bibfnamefont {D.~A.~B.}\
  \bibnamefont {Iozzo}}, \bibinfo {author} {\bibfnamefont {M.}~\bibnamefont
  {Boyle}}, \bibinfo {author} {\bibfnamefont {N.}~\bibnamefont {Deppe}},
  \bibinfo {author} {\bibfnamefont {J.}~\bibnamefont {Moxon}}, \bibinfo
  {author} {\bibfnamefont {M.~A.}\ \bibnamefont {Scheel}}, \bibinfo {author}
  {\bibfnamefont {L.~E.}\ \bibnamefont {Kidder}}, \bibinfo {author}
  {\bibfnamefont {H.~P.}\ \bibnamefont {Pfeiffer}}, \ and\ \bibinfo {author}
  {\bibfnamefont {S.~A.}\ \bibnamefont {Teukolsky}},\ }\href {\doibase
  10.1103/PhysRevD.103.024039} {\bibfield  {journal} {\bibinfo  {journal}
  {Phys. Rev. D}\ }\textbf {\bibinfo {volume} {103}},\ \bibinfo {pages}
  {024039} (\bibinfo {year} {2021}{\natexlab{b}})},\ \Eprint
  {http://arxiv.org/abs/2010.15200} {arXiv:2010.15200 [gr-qc]} \BibitemShut
  {NoStop}%
\bibitem [{\citenamefont {Mitman}\ \emph {et~al.}(2020)\citenamefont {Mitman},
  \citenamefont {Moxon}, \citenamefont {Scheel}, \citenamefont {Teukolsky},
  \citenamefont {Boyle}, \citenamefont {Deppe}, \citenamefont {Kidder},\ and\
  \citenamefont {Throwe}}]{MitMoxSch21}%
  \BibitemOpen
  \bibfield  {author} {\bibinfo {author} {\bibfnamefont {K.}~\bibnamefont
  {Mitman}}, \bibinfo {author} {\bibfnamefont {J.}~\bibnamefont {Moxon}},
  \bibinfo {author} {\bibfnamefont {M.~A.}\ \bibnamefont {Scheel}}, \bibinfo
  {author} {\bibfnamefont {S.~A.}\ \bibnamefont {Teukolsky}}, \bibinfo {author}
  {\bibfnamefont {M.}~\bibnamefont {Boyle}}, \bibinfo {author} {\bibfnamefont
  {N.}~\bibnamefont {Deppe}}, \bibinfo {author} {\bibfnamefont {L.~E.}\
  \bibnamefont {Kidder}}, \ and\ \bibinfo {author} {\bibfnamefont
  {W.}~\bibnamefont {Throwe}},\ }\href {\doibase 10.1103/PhysRevD.102.104007}
  {\bibfield  {journal} {\bibinfo  {journal} {Phys. Rev. D}\ }\textbf {\bibinfo
  {volume} {102}},\ \bibinfo {pages} {104007} (\bibinfo {year} {2020})},\
  \Eprint {http://arxiv.org/abs/2007.11562} {arXiv:2007.11562 [gr-qc]}
  \BibitemShut {NoStop}%
\bibitem [{\citenamefont {Foucart}\ \emph {et~al.}(2021)\citenamefont {Foucart}
  \emph {et~al.}}]{Fou21}%
  \BibitemOpen
  \bibfield  {author} {\bibinfo {author} {\bibfnamefont {F.}~\bibnamefont
  {Foucart}} \emph {et~al.},\ }\href {\doibase 10.1103/PhysRevD.103.064007}
  {\bibfield  {journal} {\bibinfo  {journal} {Phys. Rev. D}\ }\textbf {\bibinfo
  {volume} {103}},\ \bibinfo {pages} {064007} (\bibinfo {year} {2021})},\
  \Eprint {http://arxiv.org/abs/2010.14518} {arXiv:2010.14518 [gr-qc]}
  \BibitemShut {NoStop}%
\bibitem [{\citenamefont {Moxon}\ \emph {et~al.}(2021)\citenamefont {Moxon},
  \citenamefont {Scheel}, \citenamefont {Teukolsky}, \citenamefont {Deppe},
  \citenamefont {Fischer}, \citenamefont {H\'ebert}, \citenamefont {Kidder},\
  and\ \citenamefont {Throwe}}]{MoxSchTeu21}%
  \BibitemOpen
  \bibfield  {author} {\bibinfo {author} {\bibfnamefont {J.}~\bibnamefont
  {Moxon}}, \bibinfo {author} {\bibfnamefont {M.~A.}\ \bibnamefont {Scheel}},
  \bibinfo {author} {\bibfnamefont {S.~A.}\ \bibnamefont {Teukolsky}}, \bibinfo
  {author} {\bibfnamefont {N.}~\bibnamefont {Deppe}}, \bibinfo {author}
  {\bibfnamefont {N.}~\bibnamefont {Fischer}}, \bibinfo {author} {\bibfnamefont
  {F.}~\bibnamefont {H\'ebert}}, \bibinfo {author} {\bibfnamefont {L.~E.}\
  \bibnamefont {Kidder}}, \ and\ \bibinfo {author} {\bibfnamefont
  {W.}~\bibnamefont {Throwe}},\ }\href@noop {} {\  (\bibinfo {year} {2021})},\
  \Eprint {http://arxiv.org/abs/2110.08635} {arXiv:2110.08635 [gr-qc]}
  \BibitemShut {NoStop}%
\bibitem [{\citenamefont {Taylor}\ \emph {et~al.}(2013)\citenamefont {Taylor},
  \citenamefont {Boyle}, \citenamefont {Reisswig}, \citenamefont {Scheel},
  \citenamefont {Chu}, \citenamefont {Kidder},\ and\ \citenamefont
  {Szilágyi}}]{TayBoyRei13}%
  \BibitemOpen
  \bibfield  {author} {\bibinfo {author} {\bibfnamefont {N.~W.}\ \bibnamefont
  {Taylor}}, \bibinfo {author} {\bibfnamefont {M.}~\bibnamefont {Boyle}},
  \bibinfo {author} {\bibfnamefont {C.}~\bibnamefont {Reisswig}}, \bibinfo
  {author} {\bibfnamefont {M.~A.}\ \bibnamefont {Scheel}}, \bibinfo {author}
  {\bibfnamefont {T.}~\bibnamefont {Chu}}, \bibinfo {author} {\bibfnamefont
  {L.~E.}\ \bibnamefont {Kidder}}, \ and\ \bibinfo {author} {\bibfnamefont
  {B.}~\bibnamefont {Szilágyi}},\ }\href {\doibase 10.1103/PhysRevD.88.124010}
  {\bibfield  {journal} {\bibinfo  {journal} {Phys. Rev.}\ }\textbf {\bibinfo
  {volume} {D88}},\ \bibinfo {pages} {124010} (\bibinfo {year} {2013})},\
  \Eprint {http://arxiv.org/abs/1309.3605} {arXiv:1309.3605 [gr-qc]}
  \BibitemShut {NoStop}%
\bibitem [{\citenamefont {Bishop}\ \emph
  {et~al.}(1996{\natexlab{b}})\citenamefont {Bishop}, \citenamefont
  {G{\'o}mez}, \citenamefont {Holvorcem}, \citenamefont {Matzner},
  \citenamefont {Papadopoulos},\ and\ \citenamefont {Winicour}}]{BisGomHol96a}%
  \BibitemOpen
  \bibfield  {author} {\bibinfo {author} {\bibfnamefont {N.~T.}\ \bibnamefont
  {Bishop}}, \bibinfo {author} {\bibfnamefont {R.}~\bibnamefont {G{\'o}mez}},
  \bibinfo {author} {\bibfnamefont {P.~R.}\ \bibnamefont {Holvorcem}}, \bibinfo
  {author} {\bibfnamefont {R.~A.}\ \bibnamefont {Matzner}}, \bibinfo {author}
  {\bibfnamefont {P.}~\bibnamefont {Papadopoulos}}, \ and\ \bibinfo {author}
  {\bibfnamefont {J.}~\bibnamefont {Winicour}},\ }\href@noop {} {\bibfield
  {journal} {\bibinfo  {journal} {Phys. Rev. Lett.}\ }\textbf {\bibinfo
  {volume} {76}},\ \bibinfo {pages} {4303} (\bibinfo {year}
  {1996}{\natexlab{b}})}\BibitemShut {NoStop}%
\bibitem [{\citenamefont {Bishop}\ \emph {et~al.}(1998)\citenamefont {Bishop},
  \citenamefont {G{\'o}mez}, \citenamefont {Isaacson}, \citenamefont {Lehner},
  \citenamefont {Szilagyi},\ and\ \citenamefont {Winicour}}]{BisGomIsa98}%
  \BibitemOpen
  \bibfield  {author} {\bibinfo {author} {\bibfnamefont {N.~T.}\ \bibnamefont
  {Bishop}}, \bibinfo {author} {\bibfnamefont {R.}~\bibnamefont {G{\'o}mez}},
  \bibinfo {author} {\bibfnamefont {R.~A.}\ \bibnamefont {Isaacson}}, \bibinfo
  {author} {\bibfnamefont {L.}~\bibnamefont {Lehner}}, \bibinfo {author}
  {\bibfnamefont {B.}~\bibnamefont {Szilagyi}}, \ and\ \bibinfo {author}
  {\bibfnamefont {J.}~\bibnamefont {Winicour}},\ }in\ \href@noop {} {\emph
  {\bibinfo {booktitle} {On the Black Hole Trail}}},\ \bibinfo {editor} {edited
  by\ \bibinfo {editor} {\bibfnamefont {B.}~\bibnamefont {Bhawal}}\ and\
  \bibinfo {editor} {\bibfnamefont {B.~R.}\ \bibnamefont {Iyer}}}\ (\bibinfo
  {publisher} {Kluwer, Dordrecht},\ \bibinfo {year} {1998})\ Chap.~\bibinfo
  {chapter} {24}, p.\ \bibinfo {pages} {383}\BibitemShut {NoStop}%
\bibitem [{\citenamefont {Giannakopoulos}\ \emph {et~al.}(2020)\citenamefont
  {Giannakopoulos}, \citenamefont {Hilditch},\ and\ \citenamefont
  {Zilh\~ao}}]{GiaHilZil20}%
  \BibitemOpen
  \bibfield  {author} {\bibinfo {author} {\bibfnamefont {T.}~\bibnamefont
  {Giannakopoulos}}, \bibinfo {author} {\bibfnamefont {D.}~\bibnamefont
  {Hilditch}}, \ and\ \bibinfo {author} {\bibfnamefont {M.}~\bibnamefont
  {Zilh\~ao}},\ }\href {\doibase 10.1103/PhysRevD.102.064035} {\bibfield
  {journal} {\bibinfo  {journal} {Phys. Rev. D}\ }\textbf {\bibinfo {volume}
  {102}},\ \bibinfo {pages} {064035} (\bibinfo {year} {2020})},\ \Eprint
  {http://arxiv.org/abs/2007.06419} {arXiv:2007.06419 [gr-qc]} \BibitemShut
  {NoStop}%
\bibitem [{\citenamefont {Giannakopoulos}\ \emph {et~al.}(2022)\citenamefont
  {Giannakopoulos}, \citenamefont {Bishop}, \citenamefont {Hilditch},
  \citenamefont {Pollney},\ and\ \citenamefont {Zilh\~ao}}]{GiaBisHil21}%
  \BibitemOpen
  \bibfield  {author} {\bibinfo {author} {\bibfnamefont {T.}~\bibnamefont
  {Giannakopoulos}}, \bibinfo {author} {\bibfnamefont {N.~T.}\ \bibnamefont
  {Bishop}}, \bibinfo {author} {\bibfnamefont {D.}~\bibnamefont {Hilditch}},
  \bibinfo {author} {\bibfnamefont {D.}~\bibnamefont {Pollney}}, \ and\
  \bibinfo {author} {\bibfnamefont {M.}~\bibnamefont {Zilh\~ao}},\ }\href
  {\doibase 10.1103/PhysRevD.105.084055} {\bibfield  {journal} {\bibinfo
  {journal} {Phys. Rev. D}\ }\textbf {\bibinfo {volume} {105}},\ \bibinfo
  {pages} {084055} (\bibinfo {year} {2022})},\ \Eprint
  {http://arxiv.org/abs/2111.14794} {arXiv:2111.14794 [gr-qc]} \BibitemShut
  {NoStop}%
\bibitem [{\citenamefont {Bondi}\ \emph {et~al.}(1962)\citenamefont {Bondi},
  \citenamefont {van~der Burg},\ and\ \citenamefont {Metzner}}]{BonBurMet62a}%
  \BibitemOpen
  \bibfield  {author} {\bibinfo {author} {\bibfnamefont {H.}~\bibnamefont
  {Bondi}}, \bibinfo {author} {\bibfnamefont {M.~G.~J.}\ \bibnamefont {van~der
  Burg}}, \ and\ \bibinfo {author} {\bibfnamefont {A.~W.~K.}\ \bibnamefont
  {Metzner}},\ }\href {\doibase 10.1098/rspa.1962.0161} {\bibfield  {journal}
  {\bibinfo  {journal} {Proc. R. Soc. London}\ }\textbf {\bibinfo {volume}
  {A269}},\ \bibinfo {pages} {21} (\bibinfo {year} {1962})}\BibitemShut
  {NoStop}%
\bibitem [{\citenamefont {Cao}\ and\ \citenamefont {He}(2013)}]{CaoHe13}%
  \BibitemOpen
  \bibfield  {author} {\bibinfo {author} {\bibfnamefont {Z.}~\bibnamefont
  {Cao}}\ and\ \bibinfo {author} {\bibfnamefont {X.}~\bibnamefont {He}},\
  }\href {\doibase 10.1103/PhysRevD.88.104002} {\bibfield  {journal} {\bibinfo
  {journal} {Phys. Rev. D}\ }\textbf {\bibinfo {volume} {88}},\ \bibinfo
  {pages} {104002} (\bibinfo {year} {2013})}\BibitemShut {NoStop}%
\bibitem [{\citenamefont {Pretorius}(2005)}]{Pre05}%
  \BibitemOpen
  \bibfield  {author} {\bibinfo {author} {\bibfnamefont {F.}~\bibnamefont
  {Pretorius}},\ }\href@noop {} {\bibfield  {journal} {\bibinfo  {journal}
  {Phys. Rev. Lett.}\ }\textbf {\bibinfo {volume} {95}},\ \bibinfo {pages}
  {121101} (\bibinfo {year} {2005})},\ \Eprint
  {http://arxiv.org/abs/gr-qc/0507014} {gr-qc/0507014} \BibitemShut {NoStop}%
\bibitem [{\citenamefont {Lindblom}\ \emph {et~al.}(2006)\citenamefont
  {Lindblom}, \citenamefont {Scheel}, \citenamefont {Kidder}, \citenamefont
  {Owen},\ and\ \citenamefont {Rinne}}]{LinSchKid05}%
  \BibitemOpen
  \bibfield  {author} {\bibinfo {author} {\bibfnamefont {L.}~\bibnamefont
  {Lindblom}}, \bibinfo {author} {\bibfnamefont {M.~A.}\ \bibnamefont
  {Scheel}}, \bibinfo {author} {\bibfnamefont {L.~E.}\ \bibnamefont {Kidder}},
  \bibinfo {author} {\bibfnamefont {R.}~\bibnamefont {Owen}}, \ and\ \bibinfo
  {author} {\bibfnamefont {O.}~\bibnamefont {Rinne}},\ }\href@noop {}
  {\bibfield  {journal} {\bibinfo  {journal} {Class. Quant. Grav.}\ }\textbf
  {\bibinfo {volume} {23}},\ \bibinfo {pages} {S447} (\bibinfo {year}
  {2006})},\ \Eprint {http://arxiv.org/abs/gr-qc/0512093} {gr-qc/0512093}
  \BibitemShut {NoStop}%
\bibitem [{\citenamefont {Lindblom}\ \emph {et~al.}(2008)\citenamefont
  {Lindblom}, \citenamefont {Matthews}, \citenamefont {Rinne},\ and\
  \citenamefont {Scheel}}]{LinMatRin07}%
  \BibitemOpen
  \bibfield  {author} {\bibinfo {author} {\bibfnamefont {L.}~\bibnamefont
  {Lindblom}}, \bibinfo {author} {\bibfnamefont {K.~D.}\ \bibnamefont
  {Matthews}}, \bibinfo {author} {\bibfnamefont {O.}~\bibnamefont {Rinne}}, \
  and\ \bibinfo {author} {\bibfnamefont {M.~A.}\ \bibnamefont {Scheel}},\
  }\href {\doibase 10.1103/PhysRevD.77.084001} {\bibfield  {journal} {\bibinfo
  {journal} {Phys. Rev.}\ }\textbf {\bibinfo {volume} {D77}},\ \bibinfo {pages}
  {084001} (\bibinfo {year} {2008})},\ \Eprint {http://arxiv.org/abs/0711.2084}
  {arXiv:0711.2084 [gr-qc]} \BibitemShut {NoStop}%
\bibitem [{\citenamefont {G{\'o}mez}\ \emph {et~al.}(1998)\citenamefont
  {G{\'o}mez}, \citenamefont {Lehner}, \citenamefont {Marsa}, \citenamefont
  {Winicour}, \citenamefont {Abrahams}, \citenamefont {Anderson}, \citenamefont
  {Anninos}, \citenamefont {Baumgarte}, \citenamefont {Bishop}, \citenamefont
  {Brandt}, \citenamefont {Browne}, \citenamefont {Camarda}, \citenamefont
  {Choptuik}, \citenamefont {Cook}, \citenamefont {Correll}, \citenamefont
  {Evans}, \citenamefont {Finn}, \citenamefont {Fox}, \citenamefont {Haupt},
  \citenamefont {Huq}, \citenamefont {Kidder}, \citenamefont {Klasky},
  \citenamefont {Laguna}, \citenamefont {Landry}, \citenamefont {Lenaghan},
  \citenamefont {Masso}, \citenamefont {Matzner}, \citenamefont {Mitra},
  \citenamefont {Papadopoulos}, \citenamefont {Parashar}, \citenamefont
  {Rezzolla}, \citenamefont {Rupright}, \citenamefont {Saied}, \citenamefont
  {Saylor}, \citenamefont {Scheel}, \citenamefont {Seidel}, \citenamefont
  {Shapiro}, \citenamefont {Shoemaker}, \citenamefont {Smarr}, \citenamefont
  {Szil{\'a}gyi}, \citenamefont {Teukolsky}, \citenamefont {van Putten},
  \citenamefont {Walker},\ and\ \citenamefont {{York Jr}}}]{GomLehMar98a}%
  \BibitemOpen
  \bibfield  {author} {\bibinfo {author} {\bibfnamefont {R.}~\bibnamefont
  {G{\'o}mez}}, \bibinfo {author} {\bibfnamefont {L.}~\bibnamefont {Lehner}},
  \bibinfo {author} {\bibfnamefont {R.}~\bibnamefont {Marsa}}, \bibinfo
  {author} {\bibfnamefont {J.}~\bibnamefont {Winicour}}, \bibinfo {author}
  {\bibfnamefont {A.~M.}\ \bibnamefont {Abrahams}}, \bibinfo {author}
  {\bibfnamefont {A.}~\bibnamefont {Anderson}}, \bibinfo {author}
  {\bibfnamefont {P.}~\bibnamefont {Anninos}}, \bibinfo {author} {\bibfnamefont
  {T.~W.}\ \bibnamefont {Baumgarte}}, \bibinfo {author} {\bibfnamefont {N.~T.}\
  \bibnamefont {Bishop}}, \bibinfo {author} {\bibfnamefont {S.~R.}\
  \bibnamefont {Brandt}}, \bibinfo {author} {\bibfnamefont {J.~C.}\
  \bibnamefont {Browne}}, \bibinfo {author} {\bibfnamefont {K.}~\bibnamefont
  {Camarda}}, \bibinfo {author} {\bibfnamefont {M.~W.}\ \bibnamefont
  {Choptuik}}, \bibinfo {author} {\bibfnamefont {G.~B.}\ \bibnamefont {Cook}},
  \bibinfo {author} {\bibfnamefont {R.}~\bibnamefont {Correll}}, \bibinfo
  {author} {\bibfnamefont {C.~R.}\ \bibnamefont {Evans}}, \bibinfo {author}
  {\bibfnamefont {L.~S.}\ \bibnamefont {Finn}}, \bibinfo {author}
  {\bibfnamefont {G.~C.}\ \bibnamefont {Fox}}, \bibinfo {author} {\bibfnamefont
  {T.}~\bibnamefont {Haupt}}, \bibinfo {author} {\bibfnamefont {M.~F.}\
  \bibnamefont {Huq}}, \bibinfo {author} {\bibfnamefont {L.~E.}\ \bibnamefont
  {Kidder}}, \bibinfo {author} {\bibfnamefont {S.~A.}\ \bibnamefont {Klasky}},
  \bibinfo {author} {\bibfnamefont {P.}~\bibnamefont {Laguna}}, \bibinfo
  {author} {\bibfnamefont {W.}~\bibnamefont {Landry}}, \bibinfo {author}
  {\bibfnamefont {J.}~\bibnamefont {Lenaghan}}, \bibinfo {author}
  {\bibfnamefont {J.}~\bibnamefont {Masso}}, \bibinfo {author} {\bibfnamefont
  {R.~A.}\ \bibnamefont {Matzner}}, \bibinfo {author} {\bibfnamefont
  {S.}~\bibnamefont {Mitra}}, \bibinfo {author} {\bibfnamefont
  {P.}~\bibnamefont {Papadopoulos}}, \bibinfo {author} {\bibfnamefont
  {M.}~\bibnamefont {Parashar}}, \bibinfo {author} {\bibfnamefont
  {L.}~\bibnamefont {Rezzolla}}, \bibinfo {author} {\bibfnamefont {M.~E.}\
  \bibnamefont {Rupright}}, \bibinfo {author} {\bibfnamefont {F.}~\bibnamefont
  {Saied}}, \bibinfo {author} {\bibfnamefont {P.~E.}\ \bibnamefont {Saylor}},
  \bibinfo {author} {\bibfnamefont {M.~A.}\ \bibnamefont {Scheel}}, \bibinfo
  {author} {\bibfnamefont {E.}~\bibnamefont {Seidel}}, \bibinfo {author}
  {\bibfnamefont {S.~L.}\ \bibnamefont {Shapiro}}, \bibinfo {author}
  {\bibfnamefont {D.}~\bibnamefont {Shoemaker}}, \bibinfo {author}
  {\bibfnamefont {L.}~\bibnamefont {Smarr}}, \bibinfo {author} {\bibfnamefont
  {B.}~\bibnamefont {Szil{\'a}gyi}}, \bibinfo {author} {\bibfnamefont {S.~A.}\
  \bibnamefont {Teukolsky}}, \bibinfo {author} {\bibfnamefont {M.~H. P.~M.}\
  \bibnamefont {van Putten}}, \bibinfo {author} {\bibfnamefont
  {P.}~\bibnamefont {Walker}}, \ and\ \bibinfo {author} {\bibfnamefont {J.~W.}\
  \bibnamefont {{York Jr}}},\ }\href@noop {} {\bibfield  {journal} {\bibinfo
  {journal} {Phys. Rev. Lett.}\ }\textbf {\bibinfo {volume} {80}},\ \bibinfo
  {pages} {3915} (\bibinfo {year} {1998})},\ \Eprint
  {http://arxiv.org/abs/gr-qc/9801069} {gr-qc/9801069} \BibitemShut {NoStop}%
\bibitem [{\citenamefont {Papadopoulos}\ and\ \citenamefont
  {Font}(1999)}]{PapFon99}%
  \BibitemOpen
  \bibfield  {author} {\bibinfo {author} {\bibfnamefont {P.}~\bibnamefont
  {Papadopoulos}}\ and\ \bibinfo {author} {\bibfnamefont {J.~A.}\ \bibnamefont
  {Font}},\ }\href {\doibase 10.1103/PhysRevD.61.024015} {\bibfield  {journal}
  {\bibinfo  {journal} {Phys. Rev. D}\ }\textbf {\bibinfo {volume} {61}},\
  \bibinfo {pages} {024015} (\bibinfo {year} {1999})}\BibitemShut {NoStop}%
\bibitem [{\citenamefont {Siebel}\ \emph {et~al.}(2002)\citenamefont {Siebel},
  \citenamefont {Font}, \citenamefont {M{\"u}ller},\ and\ \citenamefont
  {Papadopoulos}}]{SieFonMul02}%
  \BibitemOpen
  \bibfield  {author} {\bibinfo {author} {\bibfnamefont {F.}~\bibnamefont
  {Siebel}}, \bibinfo {author} {\bibfnamefont {J.~A.}\ \bibnamefont {Font}},
  \bibinfo {author} {\bibfnamefont {E.}~\bibnamefont {M{\"u}ller}}, \ and\
  \bibinfo {author} {\bibfnamefont {P.}~\bibnamefont {Papadopoulos}},\
  }\href@noop {} {\bibfield  {journal} {\bibinfo  {journal} {Phys. Rev. D}\
  }\textbf {\bibinfo {volume} {65}},\ \bibinfo {pages} {064038} (\bibinfo
  {year} {2002})}\BibitemShut {NoStop}%
\bibitem [{\citenamefont {Garfinkle}(1995)}]{Gar95}%
  \BibitemOpen
  \bibfield  {author} {\bibinfo {author} {\bibfnamefont {D.}~\bibnamefont
  {Garfinkle}},\ }\href@noop {} {\bibfield  {journal} {\bibinfo  {journal}
  {Phys. Rev. D}\ }\textbf {\bibinfo {volume} {51}},\ \bibinfo {pages} {5558}
  (\bibinfo {year} {1995})}\BibitemShut {NoStop}%
\bibitem [{\citenamefont {Crespo}\ \emph {et~al.}(2019)\citenamefont {Crespo},
  \citenamefont {de~Oliveira},\ and\ \citenamefont {Winicour}}]{CreOliWin19}%
  \BibitemOpen
  \bibfield  {author} {\bibinfo {author} {\bibfnamefont {J.~A.}\ \bibnamefont
  {Crespo}}, \bibinfo {author} {\bibfnamefont {H.~P.}\ \bibnamefont
  {de~Oliveira}}, \ and\ \bibinfo {author} {\bibfnamefont {J.}~\bibnamefont
  {Winicour}},\ }\href {\doibase 10.1103/PhysRevD.100.104017} {\bibfield
  {journal} {\bibinfo  {journal} {Phys. Rev.}\ }\textbf {\bibinfo {volume}
  {D100}},\ \bibinfo {pages} {104017} (\bibinfo {year} {2019})},\ \Eprint
  {http://arxiv.org/abs/1910.03439} {arXiv:1910.03439 [gr-qc]} \BibitemShut
  {NoStop}%
\bibitem [{\citenamefont {Gundlach}\ \emph {et~al.}(2019)\citenamefont
  {Gundlach}, \citenamefont {Baumgarte},\ and\ \citenamefont
  {Hilditch}}]{GunBauHil19}%
  \BibitemOpen
  \bibfield  {author} {\bibinfo {author} {\bibfnamefont {C.}~\bibnamefont
  {Gundlach}}, \bibinfo {author} {\bibfnamefont {T.~W.}\ \bibnamefont
  {Baumgarte}}, \ and\ \bibinfo {author} {\bibfnamefont {D.}~\bibnamefont
  {Hilditch}},\ }\href {\doibase 10.1103/PhysRevD.100.104010} {\bibfield
  {journal} {\bibinfo  {journal} {Phys. Rev. D}\ }\textbf {\bibinfo {volume}
  {100}},\ \bibinfo {pages} {104010} (\bibinfo {year} {2019})},\ \Eprint
  {http://arxiv.org/abs/1908.05971} {arXiv:1908.05971 [gr-qc]} \BibitemShut
  {NoStop}%
\bibitem [{\citenamefont {Siebel}\ \emph {et~al.}(2003)\citenamefont {Siebel},
  \citenamefont {Font}, \citenamefont {M{\"u}ller},\ and\ \citenamefont
  {Papadopoulos}}]{SieFonMue03}%
  \BibitemOpen
  \bibfield  {author} {\bibinfo {author} {\bibfnamefont {F.}~\bibnamefont
  {Siebel}}, \bibinfo {author} {\bibfnamefont {J.~A.}\ \bibnamefont {Font}},
  \bibinfo {author} {\bibfnamefont {E.}~\bibnamefont {M{\"u}ller}}, \ and\
  \bibinfo {author} {\bibfnamefont {P.}~\bibnamefont {Papadopoulos}},\
  }\href@noop {} {\bibfield  {journal} {\bibinfo  {journal} {Phys. Rev. D}\
  }\textbf {\bibinfo {volume} {67}},\ \bibinfo {pages} {124018} (\bibinfo
  {year} {2003})}\BibitemShut {NoStop}%
\bibitem [{\citenamefont {Alcoforado}\ \emph {et~al.}(2021)\citenamefont
  {Alcoforado}, \citenamefont {Barreto},\ and\ \citenamefont
  {de~Oliveira}}]{AlcBarOli21}%
  \BibitemOpen
  \bibfield  {author} {\bibinfo {author} {\bibfnamefont {M.~A.}\ \bibnamefont
  {Alcoforado}}, \bibinfo {author} {\bibfnamefont {W.~O.}\ \bibnamefont
  {Barreto}}, \ and\ \bibinfo {author} {\bibfnamefont {H.~P.}\ \bibnamefont
  {de~Oliveira}},\ }\href {\doibase 10.1142/S0218271822500286} {\  (\bibinfo
  {year} {2021}),\ 10.1142/S0218271822500286},\ \Eprint
  {http://arxiv.org/abs/2110.09640} {arXiv:2110.09640 [gr-qc]} \BibitemShut
  {NoStop}%
\bibitem [{\citenamefont {Santos-Olivan}\ and\ \citenamefont
  {Sopuerta}(2016)}]{OliSop16}%
  \BibitemOpen
  \bibfield  {author} {\bibinfo {author} {\bibfnamefont {D.}~\bibnamefont
  {Santos-Olivan}}\ and\ \bibinfo {author} {\bibfnamefont {C.~F.}\ \bibnamefont
  {Sopuerta}},\ }\href {\doibase 10.1103/PhysRevLett.116.041101} {\bibfield
  {journal} {\bibinfo  {journal} {Phys. Rev. Lett.}\ }\textbf {\bibinfo
  {volume} {116}},\ \bibinfo {pages} {041101} (\bibinfo {year} {2016})},\
  \Eprint {http://arxiv.org/abs/1511.04344} {arXiv:1511.04344 [gr-qc]}
  \BibitemShut {NoStop}%
\bibitem [{\citenamefont {Santos-Olivan}(2017)}]{Oli17b}%
  \BibitemOpen
  \bibfield  {author} {\bibinfo {author} {\bibfnamefont {D.}~\bibnamefont
  {Santos-Olivan}},\ }\emph {\bibinfo {title} {{Numerical Relativity studies in
  Anti-de Sitter spacetimes: Gravitational Collapse and the AdS/CFT
  correspondence}}},\ \href@noop {} {Ph.D. thesis},\ \bibinfo  {school}
  {Barcelona U.} (\bibinfo {year} {2017})\BibitemShut {NoStop}%
\bibitem [{\citenamefont {Kreiss}\ and\ \citenamefont
  {Lorenz}(1989)}]{KreLor89}%
  \BibitemOpen
  \bibfield  {author} {\bibinfo {author} {\bibfnamefont {H.-O.}\ \bibnamefont
  {Kreiss}}\ and\ \bibinfo {author} {\bibfnamefont {J.}~\bibnamefont
  {Lorenz}},\ }\href@noop {} {\emph {\bibinfo {title} {Initial-boundary value
  problems and the {N}avier-{S}tokes equations}}}\ (\bibinfo  {publisher}
  {Academic Press},\ \bibinfo {address} {New York},\ \bibinfo {year}
  {1989})\BibitemShut {NoStop}%
\bibitem [{\citenamefont {Sarbach}\ and\ \citenamefont
  {Tiglio}(2012)}]{SarTig12}%
  \BibitemOpen
  \bibfield  {author} {\bibinfo {author} {\bibfnamefont {O.}~\bibnamefont
  {Sarbach}}\ and\ \bibinfo {author} {\bibfnamefont {M.}~\bibnamefont
  {Tiglio}},\ }\href {http://www.livingreviews.org/lrr-2012-9} {\bibfield
  {journal} {\bibinfo  {journal} {Living Reviews in Relativity}\ }\textbf
  {\bibinfo {volume} {15}} (\bibinfo {year} {2012})},\ \Eprint
  {http://arxiv.org/abs/1203.6443} {arXiv:1203.6443 [gr-qc]} \BibitemShut
  {NoStop}%
\bibitem [{\citenamefont {Frittelli}\ and\ \citenamefont
  {Lehner}(1999)}]{FriLeh99}%
  \BibitemOpen
  \bibfield  {author} {\bibinfo {author} {\bibfnamefont {S.}~\bibnamefont
  {Frittelli}}\ and\ \bibinfo {author} {\bibfnamefont {L.}~\bibnamefont
  {Lehner}},\ }\href {\doibase 10.1103/PhysRevD.59.084012} {\bibfield
  {journal} {\bibinfo  {journal} {Phys. Rev. D}\ }\textbf {\bibinfo {volume}
  {59}},\ \bibinfo {pages} {084012} (\bibinfo {year} {1999})}\BibitemShut
  {NoStop}%
\bibitem [{\citenamefont {Gomez}\ and\ \citenamefont
  {Frittelli}(2003)}]{GomFri03}%
  \BibitemOpen
  \bibfield  {author} {\bibinfo {author} {\bibfnamefont {R.}~\bibnamefont
  {Gomez}}\ and\ \bibinfo {author} {\bibfnamefont {S.}~\bibnamefont
  {Frittelli}},\ }\href {\doibase 10.1103/PhysRevD.68.084013} {\bibfield
  {journal} {\bibinfo  {journal} {Phys. Rev. D}\ }\textbf {\bibinfo {volume}
  {68}},\ \bibinfo {pages} {084013} (\bibinfo {year} {2003})},\ \Eprint
  {http://arxiv.org/abs/gr-qc/0303104} {arXiv:gr-qc/0303104} \BibitemShut
  {NoStop}%
\bibitem [{\citenamefont {Frittelli}(2005)}]{Fri04}%
  \BibitemOpen
  \bibfield  {author} {\bibinfo {author} {\bibfnamefont {S.}~\bibnamefont
  {Frittelli}},\ }\href {\doibase 10.1103/PhysRevD.71.024021} {\bibfield
  {journal} {\bibinfo  {journal} {Phys. Rev. D}\ }\textbf {\bibinfo {volume}
  {71}},\ \bibinfo {pages} {024021} (\bibinfo {year} {2005})},\ \Eprint
  {http://arxiv.org/abs/gr-qc/0408035x} {arXiv:gr-qc/0408035x} \BibitemShut
  {NoStop}%
\bibitem [{\citenamefont {R\'acz}(2014)}]{Rac13}%
  \BibitemOpen
  \bibfield  {author} {\bibinfo {author} {\bibfnamefont {I.}~\bibnamefont
  {R\'acz}},\ }\href {\doibase 10.1088/0264-9381/31/3/035006} {\bibfield
  {journal} {\bibinfo  {journal} {Class. Quant. Grav.}\ }\textbf {\bibinfo
  {volume} {31}},\ \bibinfo {pages} {035006} (\bibinfo {year} {2014})},\
  \Eprint {http://arxiv.org/abs/1307.1683} {arXiv:1307.1683 [gr-qc]}
  \BibitemShut {NoStop}%
\bibitem [{\citenamefont {Cabet}\ \emph {et~al.}(2014)\citenamefont {Cabet},
  \citenamefont {Chru\'sciel},\ and\ \citenamefont {Wafo}}]{CabChrTag14}%
  \BibitemOpen
  \bibfield  {author} {\bibinfo {author} {\bibfnamefont {A.}~\bibnamefont
  {Cabet}}, \bibinfo {author} {\bibfnamefont {P.~T.}\ \bibnamefont
  {Chru\'sciel}}, \ and\ \bibinfo {author} {\bibfnamefont {R.~T.}\ \bibnamefont
  {Wafo}},\ }\href@noop {} {\  (\bibinfo {year} {2014})},\ \Eprint
  {http://arxiv.org/abs/1406.3009} {arXiv:1406.3009 [gr-qc]} \BibitemShut
  {NoStop}%
\bibitem [{\citenamefont {Hilditch}\ \emph {et~al.}(2020)\citenamefont
  {Hilditch}, \citenamefont {Valiente~Kroon},\ and\ \citenamefont
  {Zhao}}]{HilValZha19}%
  \BibitemOpen
  \bibfield  {author} {\bibinfo {author} {\bibfnamefont {D.}~\bibnamefont
  {Hilditch}}, \bibinfo {author} {\bibfnamefont {J.~A.}\ \bibnamefont
  {Valiente~Kroon}}, \ and\ \bibinfo {author} {\bibfnamefont {P.}~\bibnamefont
  {Zhao}},\ }\href {\doibase 10.1007/s10714-020-02747-2} {\bibfield  {journal}
  {\bibinfo  {journal} {Gen. Rel. Grav.}\ }\textbf {\bibinfo {volume} {52}},\
  \bibinfo {pages} {99} (\bibinfo {year} {2020})},\ \Eprint
  {http://arxiv.org/abs/1911.00047} {arXiv:1911.00047 [gr-qc]} \BibitemShut
  {NoStop}%
\bibitem [{\citenamefont {Ripley}(2021)}]{Rip21}%
  \BibitemOpen
  \bibfield  {author} {\bibinfo {author} {\bibfnamefont {J.~L.}\ \bibnamefont
  {Ripley}},\ }\href {\doibase 10.1063/5.0055561} {\bibfield  {journal}
  {\bibinfo  {journal} {J. Math. Phys.}\ }\textbf {\bibinfo {volume} {62}},\
  \bibinfo {pages} {062501} (\bibinfo {year} {2021})},\ \Eprint
  {http://arxiv.org/abs/2104.09972} {arXiv:2104.09972 [gr-qc]} \BibitemShut
  {NoStop}%
\bibitem [{\citenamefont {{Bezanson, Jeff and Edelman, Alan and Karpinski,
  Stefan and Shah, Viral B}}(2017)}]{BezEdeKar17}%
  \BibitemOpen
  \bibfield  {author} {\bibinfo {author} {\bibnamefont {{Bezanson, Jeff and
  Edelman, Alan and Karpinski, Stefan and Shah, Viral B}}},\ }\href {\doibase
  10.1137/141000671} {\bibfield  {journal} {\bibinfo  {journal} {SIAM Review}\
  }\textbf {\bibinfo {volume} {59}},\ \bibinfo {pages} {65} (\bibinfo {year}
  {2017})}\BibitemShut {NoStop}%
\bibitem [{\citenamefont {Giannakopoulos}\ \emph
  {et~al.}(2023{\natexlab{a}})\citenamefont {Giannakopoulos}, \citenamefont
  {Bishop}, \citenamefont {Hilditch}, \citenamefont {Pollney},\ and\
  \citenamefont {Zilh\~ao}}]{GiaBisHil23_github}%
  \BibitemOpen
  \bibfield  {author} {\bibinfo {author} {\bibfnamefont {T.}~\bibnamefont
  {Giannakopoulos}}, \bibinfo {author} {\bibfnamefont {N.~T.}\ \bibnamefont
  {Bishop}}, \bibinfo {author} {\bibfnamefont {D.}~\bibnamefont {Hilditch}},
  \bibinfo {author} {\bibfnamefont {D.}~\bibnamefont {Pollney}}, \ and\
  \bibinfo {author} {\bibfnamefont {M.}~\bibnamefont {Zilh\~ao}},\ }\href
  {https://github.com/ThanasisGiannakopoulos/model_CCE_CCM_public} {\enquote
  {\bibinfo {title} {{Model CCE \& CCM (open code)}},}\ }\bibinfo
  {howpublished}
  {\url{https://github.com/ThanasisGiannakopoulos/model_CCE_CCM_public}}
  (\bibinfo {year} {2023}{\natexlab{a}})\BibitemShut {NoStop}%
\bibitem [{\citenamefont {Calabrese}\ \emph {et~al.}(2002)\citenamefont
  {Calabrese}, \citenamefont {Pullin}, \citenamefont {Sarbach},\ and\
  \citenamefont {Tiglio}}]{CalPulSar02}%
  \BibitemOpen
  \bibfield  {author} {\bibinfo {author} {\bibfnamefont {G.}~\bibnamefont
  {Calabrese}}, \bibinfo {author} {\bibfnamefont {J.}~\bibnamefont {Pullin}},
  \bibinfo {author} {\bibfnamefont {O.}~\bibnamefont {Sarbach}}, \ and\
  \bibinfo {author} {\bibfnamefont {M.}~\bibnamefont {Tiglio}},\ }\href@noop {}
  {\bibfield  {journal} {\bibinfo  {journal} {Phys. Rev. D}\ }\textbf {\bibinfo
  {volume} {66}},\ \bibinfo {pages} {041501} (\bibinfo {year} {2002})},\
  \Eprint {http://arxiv.org/abs/gr-qc/0207018} {gr-qc/0207018} \BibitemShut
  {NoStop}%
\bibitem [{\citenamefont {Calabrese}\ \emph {et~al.}(2006)\citenamefont
  {Calabrese}, \citenamefont {Hinder},\ and\ \citenamefont
  {Husa}}]{CalHinHus05}%
  \BibitemOpen
  \bibfield  {author} {\bibinfo {author} {\bibfnamefont {G.}~\bibnamefont
  {Calabrese}}, \bibinfo {author} {\bibfnamefont {I.}~\bibnamefont {Hinder}}, \
  and\ \bibinfo {author} {\bibfnamefont {S.}~\bibnamefont {Husa}},\ }\href@noop
  {} {\bibfield  {journal} {\bibinfo  {journal} {J. Comp. Phys.}\ }\textbf
  {\bibinfo {volume} {218}},\ \bibinfo {pages} {607} (\bibinfo {year}
  {2006})},\ \Eprint {http://arxiv.org/abs/gr-qc/0503056} {gr-qc/0503056}
  \BibitemShut {NoStop}%
\bibitem [{\citenamefont {Cao}\ and\ \citenamefont
  {Hilditch}(2012)}]{CaoHil11}%
  \BibitemOpen
  \bibfield  {author} {\bibinfo {author} {\bibfnamefont {Z.}~\bibnamefont
  {Cao}}\ and\ \bibinfo {author} {\bibfnamefont {D.}~\bibnamefont {Hilditch}},\
  }\href {\doibase 10.1103/PhysRevD.85.124032} {\bibfield  {journal} {\bibinfo
  {journal} {Phys. Rev. D}\ }\textbf {\bibinfo {volume} {85}},\ \bibinfo
  {pages} {124032} (\bibinfo {year} {2012})},\ \Eprint
  {http://arxiv.org/abs/1111.2177} {arXiv:1111.2177 [gr-qc]} \BibitemShut
  {NoStop}%
\bibitem [{\citenamefont {Giannakopoulos}\ \emph
  {et~al.}(2023{\natexlab{b}})\citenamefont {Giannakopoulos}, \citenamefont
  {Bishop}, \citenamefont {Hilditch}, \citenamefont {Pollney},\ and\
  \citenamefont {Zilh\~ao}}]{GiaBisHil23_zenodo}%
  \BibitemOpen
  \bibfield  {author} {\bibinfo {author} {\bibfnamefont {T.}~\bibnamefont
  {Giannakopoulos}}, \bibinfo {author} {\bibfnamefont {N.~T.}\ \bibnamefont
  {Bishop}}, \bibinfo {author} {\bibfnamefont {D.}~\bibnamefont {Hilditch}},
  \bibinfo {author} {\bibfnamefont {D.}~\bibnamefont {Pollney}}, \ and\
  \bibinfo {author} {\bibfnamefont {M.}~\bibnamefont {Zilh\~ao}},\ }\href
  {\doibase 10.5281/zenodo.7981429} {\enquote {\bibinfo {title} {{Numerical
  convergence of model Cauchy-Characteristic Extraction and Matching
  (data)}},}\ }\bibinfo {howpublished}
  {\url{https://zenodo.org/record/7981429}} (\bibinfo {year}
  {2023}{\natexlab{b}})\BibitemShut {NoStop}%
\bibitem [{\citenamefont {Stewart}\ and\ \citenamefont
  {Friedrich}(1982)}]{SteFri82}%
  \BibitemOpen
  \bibfield  {author} {\bibinfo {author} {\bibfnamefont {J.~M.}\ \bibnamefont
  {Stewart}}\ and\ \bibinfo {author} {\bibfnamefont {H.}~\bibnamefont
  {Friedrich}},\ }\href {\doibase 10.1098/rspa.1982.0166} {\bibfield  {journal}
  {\bibinfo  {journal} {Proc. Roy. Soc. Lond. A}\ }\textbf {\bibinfo {volume}
  {384}},\ \bibinfo {pages} {427} (\bibinfo {year} {1982})}\BibitemShut
  {NoStop}%
\bibitem [{\citenamefont {Moncrief}\ and\ \citenamefont
  {Rinne}(2009)}]{MonRin08}%
  \BibitemOpen
  \bibfield  {author} {\bibinfo {author} {\bibfnamefont {V.}~\bibnamefont
  {Moncrief}}\ and\ \bibinfo {author} {\bibfnamefont {O.}~\bibnamefont
  {Rinne}},\ }\href {\doibase 10.1088/0264-9381/26/12/125010} {\bibfield
  {journal} {\bibinfo  {journal} {Class.Quant.Grav.}\ }\textbf {\bibinfo
  {volume} {26}},\ \bibinfo {pages} {125010} (\bibinfo {year} {2009})},\
  \Eprint {http://arxiv.org/abs/0811.4109} {arXiv:0811.4109 [gr-qc]}
  \BibitemShut {NoStop}%
\bibitem [{\citenamefont {Zenginoglu}(2011)}]{Zen10}%
  \BibitemOpen
  \bibfield  {author} {\bibinfo {author} {\bibfnamefont {A.}~\bibnamefont
  {Zenginoglu}},\ }\href {\doibase 10.1016/j.jcp.2010.12.016} {\bibfield
  {journal} {\bibinfo  {journal} {J. Comput. Phys.}\ }\textbf {\bibinfo
  {volume} {230}},\ \bibinfo {pages} {2286} (\bibinfo {year} {2011})},\ \Eprint
  {http://arxiv.org/abs/1008.3809} {arXiv:1008.3809 [math.NA]} \BibitemShut
  {NoStop}%
\bibitem [{\citenamefont {Bardeen}\ \emph {et~al.}(2011)\citenamefont
  {Bardeen}, \citenamefont {Sarbach},\ and\ \citenamefont
  {Buchman}}]{BarSarBuc11}%
  \BibitemOpen
  \bibfield  {author} {\bibinfo {author} {\bibfnamefont {J.~M.}\ \bibnamefont
  {Bardeen}}, \bibinfo {author} {\bibfnamefont {O.}~\bibnamefont {Sarbach}}, \
  and\ \bibinfo {author} {\bibfnamefont {L.~T.}\ \bibnamefont {Buchman}},\
  }\href {\doibase 10.1103/PhysRevD.83.104045} {\bibfield  {journal} {\bibinfo
  {journal} {Phys. Rev.}\ }\textbf {\bibinfo {volume} {D83}},\ \bibinfo {pages}
  {104045} (\bibinfo {year} {2011})},\ \Eprint {http://arxiv.org/abs/1101.5479}
  {arXiv:1101.5479 [gr-qc]} \BibitemShut {NoStop}%
\bibitem [{\citenamefont {Va{\~n}{\'o}-Vi{\~n}uales}\ \emph
  {et~al.}(2015)\citenamefont {Va{\~n}{\'o}-Vi{\~n}uales}, \citenamefont
  {Husa},\ and\ \citenamefont {Hilditch}}]{VanHusHil14}%
  \BibitemOpen
  \bibfield  {author} {\bibinfo {author} {\bibfnamefont {A.}~\bibnamefont
  {Va{\~n}{\'o}-Vi{\~n}uales}}, \bibinfo {author} {\bibfnamefont
  {S.}~\bibnamefont {Husa}}, \ and\ \bibinfo {author} {\bibfnamefont
  {D.}~\bibnamefont {Hilditch}},\ }\href {\doibase
  10.1088/0264-9381/32/17/175010} {\bibfield  {journal} {\bibinfo  {journal}
  {Class. Quant. Grav.}\ }\textbf {\bibinfo {volume} {32}},\ \bibinfo {pages}
  {175010} (\bibinfo {year} {2015})},\ \Eprint {http://arxiv.org/abs/1412.3827}
  {arXiv:1412.3827 [gr-qc]} \BibitemShut {NoStop}%
\bibitem [{\citenamefont {Hilditch}\ \emph {et~al.}(2018)\citenamefont
  {Hilditch}, \citenamefont {Harms}, \citenamefont {Bugner}, \citenamefont
  {R{\"u}ter},\ and\ \citenamefont {Br{\"u}gmann}}]{HilHarBug16}%
  \BibitemOpen
  \bibfield  {author} {\bibinfo {author} {\bibfnamefont {D.}~\bibnamefont
  {Hilditch}}, \bibinfo {author} {\bibfnamefont {E.}~\bibnamefont {Harms}},
  \bibinfo {author} {\bibfnamefont {M.}~\bibnamefont {Bugner}}, \bibinfo
  {author} {\bibfnamefont {H.}~\bibnamefont {R{\"u}ter}}, \ and\ \bibinfo
  {author} {\bibfnamefont {B.}~\bibnamefont {Br{\"u}gmann}},\ }\href {\doibase
  10.1088/1361-6382/aaa4ac} {\bibfield  {journal} {\bibinfo  {journal} {Class.
  Quant. Grav.}\ }\textbf {\bibinfo {volume} {35}},\ \bibinfo {pages} {055003}
  (\bibinfo {year} {2018})},\ \Eprint {http://arxiv.org/abs/1609.08949}
  {arXiv:1609.08949 [gr-qc]} \BibitemShut {NoStop}%
\bibitem [{\citenamefont {Ansorg}\ and\ \citenamefont
  {Panosso~Macedo}(2016)}]{AnsMac16}%
  \BibitemOpen
  \bibfield  {author} {\bibinfo {author} {\bibfnamefont {M.}~\bibnamefont
  {Ansorg}}\ and\ \bibinfo {author} {\bibfnamefont {R.}~\bibnamefont
  {Panosso~Macedo}},\ }\href@noop {} {\bibfield  {journal} {\bibinfo  {journal}
  {Phys. Rev.}\ }\textbf {\bibinfo {volume} {D93}},\ \bibinfo {pages} {124016}
  (\bibinfo {year} {2016})},\ \Eprint {http://arxiv.org/abs/1604.02261}
  {arXiv:1604.02261 [gr-qc]} \BibitemShut {NoStop}%
\bibitem [{\citenamefont {Vañó-Viñuales}\ and\ \citenamefont
  {Husa}(2018)}]{VanHus17}%
  \BibitemOpen
  \bibfield  {author} {\bibinfo {author} {\bibfnamefont {A.}~\bibnamefont
  {Vañó-Viñuales}}\ and\ \bibinfo {author} {\bibfnamefont {S.}~\bibnamefont
  {Husa}},\ }\href {\doibase 10.1088/1361-6382/aaa4e2} {\bibfield  {journal}
  {\bibinfo  {journal} {Class. Quant. Grav.}\ }\textbf {\bibinfo {volume}
  {35}},\ \bibinfo {pages} {045014} (\bibinfo {year} {2018})},\ \Eprint
  {http://arxiv.org/abs/1705.06298} {arXiv:1705.06298 [gr-qc]} \BibitemShut
  {NoStop}%
\bibitem [{\citenamefont {H{\"u}bner}(1999)}]{Hub99}%
  \BibitemOpen
  \bibfield  {author} {\bibinfo {author} {\bibfnamefont {P.}~\bibnamefont
  {H{\"u}bner}},\ }\href@noop {} {\bibfield  {journal} {\bibinfo  {journal}
  {Class. Quantum Grav.}\ }\textbf {\bibinfo {volume} {16}},\ \bibinfo {pages}
  {2823} (\bibinfo {year} {1999})}\BibitemShut {NoStop}%
\bibitem [{\citenamefont {Doulis}\ and\ \citenamefont
  {Frauendiener}(2017)}]{DouFra16}%
  \BibitemOpen
  \bibfield  {author} {\bibinfo {author} {\bibfnamefont {G.}~\bibnamefont
  {Doulis}}\ and\ \bibinfo {author} {\bibfnamefont {J.}~\bibnamefont
  {Frauendiener}},\ }\href {\doibase 10.1103/PhysRevD.95.024035} {\bibfield
  {journal} {\bibinfo  {journal} {Phys. Rev. D}\ }\textbf {\bibinfo {volume}
  {95}},\ \bibinfo {pages} {024035} (\bibinfo {year} {2017})},\ \Eprint
  {http://arxiv.org/abs/1609.03584} {arXiv:1609.03584 [gr-qc]} \BibitemShut
  {NoStop}%
\bibitem [{\citenamefont {Doulis}\ \emph {et~al.}(2019)\citenamefont {Doulis},
  \citenamefont {Frauendiener}, \citenamefont {Stevens},\ and\ \citenamefont
  {Whale}}]{DouFraSte19}%
  \BibitemOpen
  \bibfield  {author} {\bibinfo {author} {\bibfnamefont {G.}~\bibnamefont
  {Doulis}}, \bibinfo {author} {\bibfnamefont {J.}~\bibnamefont
  {Frauendiener}}, \bibinfo {author} {\bibfnamefont {C.}~\bibnamefont
  {Stevens}}, \ and\ \bibinfo {author} {\bibfnamefont {B.}~\bibnamefont
  {Whale}},\ }\href@noop {} {\  (\bibinfo {year} {2019})},\ \Eprint
  {http://arxiv.org/abs/1903.12482} {arXiv:1903.12482 [cs.MS]} \BibitemShut
  {NoStop}%
\end{thebibliography}%

\end{document}